\documentclass{aa}

\usepackage{mathtools}
\usepackage{graphicx}
\usepackage{caption}
\usepackage{subcaption}
\usepackage[export]{adjustbox}
\usepackage{multirow}
\usepackage[usenames,dvipsnames]{color}
\usepackage{xcolor}
\usepackage{float}
\usepackage{tikz,tikz-3dplot}\usetikzlibrary{calc}
\usepackage{pgfplots}
\usepackage{txfonts}
\usepackage{hyperref}
\hypersetup{colorlinks,breaklinks,
  linkcolor=[rgb]{0.368417, 0.506779, 0.709798},
  citecolor=[rgb]{0.368417, 0.506779, 0.709798},
  urlcolor=[rgb]{0.368417, 0.506779, 0.709798}}
\usepackage{enumerate}
\usepackage{natbib}

\graphicspath {{figures/}}

\usepackage{amstext}

\begin{document}

\title{The dark mass signature in the orbit of S2}
%\subtitle{I. Overviewing the $\kappa$-mechanism}

\author{G.~Hei\ss el \thanks{Corresponding author.} \and T.~Paumard \and G.~Perrin \and F. Vincent %\fnmsep\thanks{ }
  }

\institute{LESIA, Observatoire de Paris, Université PSL, CNRS, Sorbonne Université, Université de Paris, 5 place Jules Janssen, 92195 Meudon France\\ \email{\href{mailto:gernot.heissel@obspm.fr}{gernot.heissel@obspm.fr}} %\and
%\thanks{}
  }

\date{Received August 15, 2021; accepted November 6, 2021}
 
\abstract
% context (optional)
{The Schwarzschild precession of star S2, which orbits the massive black hole at the centre of the Milky Way, has recently been detected with the result of $\sim12\text{\,arcminutes}$ per orbit. The same study also improved the $1\sigma$ upper bound on a possibly present dark continuous extended mass distribution (e.g. faint stars, stellar remnants, stellar mass black holes, or dark matter) within the orbit of S2 to $\sim 4000\,M_\odot$. The secular (i.e. net) effect of an extended mass onto a stellar orbit is known as mass precession, and it runs counter to the Schwarzschild precession.}
% aims (mandatory)
{We explore a strategy for how the Schwarzschild and mass precessions can be separated from each other despite their secular interference, by pinpointing their signatures within a single orbit. From these insights, we then seek to assess the prospects for improving the dark mass constraints in the coming years.}
% methods (mandatory)
{We analysed the dependence of the osculating orbital elements and of the observables on true anomaly, and we compared these functions for models with and without extended mass. We then translated the maximum astrometric impacts within one orbit to detection thresholds given hypothetical data of different accuracies. These theoretical investigations were then supported and complemented by an extensive mock-data fitting analysis.}
% results (mandatory)
{We have four main results. 1. While the mass precession almost exclusively impacts the orbit in the apocentre half, the Schwarzschild precession almost exclusively impacts it in the pericentre half, allowing for a clear separation of the effects. 2. Data that are limited to the pericentre half are not sensitive to a dark mass, while data limited to the apocentre half are, but only to a limited extent. 3. A full orbit of data is required to substantially constrain a dark mass. 4. For a full orbit of astrometric and spectroscopic data, the astrometric component in the pericentre half plays the stronger role in constraining the dark mass than the astrometric data in the apocentre half. Furthermore, we determine the $1\sigma$ dark mass detection thresholds given different datasets on one full orbit. In particular, with a full orbit of data of $50\text{\,microarcseconds}$ (VLTI/GRAVITY) and $10\mathrm{\,km/s}$ (VLT/SINFONI) precision, the $1\sigma$ bound would improve to $\sim1000\,M_\odot$, for example.}
% conclusions (optional)
{The current upper dark mass bound of $\sim 4000\,M_\odot$ has mainly been obtained from a combination of GRAVITY and VLT/NACO astrometric data, as well as from SINFONI spectroscopic data, where the GRAVITY data were limited to the pericentre half. From our results 3 and 4, we know that all components were thereby crucial, but also that the GRAVITY data were dominant in the astrometric components in constraining the dark mass. From results 1 and 2, we deduce that a future population of the apocentre half with GRAVITY data points will substantially further improve the dark mass sensitivity of the dataset, and we note that at the time of publication, we already entered this regime. In the context of the larger picture, our analysis demonstrates how precession effects that interfere on secular timescales can clearly be distinguished from each other based on their distinct astrometric signatures within a single orbit. The extension of our analysis to the Lense-Thirring precession should thus be of value in order to assess future spin detection prospects for the galactic centre massive black hole.}


\maketitle

\section{Introduction}\label{S: Intro}

The motions of the S-stars, which make up the nuclear cluster at the centre of the Milky Way, have been monitored closely now for almost three decades with respect to astrometry and radial velocity \citep{EckartGenzel1996, Ghez+98, Ghez+03, Ghez+08, Schoedel+02, GillessenEtAl2009, GillessenEtAl2017}. Their trajectories revealed the existence of a dark compact $\sim4\times10^6M_\odot$ body at the cluster centre, in coincidence with the location of the radio source Sgr~$A^*$ \citep{Reid+09, Plewa+15, ReidBrunthaler04, ReidBrunthaler20}. Observations are currently in best agreement with the notion that the compact body is a massive black hole (MBH) \citep[Sect.~IV]{GenzelEtAl2010}. The combination of the high-eccentricity and low-pericentre distance of the orbit of star S2, as well as its sufficiently bright magnitude of $\sim14$ in the K band, predestined this object as the prime probe for the task of constraining the MBH parameters and observing relativistic effects \citep{Grould+17, Waisberg+18}. Flares of radiation near the innermost stable circular orbit of the MBH are of similar importance in this respect \citep{GRAVITY+18_flares}. With regard to S2, one milestone has been the detection of the gravitational redshift together with the special relativistic transverse Doppler shift during the time of pericentre passage in 2018 \citep{GRAVITY+18_redshift, SaidaEtAl2019, Do+19}. More recently, the Schwarzschild precession of the stellar orbit could be measured at $12$~arcminutes (arcmin) per orbital period by \citet{GRAVITY+20_Schwarzschild_prec}. Schwarzschild precession thereby refers to the relativistic pericentre advance in the orbital plane due to the mass of a non-spinning black hole \citep{Merritt2013, PoissonWill2014}.

In general, an in-plane orbital precession can also be caused by other influences. In particular, if the star is moving through a (dark) extended continuous mass distribution (in the following, simply dark mass or extended mass), for instance, one that consists of faint stars and stellar remnants or dark matter, then the acceleration imposed by this extended mass causes a pericentre shift even from a Newtonian perspective \citep[][Sect.~4.4]{JiangLin85, RubilarEckart2001, Merritt2013}. This phenomenon is commonly referred to as mass precession, and it has been demonstrated that it can be of the same order as or may even dominate the Schwarzschild precession for physically reasonable density distributions \citep[][p.~208]{RubilarEckart2001, Merritt2013}. For the case of the galactic centre of the Milky Way, the uncertainties of current observational data allow one to give upper bounds on a possibly present extended mass within the apocentre of S2. For instance, a $0.3\,\text{arcseconds (as)}$ Plummer profile mass is limited to below $0.1\%$ of the mass of the MBH, which translates into $\sim4000\,M_\odot$ \citep[][Sect.~4]{GRAVITY+20_Schwarzschild_prec}.\footnote{For previous bounds see \citep{Boehle+16, Gillessen+17, GRAVITY+18_redshift, Do+19}.} For completeness we note that the related search for compact (as opposed to extended) masses near the MBH is also of great interest, in particular regarding potential gravitational wave sources from extreme mass ratio inspirals for Laser Interferometer Space Antenna (LISA) \citep{AmaroSeoane+17, Gourgoulhon+19}.

There are several motivations to continue to study, constrain, and if possible, detect an extended mass within the orbit of S2. Firstly, from this we can directly infer details about the nature of the matter content in the immediate vicinity of the MBH. Secondly, more data and a refined analysis may help to place constraints on or rule out central object models based on continuous dark matter distributions that have been proposed in addition, or as alternatives, to the MBH paradigm \citep{Lacroix2018, GRAVITY+19_Scalar_fields, BecerraVergara+20, BecerraVergara+21a, BecerraVergara+21b, Nampalliwar21}. Finally, beyond the interest in its own right, it is also important to study the mass precession as perturbation for the measurement of other effects. On secular timescales, this concerns in particular other precession effects such as the Schwarzschild and Lense-Thirring precessions. The latter is caused by the frame-dragging due to the spin of the black hole \citep{WexKopeikin1999, Merritt2013, Zhang+15, Yu+16, Grould+17, Waisberg+18, Qi+21}. In contrast to the Schwarzschild precession, the spin generally precesses the orbit within its orbital plane as well as out of its orbital plane. However, so does the mass precession when the distribution is not spherically symmetric \citep{Merritt2013}. Because of this interference, a separation of these effects solely based on the measurement of the overall accumulated secular precession of a single stellar orbit may be difficult.\footnote{For approaches based on using multiple stellar orbits see \citep{Boehle+16, GRAVITY+21_mass_distribution}.}

We restrict our considerations to spherically symmetric extended masses that cause a pure in-plane precession, just like the Schwarzschild precession. We study how both effects can be clearly separated from each other based on their distinct astrometric signatures within a single orbit, even though their secular precessions directly interfere with each other. To this end, we first study the functional form of the argument of pericentre and of other osculating orbital elements versus true anomaly over one orbit. We underline the differences in these functional forms and identify the orbital sections in which either effect is mostly active or inactive respectively. We then extend this investigation to the domain of the observables: astrometry and radial velocity. We also support and complement our theoretical investigations by an extensive mock-data analysis. Although we do consider radial velocity, our key insights regard astrometry. A complementary approach with a focus on the effects on gravitational redshift, radial velocity, and the timing of the pericentre passage is taken in the recent work of~\citet{Takamori+20}. The important topic of the detectability and separability of precession effects on timescales of a single orbit has also previously been investigated in earlier work through different lenses and by different techniques \citep[e.g.][]{AngelilSaha2014, Parsa+17}.

We start in Sect.~\ref{S: model} with a description of our MBH + star + extended mass model. In~Sect.~\ref{S: effects} we apply this model in order to discuss the different effects of the relativistic correction and of the extended mass onto the stellar orbit. We also give prospects on the ability to improve on current upper bounds or to detect a dark mass if present. After this, we proceed with a mock-data analysis to support these claims in~Sect.~\ref{S: mock}. Finally, we conclude with a discussion and prospects in Sect.~\ref{S: discussion}.

%--------------------------------------------------------------------

\section{Model for an MBH + star + extended mass}\label{S: model}

Below we introduce our theoretical model as a perturbed Kepler problem (Sect.~\ref{SS: perturbed Kepler}) and give some background about the osculating equations as a formalism tailored to the quasi-Keplerian nature of the problem (Sect.~\ref{SS: osculating equations}). In this framework, we then give the explicit perturbative accelerations for the modelled effects (Sects.~\ref{SS: form of a_1PN}-\ref{SS: form of a_XM}). Some further general remarks about the model can be found in Appendix~\ref{App: model remarks}.

\subsection{Perturbed Kepler problem}\label{SS: perturbed Kepler}

With its high mass, the MBH dominates the gravitational field in the S-star cluster. Consequently, the motion of each star in its vicinity is approximately governed by the Newtonian two-body problem MBH + star, such that the orbits are quasi-Keplerian (\citeauthor{Merritt2013}~\citeyear{Merritt2013}, Sect.~4.2; \citeauthor{PoissonWill2014}~\citeyear{PoissonWill2014}, Sect.~3.3). The deviation from Keplerian motion is thereby due to relativistic corrections of the two-body problem on the one hand, and due to the perturbative gravitational forces of other bodies on the other hand. The latter may include other compact objects, but also continuous extended mass distributions, which is what we focus on here.

\begin{figure}%[H]

\tdplotsetmaincoords{70}{196} % point of view. args: \theta, \phi

% orbital rotation angles
\pgfmathsetmacro{\Om}{60} % about Z axis
\pgfmathsetmacro{\inc}{30} % about new X axis
\pgfmathsetmacro{\om}{75} % about new Z axis
\pgfmathsetmacro{\f}{35} % about (new) Z axis
\pgfmathsetmacro{\ompf}{110} % about (new) Z axis; has to be \om + \f

\pgfmathsetmacro{\Ompnz}{150} % has to be 90 + \Om
\pgfmathsetmacro{\incmeps}{27} % has to be \inc - 3

\pgfmathsetmacro{\drawOm}{30} % has to be 90 - \Om
\pgfmathsetmacro{\drawOmp}{175} % has to be 90 + \Om + 25
\pgfmathsetmacro{\drawom}{15} % has to be 90 - \om
\pgfmathsetmacro{\drawinc}{60} % has to be 90 - \inc
\pgfmathsetmacro{\drawf}{125} % has to be 90 + \f

\pgfmathsetmacro{\aa}{.6} % semi-major axis, a
\pgfmathsetmacro{\bb}{.5} % semi-minor axis, b
\pgfmathsetmacro{\cc}{.33166} % centre to focus, c
\pgfmathsetmacro{\dd}{.26833} % focus to pericentre, c

\pgfmathsetmacro{\rMBH}{.03}    % MBH radius
\pgfmathsetmacro{\rstar}{.01}   % star radius

% def lengths of coordinate axes
\pgfmathsetmacro{\Xlength}{.31} \pgfmathsetmacro{\Xlabel}{.36}
\pgfmathsetmacro{\Ylength}{.31} \pgfmathsetmacro{\Ylabel}{.36}
\pgfmathsetmacro{\Zlength}{.31} \pgfmathsetmacro{\Zlabel}{.36}
\pgfmathsetmacro{\xlength}{.3} \pgfmathsetmacro{\xlabel}{.31}
\pgfmathsetmacro{\ylength}{.3} \pgfmathsetmacro{\ylabel}{.31}
\pgfmathsetmacro{\zlength}{.2} \pgfmathsetmacro{\zlabel}{.21}

\begin{tikzpicture}[scale=11.6,tdplot_main_coords]

% choose clipping
\clip[tdplot_screen_coords] (-.37,-.22) rectangle (.40,.29);

% def colors
\definecolor{mmblue}{rgb}{0.368417, 0.506779, 0.709798} % Mathematica blue
\definecolor{mmorange}{rgb}{0.880722, 0.611041, 0.142051} % Mathematica orange

% def coordinates
\coordinate (O) at (0,0,0);
\coordinate (X) at (0,\Xlength,0); \coordinate (Xlabel) at (0,\Xlabel,0);
\coordinate (Y) at (-\Ylength,0,0); \coordinate (Ylabel) at (-\Ylabel,0,0);
\coordinate (Z) at (0,0,\Zlength);  \coordinate (Zlabel) at (0,0,\Zlabel);

% Y coordinate axis. X, Z drawn after MBH
\draw [-stealth,ultra thick,mmblue] (-\rMBH,0,0) -- (Y) node at (Ylabel){$Y/\mathrm{RA}$};

% rotation by \Om
\tdplotsetrotatedcoords{\Om}{0}{0}
\begin{scope}[tdplot_rotated_coords]
\tdplotdrawarc[-stealth]{(O)}{.16}{\drawOm}{90}{anchor=south}{$\Omega$} % \Om angle

\draw[dotted] (O) circle (.53);                         % plane of the sky
\node at (40:.6) {plane of the sky};
\draw[thick,dashed] (0,-.7,0) -- (0,.7,0); %node [anchor=north west] {$\mathrm{line\,of\,nodes}$};                 % line of nodes
\node at (99:-.32) [align=left] {line\\of nodes};
\draw[fill=black] (0,.365,0) circle (.005); %node [anchor=north east, align=left] {ascending\,\\node};    % ascending node
\node at (75:.372)[align=left]{ascending\\node};
\draw[fill=black] (0,-.485,0) circle (.005);% node [anchor=west] {$\;\;\;$desc.~node};         % descending node
\end{scope}

% rotation by \inc
\tdplotsetrotatedcoords{\Om}{\inc}{0}

% rotate by \om
\tdplotsetrotatedcoords{\Om}{\inc}{\om}
\tdplotdrawarc[tdplot_rotated_coords,-stealth]{(O)}{.11}{\drawom}{90}{anchor=east}{$\omega$} % \om angle

\begin{scope}[tdplot_rotated_coords]
%\draw (O) -- (\drawf:.287); % line MBH-star
%\draw [red,ultra thick,-stealth] (O) -- (\drawf:.15) node [anchor=west]{$\mathbf n$};
\tdplotdrawarc[-stealth]{(O)}{.11}{90}{\drawf}{anchor=north east}{$f$} % \f angle
\draw[ultra thick,-stealth] (\drawf:.289) -- (143:.325) node [anchor=south west]{$\mathbf v$};

\draw[thick] (0,-\cc,0) ellipse (.5 and \aa); % orbit
\draw[fill=black] (0,\dd,0) circle (.005) node [anchor=south west] {$\mathrm{pericentre}$}; % pericentre
\draw (0,\rMBH,0) -- (0,\dd,0); % line focus-pericentre
\end{scope}

% rotate by \f
\tdplotsetrotatedcoords{\Om}{\inc}{\ompf}
\begin{scope}[tdplot_rotated_coords]
\draw (0,\rMBH,0) -- (0,.287,0); % line MBH-star
\draw [red,ultra thick,-stealth,mmorange,line cap=round] (0,\rMBH,0) -- (0,.16,0) node [anchor=west]{$\mathbf n$};
\draw [red,ultra thick,-stealth,mmorange,line cap=round] (-\rMBH,0,0) -- (-.16,0,0) node [anchor=south]{$\mathbf\lambda$};
% e_z drawn after MBH ball
\end{scope}

% orbital frame, MBH and star
%\draw[tdplot_rotated_coords,red,thick,-stealth] (-\rMBH,0,0) -- (-\ylength,0,0) node at (-\ylabel,0,0){$y$};
\tdplottransformmainscreen{0}{0}{0}
\begin{scope}[tdplot_screen_coords]
\shade[ball color = black] (\tdplotresx,\tdplotresy) circle (\rMBH);    % MBH
\shade[ball color = black] (.1,.208) circle (\rstar);                           % star
\end{scope}

\draw[tdplot_rotated_coords,ultra thick,-stealth,mmorange,line cap=round] (0,0,\rMBH) -- (0,0,.16) node [anchor=south]{$\mathbf e_z$};

% draw inclination angle
\tdplotsetthetaplanecoords{0}
\tdplotdrawarc[tdplot_rotated_coords,-stealth]{(0,0,0)}{.1}{0}{\incmeps}{anchor=north}{$\iota$} % \inc angle

% fundamental frame (X,Y,Z)
\tdplotsetrotatedcoords{0}{0}{0}
\begin{scope}[tdplot_rotated_coords,ultra thick,mmblue]
\draw [-stealth,line cap=round](0,\rMBH,0) -- (X) node at (Xlabel){$X/\mathrm{DEC}$};
\draw [line cap=round](-\rMBH,0,0) -- (-.05,0,0); % only small bit of Y axis for visibility in front of MBH
\draw [-stealth,line cap=round](0,0,\rMBH) -- (Z) node [anchor=north east]{$Z/\mathrm{to\,earth}$};%at (Zlabel){$Z/\mathrm{to\,earth}$};
\end{scope}

\end{tikzpicture}

\caption{Definition of the Euler angles $(\Omega,\iota,\omega)$ giving the orbital orientation and the true anomaly $f$ that gives  the current position within the orbit. The static fundamental frame $(X,Y,Z)$ (blue) is adapted to the observables, and the co-rotated Gaussian frame $(\mathbf{n, \lambda, e}_z)$ (orange) is adapted to the symmetry of the (perturbed) two-body problem. See Fig.~\ref{F: MPE convention} for an alternative convention.}
\label{F: orientation angles}
\end{figure}
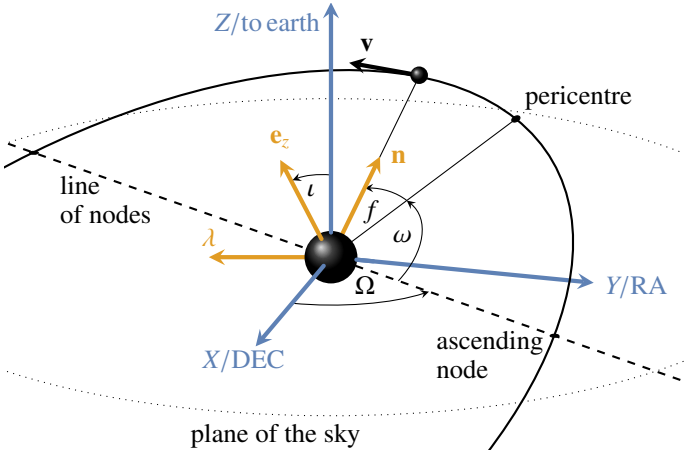

For the analysis to follow, we work with a perturbed Kepler orbit model for a star of mass $m_s$ in the gravitational field created by an MBH of mass $M_\bullet$ together with a spherically symmetric continuous extended mass distribution centred at the location of the MBH. We assume that $m_s$ is much lower than $M_\bullet$ and the extended mass. Hence we treat the star as a test particle such that formally, $m_s=0$. This also ensures consistency with our assumption that the extended mass stays centred at the MBH. Placing the origin of the coordinate system at the MBH (Fig.~\ref{F: orientation angles})
 %%%%%%%%%%%%%%%%%%%%%%%%%%%%%%%%%%%%%%%%%%%
and denoting the stellar position by $\mathbf r$, its equations of motion then take the form
\begin{align}\label{E: perturbed Kepler model}
\ddot{\mathbf r} = -\frac{GM_\bullet}{r^2}\mathbf n + \mathbf a_p,
\end{align}
with a two-component perturbative acceleration $\mathbf a_p = \mathbf a_\mathrm{1PN} + \mathbf a_\mathrm{XM}$. The first term on the right-hand side of Eq.~\eqref{E: perturbed Kepler model} denotes the dominant Newtonian two-body acceleration, which is responsible for the main Keplerian component of the stellar motion, and we denote $r=|\mathbf r|, \mathbf n=\mathbf r/r$. $G$ is the gravitational constant. The two accelerations $\mathbf a_\mathrm{1PN}, \mathbf a_\mathrm{XM}$ are responsible for perturbations from Keplerian motion. $\mathbf a_\mathrm{1PN}$ thereby denotes the relativistic correction to the Newtonian two-body equations of motion to first post-Newtonian (1PN) order \citep{Merritt2013, PoissonWill2014}. $\mathbf a_\mathrm{XM}$, on the other hand, denotes the perturbative acceleration due to the presence of an extended mass distribution. We return to the explicit form of these terms in Sects.~\ref{SS: form of a_1PN} and~\ref{SS: form of a_XM}. Before this, we switch to a formalism that is tailored to the quasi-Keplerian nature of the problem.

\subsection{Osculating equations}\label{SS: osculating equations}

While the equations of motion in the form of Newton's second law (Eq.~\eqref{E: perturbed Kepler model}) are well suited in order to demonstrate the underlying physics of the problem, other formalisms are better tailored to its quasi-Keplerian nature. One such is the post-Keplerian formalism of~\citet{DamourDeruelle1985}, which is well established in pulsar-timing research, for instance. Another is the formalism of osculating orbits, which is widely used to study celestial mechanics in the solar system, for example (\citeauthor{Merritt2013}~\citeyear{Merritt2013}, Sect.~4.2; \citeauthor{PoissonWill2014}~\citeyear{PoissonWill2014}, Sect.~3.3.2). The correspondence between the parameters of both formalisms is given in~\citet{KlionerKopeikin1994}. We use here the latter framework, whose base equations (i.e. the equivalent to Eq.~\eqref{E: perturbed Kepler model}) form a system of first-order evolution equations for a set of six Keplerian orbital elements $(\mu^a)_{a=1}^6$ in time. These are called the osculating equations. Different sets of orbital elements are equivalent. We follow the formulation of~\citet{PoissonWill2014} and pick $(p,e,\iota,\Omega,\omega,f)$. In order: semi-latus rectum, eccentricity, inclination, argument of the ascending node, argument of pericentre, and true anomaly. The last four elements are position and orientation angles (Fig.~\ref{F: orientation angles}). As auxiliary orbital element, we also consider the semi-major axis,
\begin{align}\label{E: semi-major axis}
a = p/(1-e^2).
\end{align}

Furthermore, we decompose the perturbative acceleration along its co-rotated Gaussian frame components,
\begin{align}\label{E: Gaussian components}
\mathbf a_p = \mathcal R\mathbf n + \mathcal S\mathbf\lambda + \mathcal W\mathbf e_z ,
\end{align}
where $\mathbf e_z$ is the unit vector in the direction of orbital angular momentum and $\mathbf\lambda$ completes the right-handed orthonormal basis (Fig.~\ref{F: orientation angles}). With this, the osculating equations take the form
\begin{align}\label{E: abstract osc equations}
\dot{\mu}^a(t) = F^a(\mu^b;\mathcal{R,S,W}) ,
\end{align}
and we quote the explicit form of the right-hand side functions $F^a$ from~\citet{PoissonWill2014} in Appendix~\ref{App: osc equations}. In general, $\mathcal{R,S,W}$ are functions of the $\mu^a$ themselves. Importantly, the functions $F^a$ are linear in $\mathcal{R,S,W,}$ and in particular,
\begin{align}\notag%\label{E: Kepler case}
F^a(\mu^b;0,0,0)=\begin{cases}
0 & a\in(1,\dots,5) \\
\sqrt{\frac{GM_\bullet}{p^3}}(1+e\cos f)^2 & a=6, 
\end{cases}
\end{align}
such that the unperturbed case reduces to the Kepler problem, in which the first five orbital elements are constants of motion, and the true anomaly $f$ strictly monotonically increases at the above rate. In the general perturbed case, all orbital elements may vary with time, describing a precessing or otherwise evolving orbit.

From the orbital elements, the components of the position~$\mathbf r$ and velocity~$\mathbf{v=\dot r}$ of the secondary body wit respect to the primary can be obtained by recognising that an Euler rotation around the angles $\Omega$, $\iota$, and $\omega+f$ rotates the fundamental frame $(X,Y,Z)$ into the Gaussian frame $(\mathbf n, \mathbf\lambda, \mathbf e_z)$ (Fig.~\ref{F: orientation angles}). A detailed derivation of the resulting transformations can be found in~\citet[Sect.~3.2.5]{PoissonWill2014}, from where we quote the results in Eqs.~\eqref{E: POS}-\eqref{E: VEL}.

\subsection{Explicit form of $\mathbf a_\mathrm{1PN}$}\label{SS: form of a_1PN}

With the star as test particle, the relativistic correction to the Newtonian equations of motion to 1PN order is given by
\begin{align}\label{E: 1PN}
\mathbf a_\mathrm{1PN} =  4\frac{GM_\bullet}{c^2r^2} \left( \left(\frac{GM_\bullet}{r} - \frac{v^2}{4}\right)\mathbf n + (\mathbf n\cdot\mathbf v)\mathbf v \right),
\end{align}
where $v=|\mathbf v|$ (\citeauthor{Merritt2013}~\citeyear{Merritt2013}, Eq. (4.166); \citeauthor{PoissonWill2014}~\citeyear{PoissonWill2014}, Eq.~(10.1)). Projecting this onto $(\mathbf{n,\lambda,e}_z),$ we obtain
\begin{align}
\mathcal R_\mathrm{1PN} &= \frac{G^2M_\bullet^2}{c^2p^3}(1+e\cos f)^2 \left( 3(e^2+1) + 2e\cos f - 4e^2\cos^2f \right) \label{E: R 1PN}\\
\mathcal S_\mathrm{1PN} &= \frac{G^2M_\bullet^2}{c^2p^3}4(1+e\cos f)^3e\sin f, \label{E: S 1PN}
\end{align}
and $\mathcal W_\mathrm{1PN}=0$ for the corresponding Gaussian components (\citeauthor{Merritt2013}~\citeyear{Merritt2013}, Eq.~(4.207); \citeauthor{PoissonWill2014}~\citeyear{PoissonWill2014}, Sect.~10.1.3). From the vanishing of $\mathcal W_\mathrm{1PN}$ together with Eqs.~\eqref{E: i dot}--\eqref{E: Om dot}, it is immediately clear that $\mathbf a_\mathrm{1PN}$ does not cause an out-of-plane orbital precession: the Schwarzschild precession is in-plane.

\subsection{Explicit form of $\mathbf a_\mathrm{XM}$}\label{SS: form of a_XM}

We proceed with calculating $\mathbf a_\mathrm{XM}$ for a Newtonian spherically symmetric extended mass distribution $\rho(r)$. The easiest way to do so is via Gauss' law in integral form, which in general reads
\begin{align}\label{E: Gauss}
\oint_{\partial V} a_\perp\,\mathrm d\sigma =
-4\pi G\oint_V \rho\,\mathrm dV .
\end{align}
The integrals are taken over a volume $V$ and its boundary surface $\partial V$. $\mathrm dV$ and $\mathrm d\sigma$ denote the respective volume and surface elements. $a_\perp$ is the outward-pointing normal component to $\partial V$ of the acceleration created by the mass distribution $\rho$. Exploiting the symmetry and choosing $V$ to be a sphere of radius $r,$ we have $a_\perp = \mathcal R_\mathrm{XM}(r)$ such that the left-hand side of Eq.~\eqref{E: Gauss} gives $4\pi r^2\mathcal R_\mathrm{XM}(r)$. On the right-hand side, we obtain $-(4\pi)^2G\int_0^r\rho(s)s^2\,\mathrm ds$. Equating both results, we obtain
\begin{align}\label{E: general R}
\mathcal R_\mathrm{XM}(r) = -\frac{4\pi G}{r^2}\int_0^r\rho(s)s^2\,\mathrm ds
\end{align}
for arbitrary spherically symmetric distributions $\rho(r)$. Furthermore, due to spherical symmetry, we have $\mathcal S_\mathrm{XM}=\mathcal W_\mathrm{XM}=0$. As in the 1PN case, from the vanishing of $\mathcal W_\mathrm{XM}$ , it immediately follows that a spherically symmetric extended mass does not cause an out-of-plane precession. However, in contrast to the 1PN case, the vanishing of $\mathcal S_\mathrm{XM}$ and Eq.~\eqref{E: p dot} show that here,  $p$ is also a constant of motion.

\begin{figure}%[hbt]
  \centering
  \includegraphics[clip=true,trim=0 0 0 0,scale=.49]{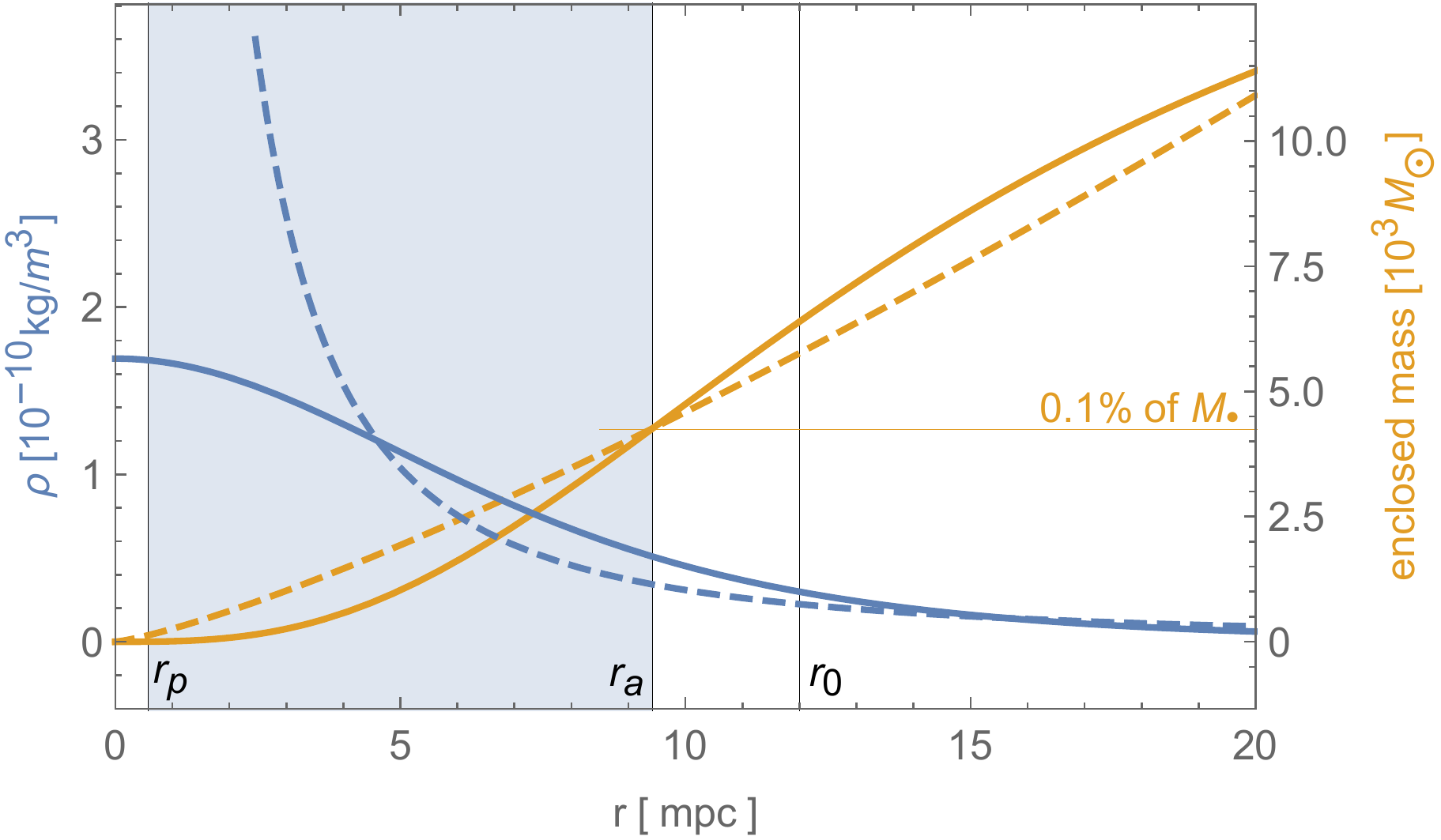}
  \caption{Density profiles (blue, left horizontal axis) and enclosed masses (orange, right horizontal axis) for the Plummer (solid) and cusp (dashed) profiles (Eq.~\eqref{E: profiles}). The profiles have a scale parameter $r_0$ given by Eq.~\eqref{E: r0} and density parameters $\rho_0$ given by Eq.~\eqref{E: rho0}. The cusp exponent $\gamma$ is given by Eq.~\eqref{E: BW gamma}. $r_\mathrm p$ and $r_\mathrm a$ denote the pericentre and apocentre distance of S2, respectively. The profiles are chosen such that $0.1\%$ of the galactic centre MBH mass reside within $r_\mathrm a$, i.e. they correspond to the current upper bound of~\citet{GRAVITY+20_Schwarzschild_prec}.}
  \label{F: profiles}
\end{figure}

In the following, we restrict our considerations to Plummer and power-law cusp profiles
\begin{align}\label{E: profiles}
\rho(r) = \begin{cases}
\rho_0\left(1+\frac{r^2}{r_0^2}\right)^{-5/2} &\text{Plummer} \\
\rho_0\left(\frac{r}{r_0}\right)^{-\gamma} & \text{cusp}
\end{cases}
\end{align}
with density and scale parameters $\rho_0, r_0$ (Fig.~\ref{F: profiles}). (Note the different interpretations of $\rho_0$ for both profiles: maximum (i.e. central) density (Plummer) versus density at $r_0$ (cusp).) We limit the cusp exponent $\gamma$  to the range $(0,3)$ for an outward falloff and integrability.
A prominent example is a Bahcall-Wolf cusp, which corresponds to
\begin{align}\label{E: BW gamma}
\gamma=7/4
\end{align}
\citep[Sect.~5.5.2]{BahcallWolf1976, Merritt2013}. This is the distribution to which a spherically symmetric nuclear star cluster should relax after a sufficiently long time. Concerning the galactic centre, the cusp profile has some theoretical and observational backing as well, although with slightly smaller exponents \citep{Schoedel+18, Baumgardt+18, GallegoCano+18}. However, we have to keep in mind that these results are concerned with the distribution of the cluster as a whole,\footnote{See also \citet{Tepp+21} for a recent related analysis on these scales.} while here we are concerned with scales of about the orbit of S2, that is, with the innermost part of the cluster in the immediate vicinity of the MBH. For these scales, a profile that (unlike a cusp) plateaus at a finite density at its centre may appear more plausible, which is why we set a slightly stronger focus on the Plummer profile in our analyses \citep{Plummer1911}. Originally, it has been found by modelling the distributions of globular clusters.

For Eq.~\eqref{E: profiles} we can perform the radial integral of Eq.~\eqref{E: general R} analytically and obtain
\begin{align}\label{E: R XM}
\mathcal R_\mathrm{XM}(r) = \begin{cases}
-\frac{4\pi G}{3}\rho_0\frac{r_0^3r}{\left(r^2+r_0^2\right)^{3/2}} & \text{Plummer}\\
-\frac{4\pi G}{3-\gamma}\rho_0r_0\left(\frac{r}{r_0}\right)^{1-\gamma} & \text{cusp}
\end{cases} .
\end{align}
Finally, to yield Eq.~\eqref{E: abstract osc equations} in closed form, we substitute $r=p/(1+e\cos f)$.

Summarising the results of this section and Sect.~\ref{SS: osculating equations}, we have a net perturbative acceleration $\mathbf a_p = \mathbf a_\mathrm{1PN} + \mathbf a_\mathrm{XM}$ with Gaussian components
\begin{align}\label{E: RSW for full model}
\mathcal R &= \mathcal R_\mathrm{1PN} + \mathcal R_\mathrm{XM} &
\mathcal S &= \mathcal S_\mathrm{1PN} &
\mathcal W &= 0
\end{align}
with the constituents given by Eqs.~\eqref{E: R 1PN}, \eqref{E: S 1PN}, and \eqref{E: R XM}. Hence we now have our osculating equations (Eq. ~\eqref{E: abstract osc equations}) of consideration in explicit and closed form (see also Eqs.~\eqref{E: p dot}-\eqref{E: f dot}). This concludes the summary of our MBH + star + extended mass model. Some further general remarks are given in Appendix~\ref{App: model remarks}.

\section{Effects of the extended mass on the stellar orbit}\label{S: effects}

In the following, we investigate the impact of an extended continuous mass distribution on the osculating orbital elements (Sect.~\ref{SS: orbital elements}) and on the observables (Sect.~\ref{SS: astrometry and velocity}) of one full orbit of S2. In Sect.~\ref{SS: setup} we present the setup to this end, for which we restrict ourselves to specific mass distributions motivated by the recent upper dark mass bounds of~\citet{GRAVITY+20_Schwarzschild_prec}. Sect.~\ref{SS: detection thresholds} is then concerned with varying the mass profile parameters in order to estimate detection thresholds. We then conclude this section with some comments about possible reservations concerning our theoretical analysis and the conclusions drawn from it in Sect.~\ref{SS: reservations}.

\subsection{Setup}\label{SS: setup}

We integrate the osculating equations (Eq.~\eqref{E: abstract osc equations}) (see also Eqs.~\eqref{E: p dot}-\eqref{E: f dot}) over a full orbit $f\in[0,2\pi]$ for the following perturbed Kepler models (Eq.~\eqref{E: perturbed Kepler model}):
\begin{align}
\mathbf a_p &= \mathbf a_\mathrm{1PN} \tag{I}\label{E: i} \\
\mathbf a_p &= \mathbf a_\mathrm{XM}\,\text{(Plummer)} \tag{IIa}\label{E: iia} \\
\mathbf a_p &= \mathbf a_\mathrm{XM}\,\text{(Bahcall-Wolf cusp)} \tag{IIb}\label{E: iib} \\
\mathbf a_p &= \mathbf a_\mathrm{1PN} + \mathbf a_\mathrm{XM}\,\text{(Plummer)} \tag{IIIa}\label{E: iiia} \\
\mathbf a_p &= \mathbf a_\mathrm{1PN} + \mathbf a_\mathrm{XM}\,\text{(Bahcall-Wolf cusp).} \tag{IIIb}\label{E: iiib}
\end{align}
The Gaussian components of these perturbative accelerations are given by the respective combinations of Eq.~\eqref{E: RSW for full model}. In picking the profile parameters for the extended masses, we orient ourselves at the recent upper bounds obtained in~\citet{GRAVITY+20_Schwarzschild_prec} and choose a scale factor
\begin{align}\label{E: r0}
r_0 = 0.3\,\mathrm{as} \approx 0.012\,\mathrm{pc} \approx 2474.01\,\mathrm{au}
\end{align}
together with a density parameter $\rho_0$ such that $0.1\%$ of $M_\bullet$ resides within the apocentre distance of S2 (Fig.~\ref{F: profiles}). (Here we use the apocentre distance of the initial osculating orbit, that is, $p_0/(1 - e_0)$ with $p_0$ and $e_0$ given by \eqref{E: ID2}.) Consequently,
\begin{align}\label{E: rho0}
\rho_0 \approx \begin{cases}
1.69\times10^{-10}\,\mathrm{kg/m^3} & \text{Plummer model} \\
2.24\times10^{-11}\,\mathrm{kg/m^3} & \text{Bahcall-Wolf cusp}
\end{cases} .
\end{align}

In all cases, we choose the $\sim2018$ pericentre of S2 as initial point and time $t_0$ at which we prescribe the osculating orbital elements. For these, as well as for the MBH mass $M_\bullet$ and the distance to earth $R_0$, we pick the recent best-fit values of~\citet[Table~E.1]{GRAVITY+20_Schwarzschild_prec}. Rounded to two digits in units of convenience, we thus have
%\footnote{We follow here the convention of standard text-books on celestial mechanics, e.g. \cite{Merritt2013, PoissonWill2014}, in which the sense of rotation of $\iota$ is right handed w.~r.~t. the line of nodes pointing towards the ascending node, and in which `ascending' is defined w.~r.~t. the $Z$-axis of the fundamental frame for $\iota\in(0,\pi)$; cf. Fig.~\ref{F: orientation angles}.}
\begin{align}
M_\bullet&\approx4.26\times10^6\,M_\odot        &       R_0&\approx8246.7\,\mathrm{pc}  &       t_0&\approx2018.38\,\mathrm{yr}
\label{E: ID1} \\
p_0&\approx224.46\,\mathrm{au}                  &       e_0&\approx0.88                                 &       \iota_0&\approx-2.35
\label{E: ID2} \\
\Omega_0&\approx3.98                                    &       \omega_0&\approx1.16                    &       f_0&=0, \label{E: ID3}
\end{align}
where the angles are given in radians and where $\mu^a_0=\mu^a(t_0)$. The negative sign of the inclination angle arises because \citet{GRAVITY+20_Schwarzschild_prec} used the opposite convention for the sense of rotation of the inclination (Appendix~\ref{App: conventions}).

\subsection{Impact on the osculating orbital elements}\label{SS: orbital elements}

\begin{figure}%[t!]
\centering
\begin{subfigure}{.481\textwidth}
  \centering
  \includegraphics[clip=true,trim=0 0 0 0,scale=.49]{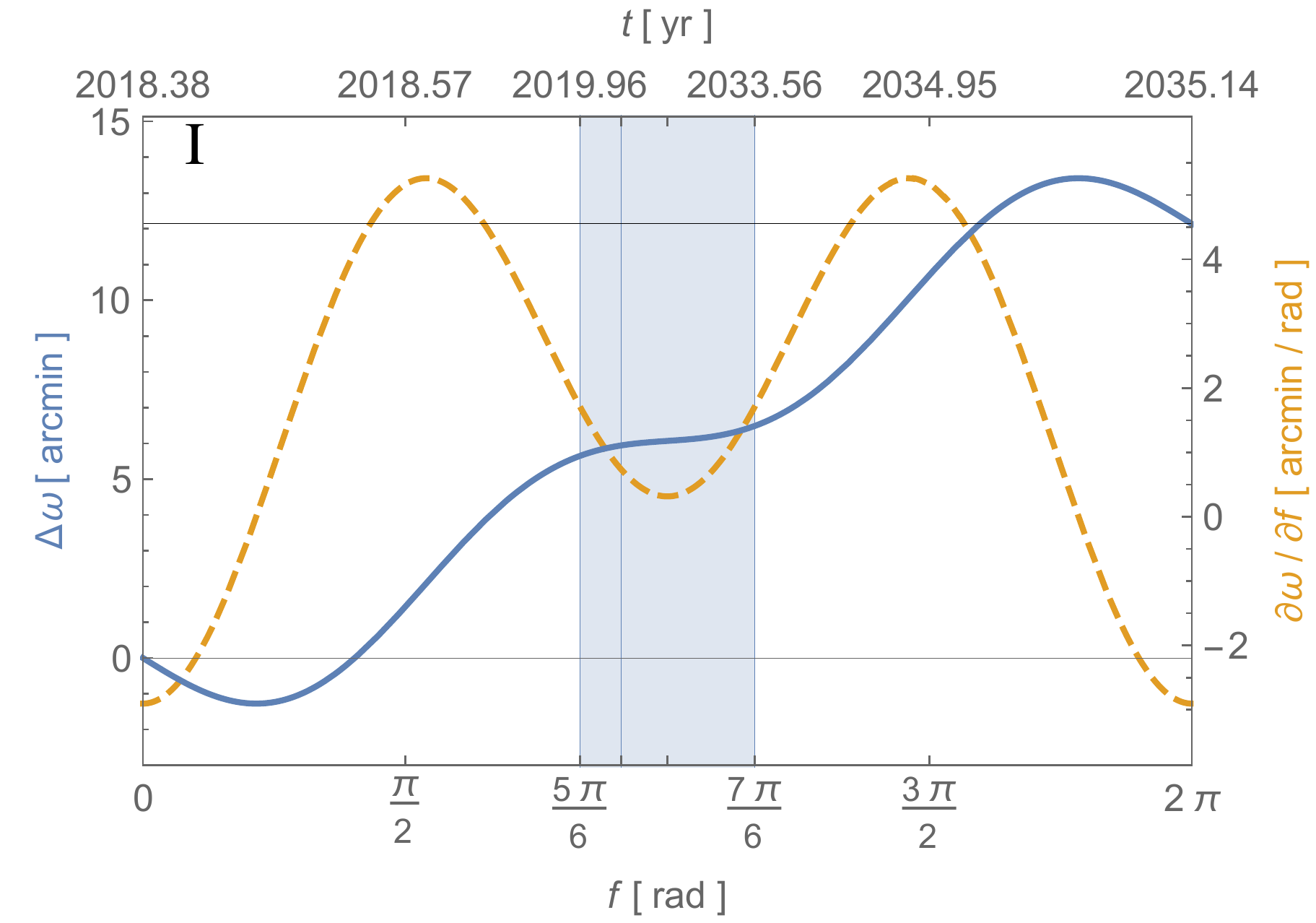}
  \caption{}
  \label{F: om i}
\end{subfigure}
\begin{subfigure}{.481\textwidth}
  \centering
  \includegraphics[clip=true,trim=0 0 0 0,scale=.49]{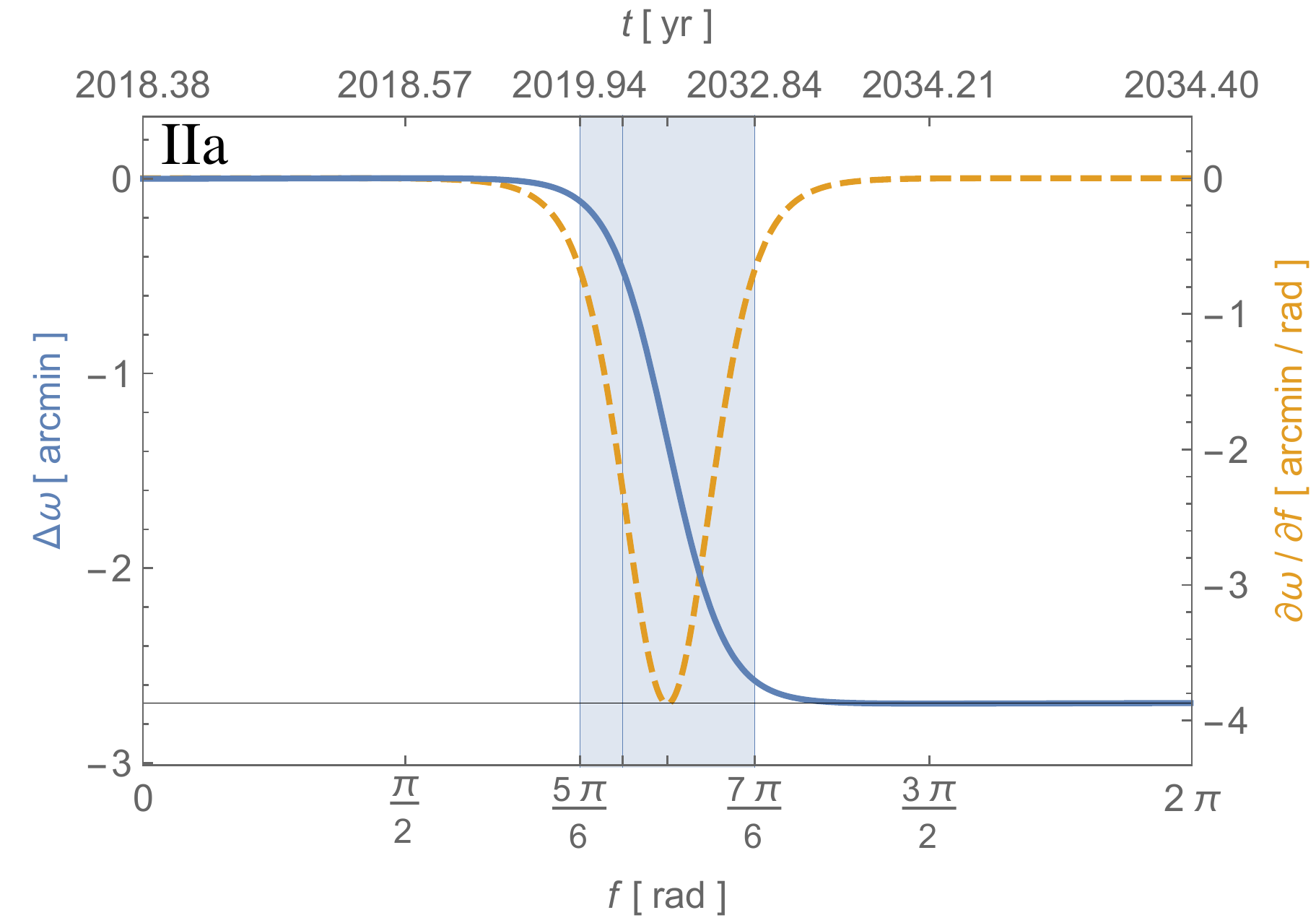}
  \caption{}
  \label{F: om iia}
\end{subfigure}
\begin{subfigure}{.481\textwidth}
  \centering
  \includegraphics[clip=true,trim=0 0 0 0,scale=.49]{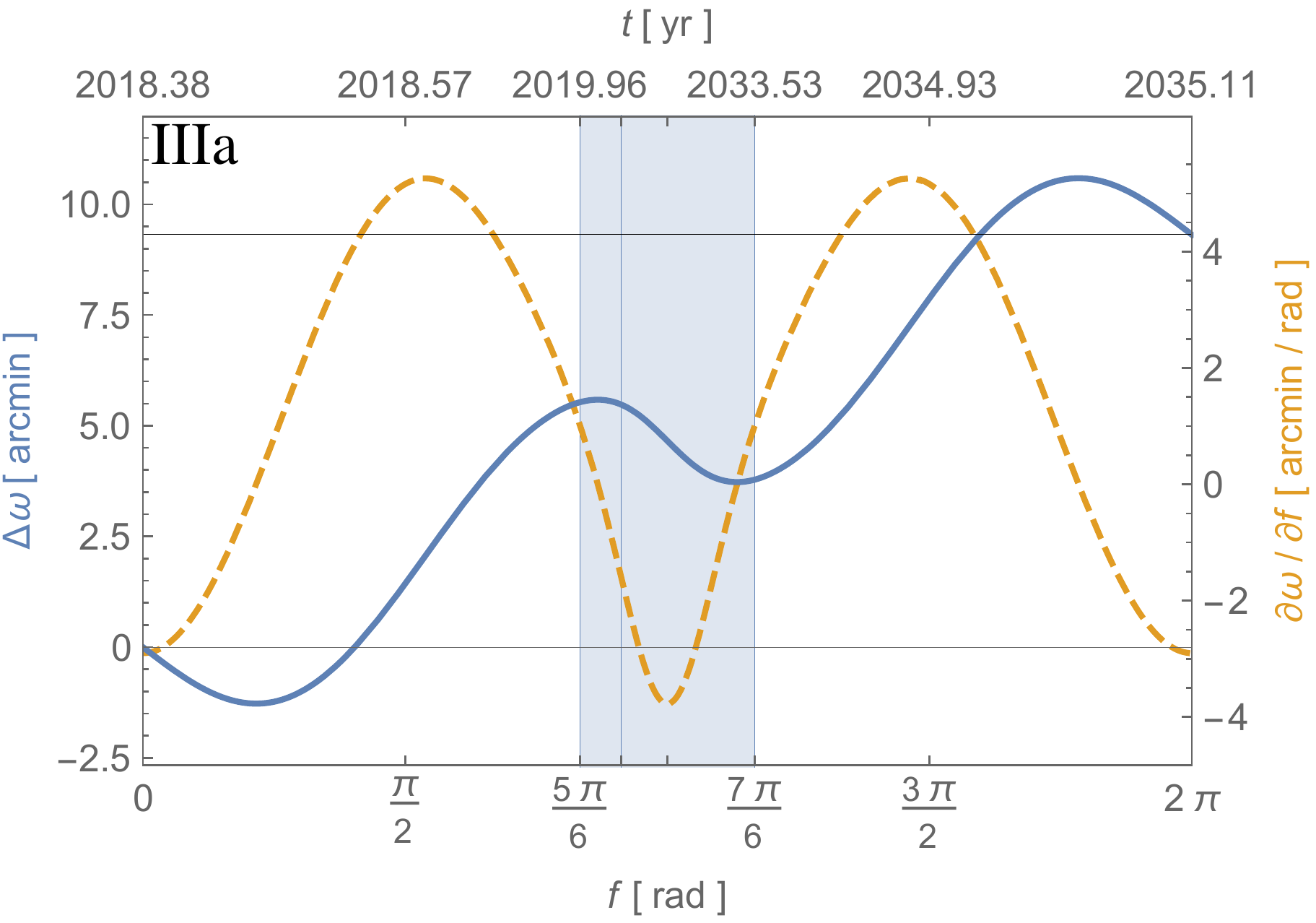}
  \caption{}
  \label{F: om iiia}
\end{subfigure}
\caption{Plots of $\Delta\omega(f)=\omega(f)-\omega_0$ together with $\omega'(f)$ for the models~\eqref{E: i}, \eqref{E: iia}, and~\eqref{E: iiia}. The shaded blue region $f\in(5\pi/6,7\pi/6)$ marks the orbital section at which the extended mass drives the orbital change, while the 1PN correction is nearly ineffective. The opposite holds for the rest of the orbit. The vertical line within the shaded blue region marks the true anomaly in $2021.96$, which is approximately the time of publication \citep{GRAVITY+21_mass_distribution}.}
\label{F: omega plots}
\end{figure}

Because it is the quantity that directly encodes the pericentre shift, we start our investigation with the evolution of the argument of pericentre $\omega$. Fig.~\ref{F: omega plots} shows plots of $\Delta\omega(f)=\omega(f)-\omega_0$ together with $\omega'(f)$ for the models~\eqref{E: i}, \eqref{E: iia}, and~\eqref{E: iiia}.
 % -----------
The plots for the cusp models~\eqref{E: iib} and~\eqref{E: iiib} are very similar in terms of qualitative features to their Plummer counterparts. We thus do not show them separately and demonstrate our arguments at the example of the Plummer model.

Firstly, comparing Figs.~\ref{F: om i} and~\subref{F: om iia}, we see that both the Schwarzschild and the mass precession have a secular effect, that is, they yield an accumulated pericentre shift over one orbit. Moreover, we see that the Schwarzschild precession is prograde $(\omega'>0)$ while the mass precession is retrograde $(\omega'<0)$, as is well known \citep{JiangLin85, RubilarEckart2001, Merritt2013}. Secondly, we emphasise as the key observation from these plots that while the Schwarzschild precession is nearly ineffective for $f\in(5\pi/6, 7\pi/6)$ and predominantly active elsewhere on the orbit, the exact opposite is true for the mass precession; see~the shaded blue region in the plots. In other words, the different perturbations are driving the orbital change over the course of distinct orbital sections, and we identify these sections spatially in Sect.~\ref{SS: astrometry and velocity} below. Consequently, Fig.~\ref{F: om iiia} shows that despite its comparatively low mass, the mass precession even dominates the Schwarzschild precession in the shaded blue region, causing a temporary retrograde precession in the combined model~\eqref{E: iiia}.\footnote{The extended mass given by Eqs.~\eqref{E: profiles}, \eqref{E: r0} required to balance $\Delta\omega$ over one S2 orbit to zero (i.e. no net precession) accounts for $\sim18\,400\,M_\odot$ or $0.43\%$ of $M_\bullet$ (Plummer) and $\sim14\,700\,M_\odot$ or $0.34\%$ of $M_\bullet$ (Bahcall-Wolf cusp) within the apocentre.} On the one hand, it does not come as a surprise that the Schwarzschild precession is strongest around pericentre and that the mass precession is strongest around apocentre. After all, relativistic effects are stronger at shorter distances, and the Newtonian pull from the spherical mass distribution is greater the more of the distribution is between the star and the MBH. On the other hand, it is not trivial that these orbital sections are as clear cut and disjoint as is evident from the plots.

\begin{figure}%[hbt]
  \centering
  \includegraphics[clip=true,trim=0 0 0 0,scale=.49]{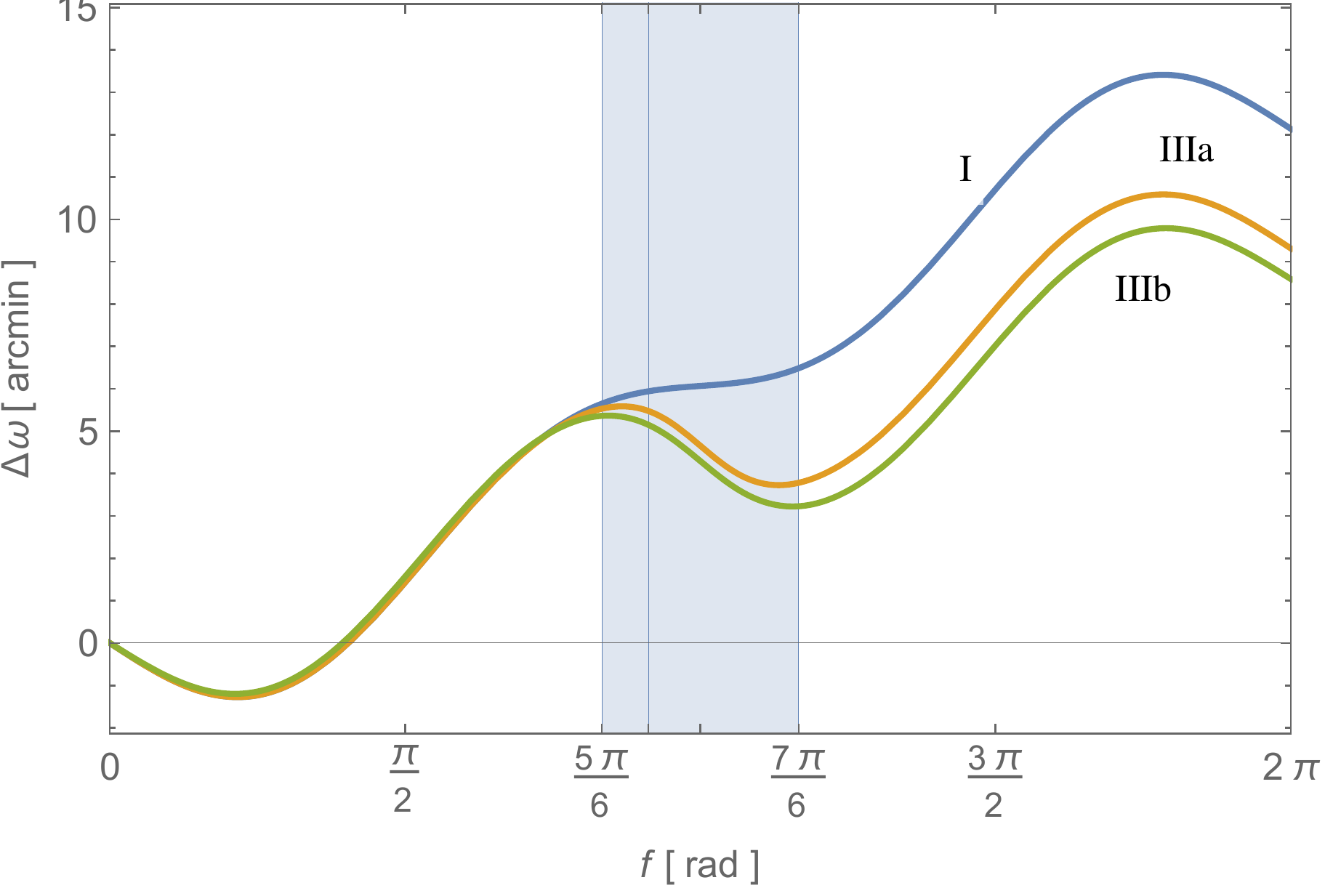}
  \caption{ $\Delta\omega(f)=\omega(f)-\omega_0$ for models~\eqref{E: i}, \eqref{E: iiia}, and~\eqref{E: iiib}. The discrepancy in pericentre shift per orbit between models~\eqref{E: iiia}-\eqref{E: iiib} with extended mass and model~\eqref{E: i} without it predominantly builds up in the shaded blue region. The vertical line within this region marks the true anomaly of model~\eqref{E: i} in $2021.96$, which is approximately the time of publication \citep{GRAVITY+21_mass_distribution}.}
  \label{F: om3}
\end{figure}

To further underline the latter point, we compare $\Delta\omega(f)$ for models~\eqref{E: i}, \eqref{E: iiia}, and~\eqref{E: iiib} in one plot (Fig.~\ref{F: om3}).
(We recall that all three models share the data of Eqs.~\eqref{E: ID1}-\eqref{E: ID3}.) From the plot it is evident that the discrepancy in the pericentre shift per orbit between models~\eqref{E: iiia}-\eqref{E: iiib} with an extended mass and model~\eqref{E: i} without one almost exclusively builds up in the shaded blue region. The curves for all three models lie almost on top of each other before this orbital section and stay at approximately constant in the separation after. This strongly emphasises the clear-cut spatial separation of both precession effects. It further demonstrates that when the observational data are limited to the $f\in(-5\pi/6,5\pi/6)$ section of the orbit, then the sensitivity to a dark mass will be very weak, and will substantially improve by additionally sampling the complementary orbital section $f\in(5\pi/6,7\pi/6)$. In the latter orbital section, data are directly sensitive do a dark mass. The plot also shows that the curves for models~\eqref{E: iiia} and~\eqref{E: iiib} do not differ significantly, which we can understand from Fig.~\ref{F: profiles}: Despite the different density profiles, the enclosed masses within a certain distance do not differ significantly, and the enclosed mass is responsible for the Newtonian acceleration.

\begin{figure}%[hbt]
\centering
\begin{subfigure}{.481\textwidth}
    \centering
    \includegraphics[clip=true,trim=0 0 0 0,scale=.49]{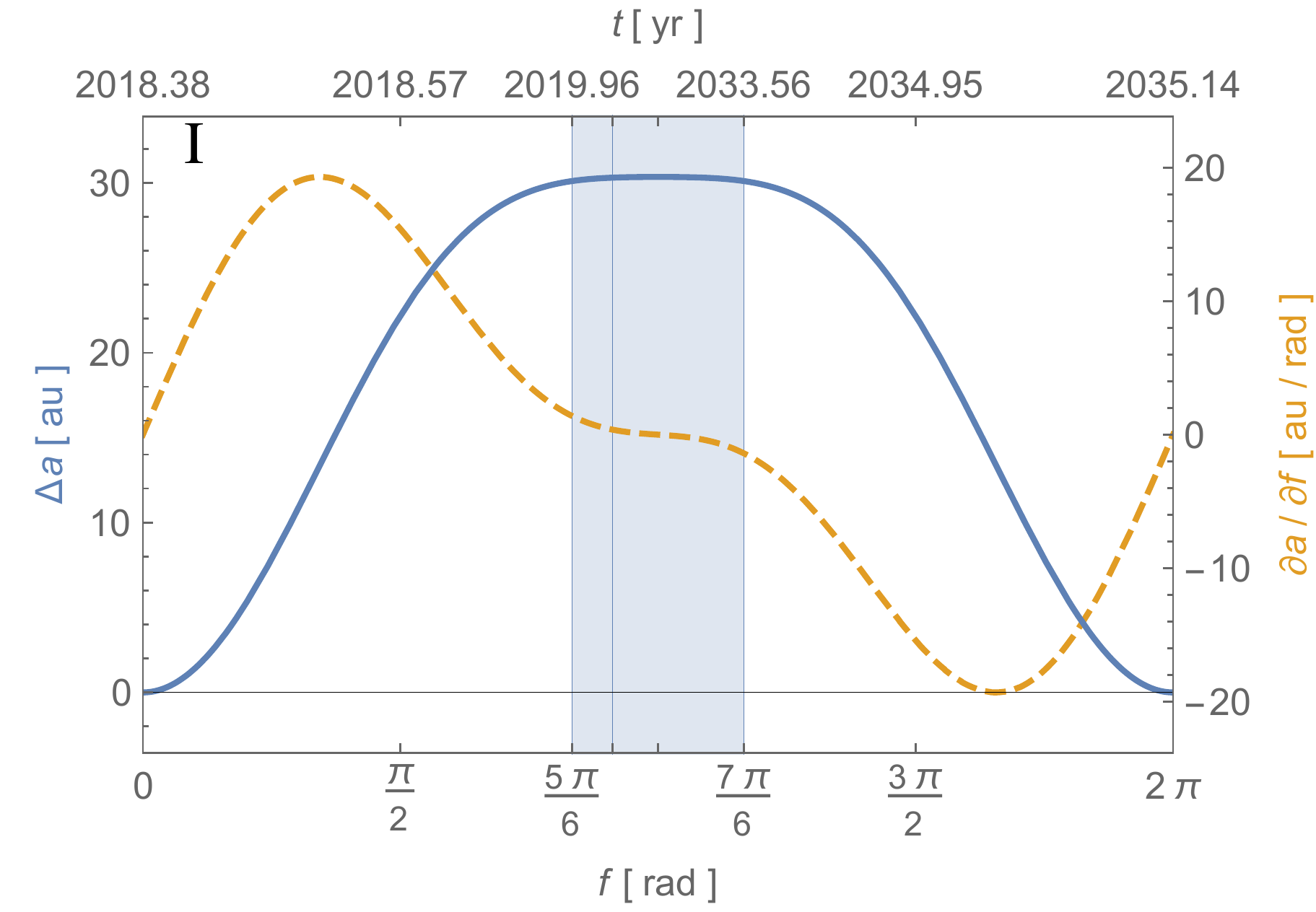}
    \caption{}
    \label{F: a i}
\end{subfigure}
\begin{subfigure}{.481\textwidth}
    \centering
    \includegraphics[clip=true,trim=0 0 0 0,scale=.49]{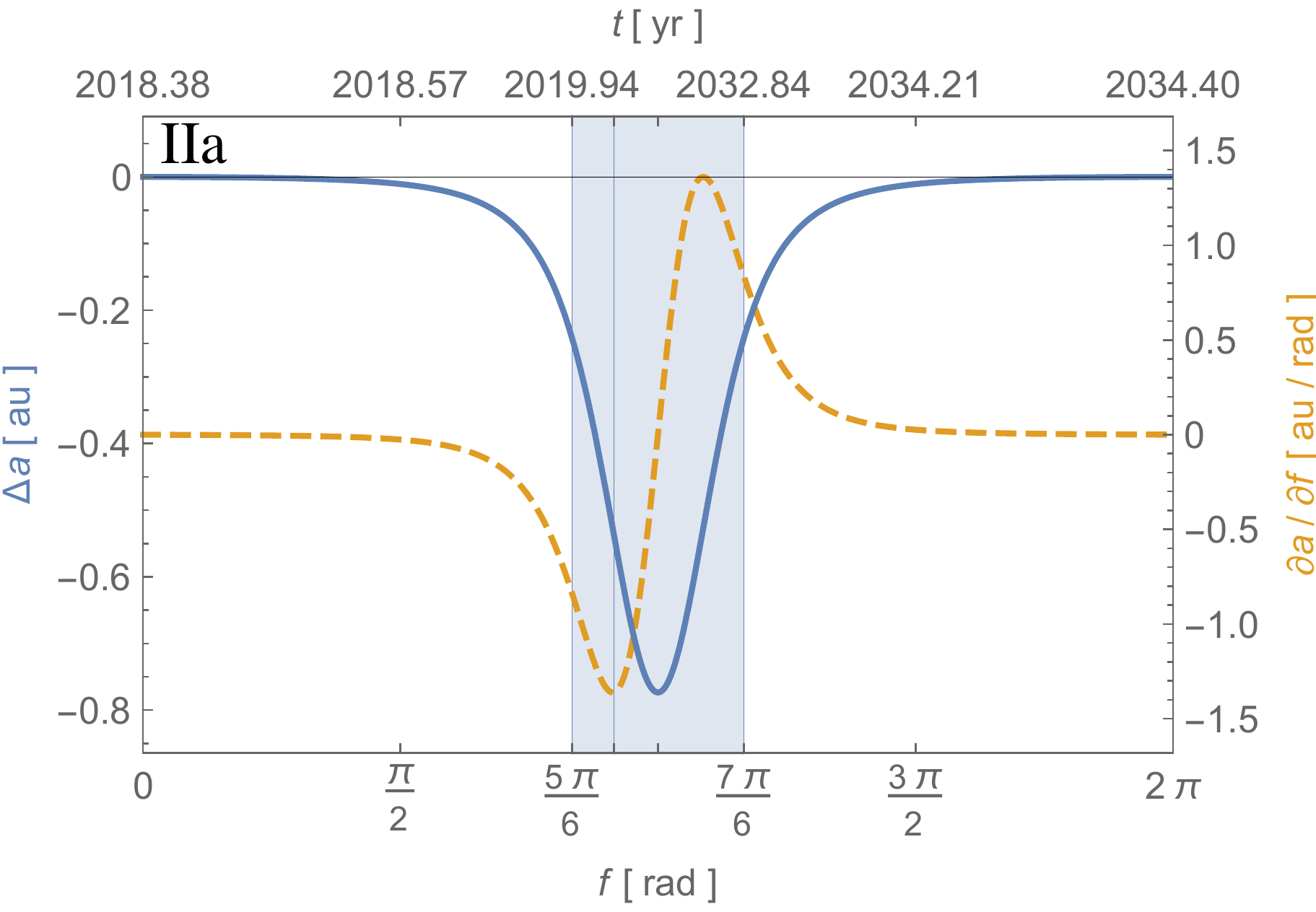}
    \caption{}
    \label{F: a iia}
\end{subfigure}
\begin{subfigure}{.481\textwidth}
  \centering
  \includegraphics[clip=true,trim=0 0 0 0,scale=.49]{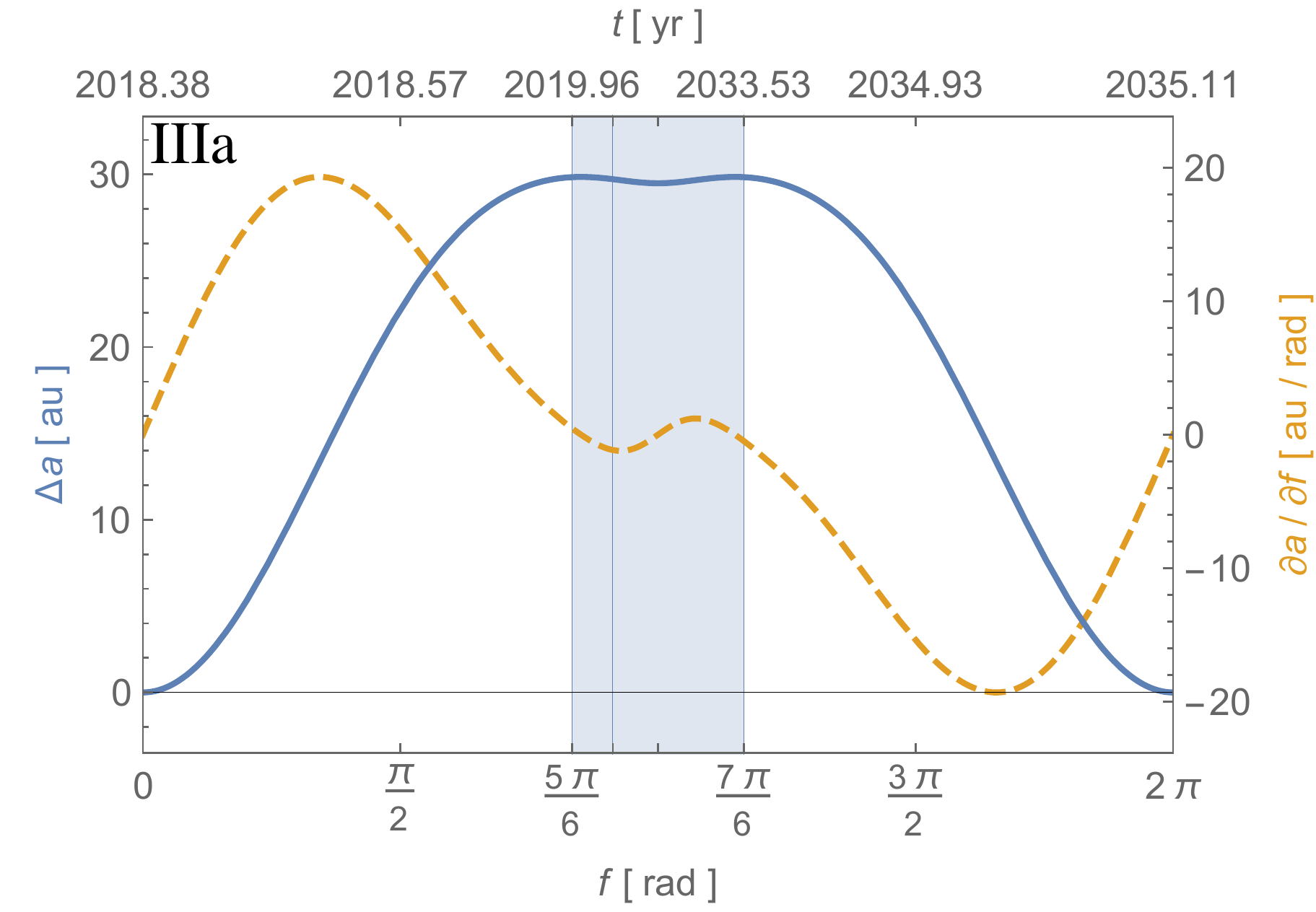}
  \caption{}
  \label{F: a iiia}
\end{subfigure}
\caption{ $\Delta a(f)=a(f)-a_0$ together with $a'(f),$ analogous to Fig.~\ref{F: omega plots}.}
\label{F: a plots}
\end{figure}

We now turn to the other orbital elements. As discussed in Sects.~\ref{SS: form of a_1PN}-\ref{SS: form of a_XM}, there is no out-of-plane precession in our cases, such that $\Omega$ and $\iota$ are constants of motion. Additionally, we established that $p$ is unaffected by a spherically symmetric extended mass, such that we would not learn about the interference of the Schwarzschild and mass precessions from investigating it. We could investigate $e$, but because $p$ is not altered by the extended mass, the semi-major axis $a$ carries the same information, as shown by Eq.~\eqref{E: semi-major axis}. We thus choose to plot $a$ together with $a'(f)$ in Fig.~\ref{F: a plots} as it is related to astrometry in a more direct manner than eccentricity.
Again, the corresponding cusp plots show the same qualitative features, and we do not plot them separately. In contrast to the pericentre precession, there is no secular effect on $a$ such that the semi-major axis returns to its initial value after a full orbit. Similarly as with the pericentre precession, the intermediate changes in $a$ mostly go in the opposite direction for both effects, and they again predominantly occur in distinct orbital sections. The extended mass has again the strongest effect for $f\in(5\pi/6,7\pi/6)$, precisely where the 1PN perturbation is nearly ineffective, and it causes a temporary contraction of $a$ with the minimum at apocentre $f=\pi$. Consequently, in the plot for the combined model~\eqref{E: iiia} (Fig.~\ref{F: a iiia}), the effect of the extended mass can be identified as the little dip in the plateau around apocentre.

Even though the osculating orbital elements are not observables, the plots of~Figs.~\ref{F: omega plots}-\ref{F: a plots} already allow us to give a first estimate of the extended mass impact on astrometry. In Figs.~\ref{F: om iia} and~\subref{F: om iiia} we identify a temporary retrograde pericentre shift of $\sim -2.7\,\mathrm{amin}$ in the  shaded blue orbital section. By trigonometry, this accounts for an astrometric shift near apocentre in the direction parallel to the minor axis of about the apocentre distance $a(\pi)(1+e(\pi))$ times this angle. Dividing by $R_0$ , this translates into $\sim191\,\mathrm{\mu as}$ angular distance modulo the projection onto the sky. On the other hand, we estimate directly from Figs.~\ref{F: a iia} and~\subref{F: a iiia}  an astrometric impact along the major axis of $\sim-0.8\,\mathrm{au}$ times $(1+e(\pi))$, which translates into $\sim-183\,\mathrm{\mu as}$ angular distance modulo the projection onto the sky. In the following section, we compare these estimates to those obtained directly from the astrometry plots.

\subsection{Impact on astrometry and radial velocity}\label{SS: astrometry and velocity}
% F: orientation angles
Motivated by the above results, we transform by Eqs.~\eqref{E: POS}-\eqref{E: VEL} from the osculating orbital elements to the position $\mathbf r$ and velocity $\mathbf v$ of the star in order to shed direct light on the observables $\mathrm{RA,DEC,\text{and }RV}$. These are right ascension, declination, and radial velocity, where the last is defined as minus the $Z$-component of $\mathbf v$.

\begin{figure}%[hbt]
\centering
\includegraphics[clip=true,trim=0 0 0 0,scale=.49]{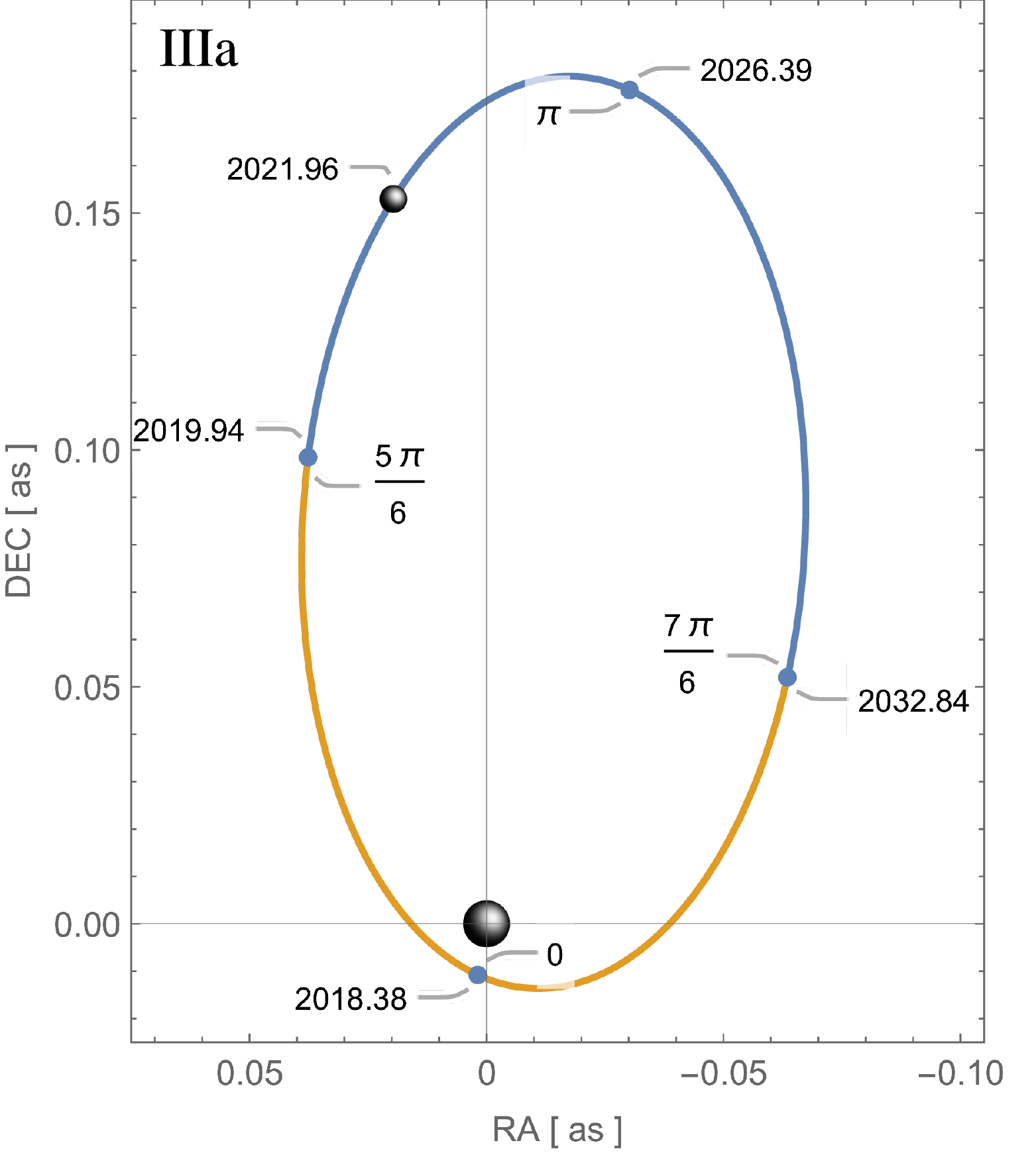}
\caption{Orbit projected onto the plane of the sky for model~\eqref{E: iiia} and~$f\in[0,2\pi]$. The position of S2 is marked in $2021.96$, which is approximately the time of publication.}
\label{F: orbit iiia}
\end{figure}

Fig.~\ref{F: orbit iiia} shows a plot of the orbit projected onto the plane of the sky for model~\eqref{E: iiia} and~$f\in[0,2\pi]$. (The respective plots for models~\eqref{E: i} and~\eqref{E: iiib} are indistinguishable from Fig.~\ref{F: orbit iiia} by naked eye, up to the time labels (Fig.~\ref{F: omega plots}).)
Although the gap at pericentre is too small to be visible by naked eye, the orbit is not closed due to the pericentre shift. Labels inside the orbit indicate the true anomaly~$f$ of the respective points in the trajectory, while labels outside the orbit give the corresponding time in years. S2 is drawn at its position in $2021.96$---ca. the time of publication. It is apparent now that the above emphasised orbital section $f\in(5\pi/6,7\pi/6)$ marks about the apocentre half of the orbit (blue section), while the rest of the orbit marks the pericentre half (orange section). Comparing this to~\citet[Sect.~2 \& Fig.~1]{GRAVITY+20_Schwarzschild_prec} we see that of the dataset used to derive the current upper dark mass bound ($0.1\%$ of $M_\bullet$ enclosed by S2) the VLTI/GRAVITY astrometric data is limited to the pericentre half $f\in(-5\pi/6,5\pi/6)$. The apocentre half is well covered by both VLT/NACO astrometric and VLT/SINFONI spectroscopic data, the former however being less accurate than GRAVITY.\footnote{Instruments references: \citeauthor{GRAVITY+17}~\citeyear{GRAVITY+17} (GRAVITY); \citeauthor{Lenzen+03}~\citeyear{Lenzen+03}; \citeauthor{Rousset+03}~\citeyear{Rousset+03} (NACO); \citeauthor{Eisenhauer+03}~\citeyear{Eisenhauer+03}; \citeauthor{Bonnet+04}~\citeyear{Bonnet+04} (SINFONI).} From the insights we gathered in Sect.~\ref{SS: orbital elements} we thus expect that a tracking of S2 with GRAVITY \citep[and ELT/MICADO; ][]{Davies+18, Davies+21} over the current apocentre half ending in ca. 2033 will allow to substantially improve the current upper dark mass bounds, or even to detect one if present and not much smaller than those considered here. Note that as evident from the figures, at the time of publication (ca. $2021.96$) we already entered this domain \citep{GRAVITY+21_mass_distribution}.

Having identified the orbital sections $f\in(-5\pi/6,5\pi/6)$ and $f\in(5\pi/6,7\pi/6)$ with the pericentre and apocentre halves respectively, we can now also gain a more intuitive understanding of the point already emphasised in Sect.~\ref{SS: orbital elements}, that the spatial separation between the orbital sections in which the extended mass is ineffective and effective is so clear cut. As established in Sect.~\ref{SS: form of a_XM}, the extended mass acceleration is given by $\mathbf a_\mathrm{XM}=\mathcal R_\mathrm{XM}(r)\mathbf n$ (Eqs. \eqref{E: Gaussian components}, \eqref{E: R XM}). For it to divert the course of the star from its otherwise Keplerian orbit, thus causing a precession, two things need to hold: Firstly, its magnitude $R_\mathrm{XM}(r)$ needs to be sufficiently high, and secondly, its direction $\mathbf n$ needs to have a sufficiently large normal component to the star's velocity vector. The former is given when there is a sufficient amount of extended mass between the star and the MBH, i.e when $r$ is sufficiently large. The latter is given in particular around both pericentre and apocentre. However from Figs.~\ref{F: orientation angles} and~\ref{F: orbit iiia} we see that both requirements are only well satisfied simultaneously in the apocentre half of the orbit.

\begin{figure}%[hbt]
\centering
\begin{subfigure}{.481\textwidth}
  \centering
  \includegraphics[clip=true,trim=0 0 0 0,scale=.49]{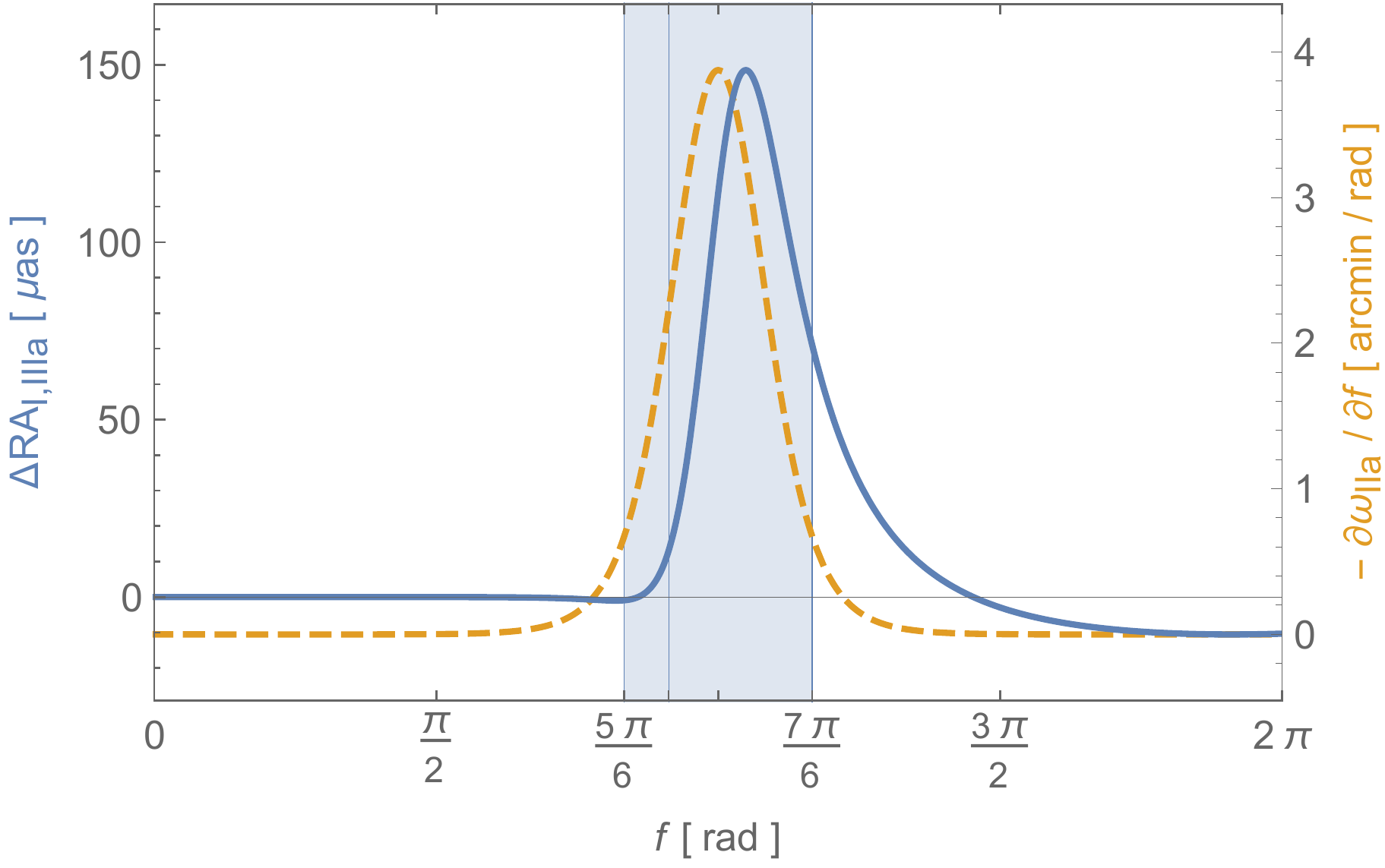}
  \caption{}
  \label{F: RA iiia}
\end{subfigure}
\begin{subfigure}{.481\textwidth}
  \centering
  \includegraphics[clip=true,trim=0 0 0 0,scale=.49]{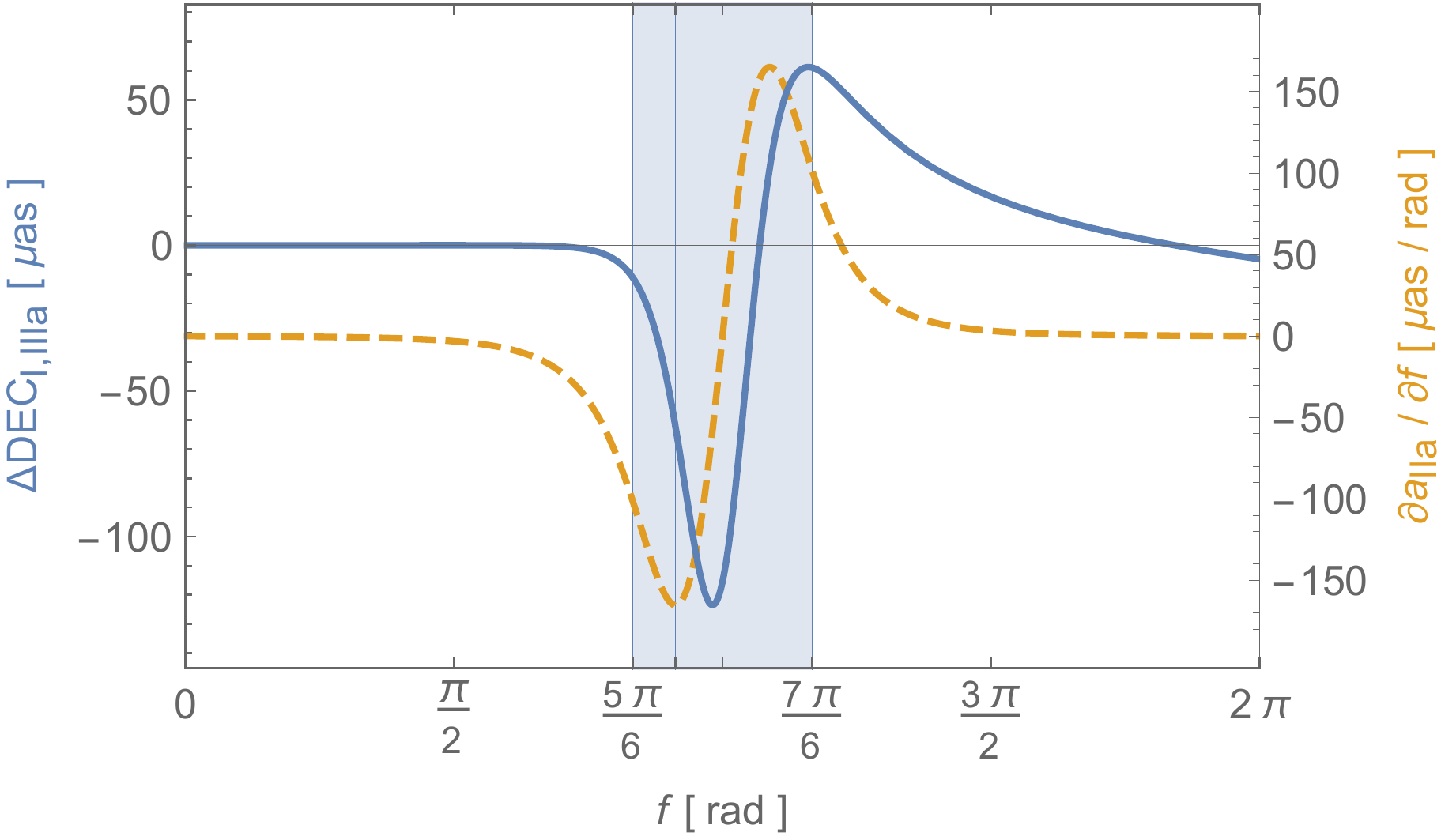}
  \caption{}
  \label{F: DEC iiia}
\end{subfigure}
\begin{subfigure}{.481\textwidth}
  \centering
  \includegraphics[clip=true,trim=0 0 0 0,scale=.49]{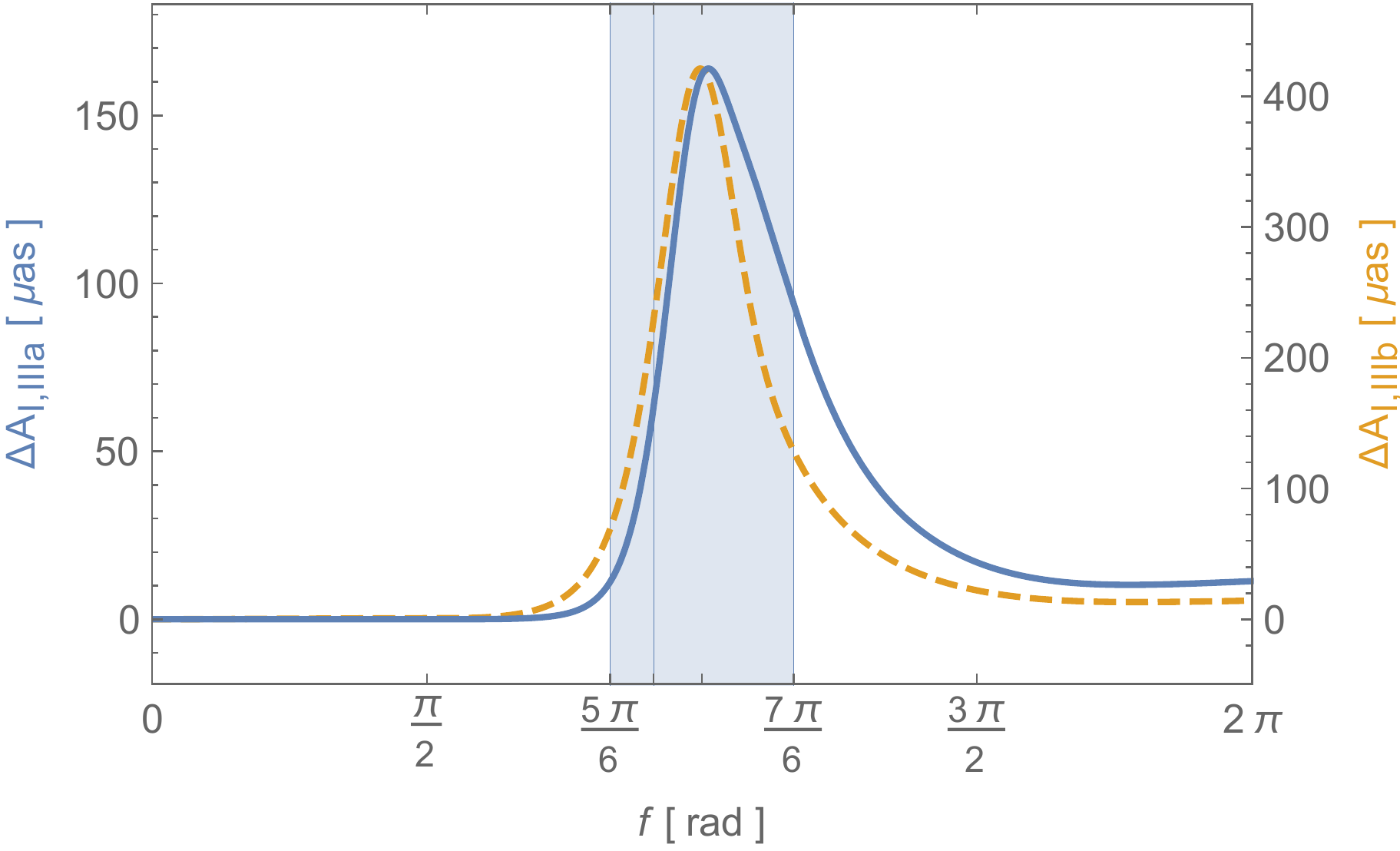}
  \caption{}
  \label{F: absAM}
\end{subfigure}
\caption{
Dark mass impact on astrometry. 
(\subref{F: RA iiia}) $\Delta\mathrm{RA_{I,IIIa}}(f) = \mathrm{RA_{IIIa}}(f) - \mathrm{RA_I}(f)$ together with $-\omega_\mathrm{IIa}'(f)$.
(\subref{F: DEC iiia})  Same for $\mathrm{DEC}$ together with $a_\mathrm{IIa}'(f)$.
(\subref{F: absAM})  $\Delta\mathrm{A_{I,IIIa}}(f) = \Delta\mathrm{RA}_\mathrm{I,IIIa}^2(f) + \Delta\mathrm{DEC}_\mathrm{I,IIIa}^2(f)$ together with the analogous quantity for model~\eqref{E: iiib}; note the different scales on the left and right vertical axes. Subscripts indicate the models to which the respective quantities correspond (Eqs.~\eqref{E: i}-\eqref{E: iiib}). The vertical line within the shaded blue region marks the true anomaly of model~\eqref{E: i} in $2021.96$, which is approximately the time of publication.}
\label{F: AM plots}
\end{figure}

In Fig.~\ref{F: AM plots} we examine the functional dependence of the astrometric quantities on $f$. Fig.~\ref{F: RA iiia} shows the difference in $\mathrm{RA}(f)$ between models~\eqref{E: i} and~\eqref{E: iiia}.
The plot is overlaid with that of $-\omega'(f)$ for model~\eqref{E: iia} in order to emphasise the connection with the estimate which we extracted from Figs.~\ref{F: om iia} and~\subref{F: om iiia} in Sect.~\ref{SS: orbital elements}.\footnote{We overlay it with $-\omega'$ and not $\omega$ because we plot the difference in $\mathrm{RA}$ between the model with and without extended mass. The minus  sign allows a direct comparison of $\mathrm{RA}$ with the, in this orbital section, counterclockwise apocentre shift.}
The orbit in the sky is oriented such that the projected minor-axis (major-axis) is roughly parallel to the $\mathrm{RA}$ ($\mathrm{DEC}$) axis and the extended mass causes a counterclockwise pericentre shift when the star is in the apocentre half (Figs.~\ref{F: orientation angles} and~\ref{F: orbit iiia}). This causes a shift of points on the orbit near apocentre in roughly $\mathrm{RA}$-direction.
%Due to the orientation of the orbit in the sky (Figs.~\ref{F: orientation angles} and~\ref{F: orbit iiia}) the counterclockwise pericentre shift results in a shift of points near apocentre in roughly $\mathrm{RA}$-direction.
The $\sim191\,\mathrm{\mu as}$ estimate of Sect.~\ref{SS: orbital elements} hence concerns the right ascension, and as a crude estimate is in agreement with the peak of $\sim150\,\mathrm{\mu as}$ of $\Delta\mathrm{RA}$ for the Plummer model close to apocentre in Fig.~\ref{F: RA iiia}. Furthermore, the plot shows that indeed the $\Delta\mathrm{RA}$ curve mimics that of $-\omega'(f)$. This is expected over the part of the curves where $\Delta\mathrm{RA}$ increases because the mass precession drives this deviation. It appears non-trivial, however, that this deviation decreases again after the peak influence of the extended mass. This demonstrates that not only is the apocentre half of the orbit the section in which the effect of the extended mass on astrometry starts to become discernible, but also that the most pronounced deviations are limited to this orbital section. This shows through yet another facet that it is indeed very important to resolve the region $f\in[5\pi/6,7\pi/6]$ well in order to capture the dark mass signature (see also Sect.~\ref{S: mock} for a support of this claim by means of a mock-data analysis).

Fig.~\ref{F: DEC iiia} shows the same as Fig.~\ref{F: RA iiia} for the declination. Here the plot for the difference in $\mathrm{DEC}(f)$ is overlaid with that of $a'(f)$ as the orientation of the orbit in the sky allows us to draw a direct connection between semi-major axis and declination. Again, we observe the mimicking of the two curves and a maximum deviation of $\sim125\,\mathrm{\mu as,}$ in agreement with our crude estimate of $\sim183\,\mathrm{\mu as}$ in Sect.~\ref{SS: orbital elements}. The deviation here also begins at the beginning of the apocentre half, then peaks during it, and declines again afterwards. Again, the cusp plots corresponding to Figs.~\ref{F: RA iiia} and~\ref{F: DEC iiia} show the same qualitative features and we do not show them separately. In terms of magnitude, the astrometric impact of the cusp is greater, however, as we show in the following.

In Fig.~\ref{F: absAM} we consider the absolute value of the astrometric deviation in the plane of the sky,
\begin{align}\label{E: absAM}
\Delta A = \sqrt{\Delta\mathrm{RA}^2 + \Delta\mathrm{DEC}^2},
\end{align}
between models~\eqref{E: i} and~\eqref{E: iiia}. To demonstrate the similar impacts of the Plummer and cusp models, we overlay the plot with the respective quantity for models~\eqref{E: i} and~\eqref{E: iiib}. As expected from Figs.~\ref{F: RA iiia} and~\subref{F: DEC iiia}, the effect for the Plummer model peaks with $\sim164\,\mathrm{\mu as}$ at about apocentre, and it is predominantly limited to the apocentre half of the orbit. In particular, the effect only kicks in once the star enters the apocentre half. The same holds for our Bahcall-Wolf cusp with $0.1\%$ of $M_\bullet$ within the apocentre of S2 (Eqs. \eqref{E: profiles}, \eqref{E: BW gamma}, \eqref{E: rho0}, \eqref{E: r0}), for which the peak is stronger, however, with $\sim420\,\mathrm{\mu as}$.

\begin{figure}%[hbt]
  \centering
  \includegraphics[clip=true,trim=0 0 0 0,scale=.49]{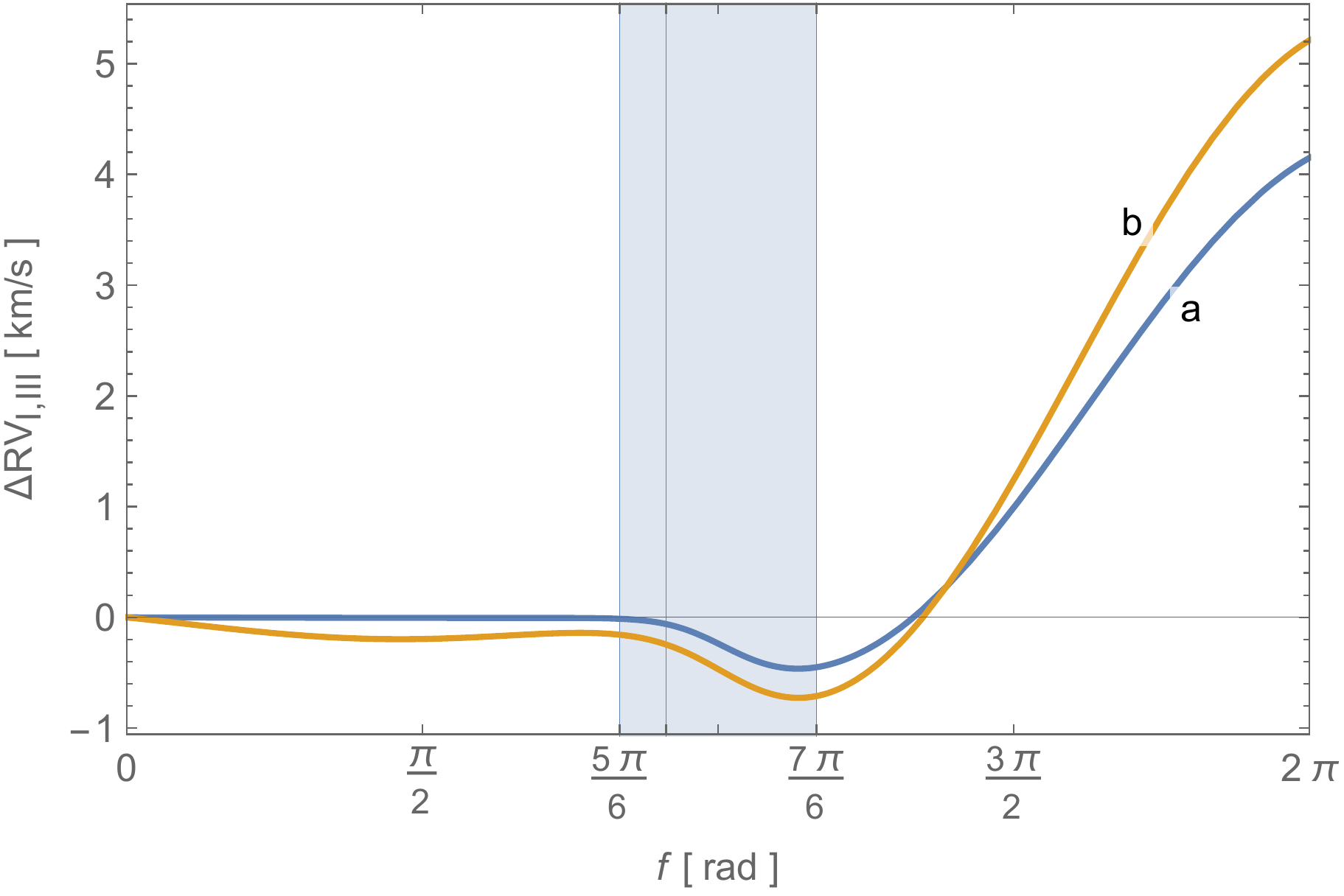}
  \caption{$\Delta\mathrm{RV}_\mathrm{I,IIIa}(f) = \mathrm{RV_{IIIa}}(f) - \mathrm{RV_{I}}(f)$ (blue) together with the analogous quantity for model~\eqref{E: iiib} (orange). The subscripts indicate the correspondence to the models~\eqref{E: i} and~\eqref{E: iiia}-\eqref{E: iiib}. The vertical line within the blue shaded region marks the true anomaly of model~\eqref{E: i} in $2021.96$---ca. the time of publication.}
  \label{F: RV}
\end{figure}

The observable left to analyse is radial velocity. In Fig.~\ref{F: RV} we plot the difference in $\mathrm{RV}(f)$ between models~\eqref{E: i} and~\eqref{E: iiia} together with the respective quantity for model~\eqref{E: iiib}.
Although in contrast to astrometry, the deviation does not decline after the main influence of the extended mass, the effect still begins in this orbital section. Consequently, the lesson of this plot is similar to that learned from Fig.~\ref{F: om3} in Sect.~\ref{SS: orbital elements}: Given data limited to the pericentre half, the sampling of the apocentre half of the orbit will clearly improve the detection prospects of a dark extended mass.

\subsection{Detection thresholds}\label{SS: detection thresholds}

In Sect.~\ref{SS: astrometry and velocity} we assessed the impact of an extended mass on the observables at the basis of the residuals between orbits with and without an extended mass component. These residuals are a priori not observables because they involve two orbits, while in an actual observation, nature only confronts us with one orbit, that is, with the actual one. The question we address now is how we can relate these residuals to observables, and how we can properly read them as detection thresholds for an extended mass based on the accuracy of the available data. For this analysis, we restrict ourselves to the Plummer model and to astrometry. It is then also part of our investigation of Sect.~\ref{S: mock} to examine the thresholds we obtained here theoretically from the perspective of a mock-data analysis.

Suppose that we are observing one full orbit in a situation with an extended mass. First we recall a key observation from the previous sections: The impact of the extended mass is negligible in the pericentre half. This insight allows us to identify this orbital section of the observed orbit with the corresponding orbital section of a hypothetical and thus unobservable orbit without extended mass. Consequently, this identification lifts residuals such as that for absolute astrometry $\Delta A$ into the domain of observables (see Eq.~\eqref{E: absAM} and Fig.~\ref{F: absAM}).

When we now restrict this to absolute astrometry, the question left to answer is the accuracy required for data in which parts of the orbit in order to resolve a certain astrometric residual $\Delta A.$  We tackle this question the other way around: Suppose that we have one orbit of (astrometrically) evenly distributed data given with astrometric accuracies of $\sigma_\mathrm p$ in the pericentre half $f\in[-5\pi/6,5\pi/6]$ and $\sigma_\mathrm a$ in the apocentre half $f\in[5\pi/6,7\pi/6]$. In a simple ansatz, we view the residual $\Delta A$ as given by the difference between two uncorrelated measurements with uncertainties $\sigma_\mathrm p$ and $\sigma_\mathrm a$, respectively, such that by error propagation, we have
\begin{align}\label{E: error prop}
\sigma_{\Delta A} = \kappa\sqrt{\sigma_\mathrm p^2 + \sigma_\mathrm a^2}
\end{align}
for the uncertainty of $\Delta A$. The proportionality constant $\kappa$ we introduced by hand in order to a posteriori account for details that we a priori neglected in our ansatz, such as that the uncertainty with which the overall orbits are determined decreases with the number of data points. We gauge $\kappa$ for the cases of our interest below. Importantly, once we have $\kappa$, then with Eq.~\eqref{E: error prop} we arrive at a relation between the astrometric accuracies of our data points and the resulting accuracy with which the astrometric residual $\Delta A$ in Fig.~\ref{F: absAM} can be resolved. In particular, we have
\begin{align}\label{E: detection threshold}
\sigma_{\Delta A} = \mathrm{max}(\Delta A)\qquad\text{at the $1\sigma$ detection threshold,}
\end{align}
and because $\Delta A$ is directly related to the extended mass parameters, we thus found a way to estimate which extended masses we should be able to detect with $1\sigma$ given our data.

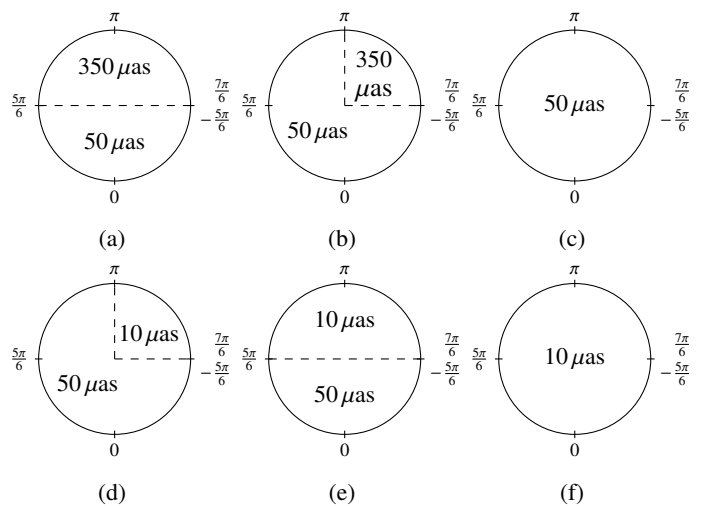
\begin{figure}%[H]
\centering
\begin{subfigure}{.3273\columnwidth}
  \centering
  \begin{tikzpicture}
  \draw (0,0) circle (1);
  \draw [dashed](-1,0) -- (1,0);
  \node at (0,.5){\small$350\,\mathrm{\mu as}$};
  \node at (0,-.5){\small$50\,\mathrm{\mu as}$};
  \draw (0,-.95) -- (0,-1.05);
  \node at (0,-1)[anchor=north]{$\scriptstyle0$};
  \draw (-1.05,0) -- (-.95,0);
  \node at (-1,0)[anchor=east]{$\scriptstyle\frac{5\pi}{6}$};
  \draw (0,1.05) -- (0,.95);
  \node at (0,1)[anchor=south]{$\scriptstyle\pi$};
  \draw (1.05,0) -- (.95,0);
  \node at (1,0)[anchor=west,align=left]{$\;\,\scriptstyle\frac{7\pi}{6}$\\$\scriptstyle-\frac{5\pi}{6}$};
  \end{tikzpicture}
  \caption{}
  \label{F: full orbit a}
\end{subfigure}
\begin{subfigure}{.3273\columnwidth}
  \centering
  \begin{tikzpicture}
  \draw (0,0) circle (1);
  \draw [dashed](0,0) -- (0,1);
  \draw [dashed](0,0) -- (1,0);
  \node at (45:.55)[align=left]{\small$350$\\$\mathrm{\mu as}$};
  \node at (225:.5){\small$50\,\mathrm{\mu as}$};
  \draw (0,-.95) -- (0,-1.05);
  \node at (0,-1)[anchor=north]{$\scriptstyle0$};
  \draw (-1.05,0) -- (-.95,0);
  \node at (-1,0)[anchor=east]{$\scriptstyle\frac{5\pi}{6}$};
  \draw (0,1.05) -- (0,.95);
  \node at (0,1)[anchor=south]{$\scriptstyle\pi$};
  \draw (1.05,0) -- (.95,0);
  \node at (1,0)[anchor=west,align=left]{$\;\,\scriptstyle\frac{7\pi}{6}$\\$\scriptstyle-\frac{5\pi}{6}$};
  \end{tikzpicture}
  \caption{}
  \label{F: full orbit b}
\end{subfigure}
\begin{subfigure}{.3273\columnwidth}
  \centering
  \begin{tikzpicture}
  \draw (0,0) circle (1);
  \node at (0,0){\small$50\,\mathrm{\mu as}$};
  \draw (0,-.95) -- (0,-1.05);
  \node at (0,-1)[anchor=north]{$\scriptstyle0$};
  \draw (-1.05,0) -- (-.95,0);
  \node at (-1,0)[anchor=east]{$\scriptstyle\frac{5\pi}{6}$};
  \draw (0,1.05) -- (0,.95);
  \node at (0,1)[anchor=south]{$\scriptstyle\pi$};
  \draw (1.05,0) -- (.95,0);
  \node at (1,0)[anchor=west,align=left]{$\;\,\scriptstyle\frac{7\pi}{6}$\\$\scriptstyle-\frac{5\pi}{6}$};
  \end{tikzpicture}
  \caption{}
  \label{F: full orbit c}
\end{subfigure}
\\
\begin{subfigure}{.3273\columnwidth}
  \centering
  \begin{tikzpicture}
  \draw (0,0) circle (1);
  \draw [dashed](0,0) -- (0,1);
  \draw [dashed](0,0) -- (1,0);
  \node at (35:.55){\small$10\,\mathrm{\mu as}$};
  \node at (225:.5){\small$50\,\mathrm{\mu as}$};
  \draw (0,-.95) -- (0,-1.05);
  \node at (0,-1)[anchor=north]{$\scriptstyle0$};
  \draw (-1.05,0) -- (-.95,0);
  \node at (-1,0)[anchor=east]{$\scriptstyle\frac{5\pi}{6}$};
  \draw (0,1.05) -- (0,.95);
  \node at (0,1)[anchor=south]{$\scriptstyle\pi$};
  \draw (1.05,0) -- (.95,0);
  \node at (1,0)[anchor=west,align=left]{$\;\,\scriptstyle\frac{7\pi}{6}$\\$\scriptstyle-\frac{5\pi}{6}$};
  \end{tikzpicture}
  \caption{}
  \label{F: full orbit d}
\end{subfigure}
\begin{subfigure}{.3273\columnwidth}
  \centering
  \begin{tikzpicture}
  \draw (0,0) circle (1);
  \draw [dashed](-1,0) -- (1,0);
  \node at (0,.5){\small$10\,\mathrm{\mu as}$};
  \node at (0,-.5){\small$50\,\mathrm{\mu as}$};
  \draw (0,-.95) -- (0,-1.05);
  \node at (0,-1)[anchor=north]{$\scriptstyle0$};
  \draw (-1.05,0) -- (-.95,0);
  \node at (-1,0)[anchor=east]{$\scriptstyle\frac{5\pi}{6}$};
  \draw (0,1.05) -- (0,.95);
  \node at (0,1)[anchor=south]{$\scriptstyle\pi$};
  \draw (1.05,0) -- (.95,0);
  \node at (1,0)[anchor=west,align=left]{$\;\,\scriptstyle\frac{7\pi}{6}$\\$\scriptstyle-\frac{5\pi}{6}$};
  \end{tikzpicture}
  \caption{}
  \label{F: full orbit e}
\end{subfigure}
\begin{subfigure}{.3273\columnwidth}
  \centering
  \begin{tikzpicture}
  \draw (0,0) circle (1);
  \node at (0,0){\small$10\,\mathrm{\mu as}$};
  \draw (0,-.95) -- (0,-1.05);
  \node at (0,-1)[anchor=north]{$\scriptstyle0$};
  \draw (-1.05,0) -- (-.95,0);
  \node at (-1,0)[anchor=east]{$\scriptstyle\frac{5\pi}{6}$};
  \draw (0,1.05) -- (0,.95);
  \node at (0,1)[anchor=south]{$\scriptstyle\pi$};
  \draw (1.05,0) -- (.95,0);
  \node at (1,0)[anchor=west,align=left]{$\;\,\scriptstyle\frac{7\pi}{6}$\\$\scriptstyle-\frac{5\pi}{6}$};
  \end{tikzpicture}
  \caption{}
  \label{F: full orbit f}
\end{subfigure}
\caption{Schematics of six cases with data over one full orbit, differing by the astrometric accuracy. The bottom semi-circles represent the pericentre half of the orbit, and the top semi-circles show the apocentre half. Outside numbers mark the true anomaly, and inside numbers indicate the astrometric accuracies of the data in the orbital sections.}
\label{F: full orbit cases}
\end{figure}

We demonstrate this with the six cases of Fig.~\ref{F: full orbit cases}, each having the same number of (astrometrically) evenly distributed data points on one full orbit $f\in[-5\pi/6,7\pi/6]$, ensuring a balanced sampling of the pericentre and apocentre halves, and consequently, of the signatures of the Schwarzschild and mass precessions (Fig.~\ref{F: orbit iiia}).
 %%%%%%%%%%%%%%%%%%%%%%%%%%%%%%%%%%%%%%%%%%%
The cases differ, however, in the accuracy of these data in different orbital sections. The case of Fig.~\ref{F: full orbit a} thereby mimics the data from which the current $1\sigma$ upper bound of $0.1\%$ of $M_\bullet$ ($4261\,M_\odot$) enclosed by S2 has been deduced, with the $50\,\mathrm{\mu as}$ errors on the pericentre half representing the GRAVITY data, and the $350\,\mathrm{\mu as}$ errors on the apocentre half representing the NACO data  \citep[Sect.~2 \& Fig.~1]{GillessenEtAl2017, GRAVITY+20_Schwarzschild_prec}. We can thus use this case to roughly gauge $\kappa$ against observation by evaluating Eq.~\eqref{E: error prop} at the detection threshold (Eq.~\eqref{E: detection threshold}) given by the peak value $\mathrm{max}(\Delta A) = 164\,\mathrm{\mu as,}$ which we found for model~\eqref{E: iiia} in Fig.~\ref{F: absAM}. This yields $\kappa\approx1/2,$ and adopting this value for all other cases, we can now calculate their $\sigma_{\Delta A}$ from Eq.~\eqref{E: error prop} and again identify them with the respective threshold peaks $\mathrm{max}(\Delta A)$ by Eq.~\eqref{E: detection threshold}.
\begin{table}%[hbt]
\caption{Estimated $1\sigma$ Plummer mass detection thresholds for the cases of Fig.~\ref{F: full orbit cases}.}
\label{Tbl: thresholds}
\centering
\begin{tabular}{l l l l}
\hline\hline
Case    & Threshold                                                                     & Density parameter                               & Enclosed mass                 \\
                & $\sigma_{\Delta A}$ $[\mathrm{\mu as}]$               & $\rho_0$ $[\mathrm{kg/m^3}]$    & $[M_\odot]$   \\
\hline
Fig.~\ref{F: full orbit a}      & 164   & $1.69\times10^{-10}$  & $4261$        \\
Fig.~\ref{F: full orbit b}      & 103   & $1.06\times10^{-10}$  & $2828$        \\
Fig.~\ref{F: full orbit c}      & 35            & $3.64\times10^{-11}$  & $969$   \\
Fig.~\ref{F: full orbit d}      & 29            & $3.00\times10^{-11}$  & $799$   \\
Fig.~\ref{F: full orbit e}      & 26            & $2.63\times10^{-11}$  & $699$   \\
Fig.~\ref{F: full orbit f}      & 7             & $7.28\times10^{-12}$  & $194$   \\
\hline
\end{tabular}
\tablefoot{The enclosed mass denotes the extended mass component within the apocentre distance of the initial osculating orbit, $p_0/(1-e_0)$ (Eq. \eqref{E: ID2}).}
\end{table}
The resulting values are given in the second column of Table~\ref{Tbl: thresholds}.

With this we can now for instance examine how low a Plummer mass of the same spatial scale of Eq.~\eqref{E: r0} we would still be able to detect with $\sim1\sigma$ for each case. To do this, we vary the density parameter $\rho_0$ of model~\eqref{E: iiia} until the peak value $\mathrm{max}(\Delta A)$ in Fig.~\ref{F: absAM} meets the thresholds we just calculated. The resulting values are given in Table~\ref{Tbl: thresholds} together with the corresponding number of solar masses within the apocentre of S2. As more data are taken with GRAVITY or MICADO in the coming years, the accuracy of the observational data will gradually improve following the analogous sequence of cases from Figs.~\ref{F: full orbit a} to~\subref{F: full orbit c}. Hence, from our estimates of Table~\ref{Tbl: thresholds}, we can await to be able to improve the current upper bound by about a factor $0.7$ until the next apocentre passage in approximately 2026, and by about a factor $0.2$ until a full orbit is sampled with GRAVITY (and MICADO) at its current accuracy performance of $50\,\mathrm{\mu as}$. During the publication process of the present work, a new upper bound has been obtained with the data gathered until August 2021, which in our scheme falls between the cases of Figs.~\ref{F: full orbit a} and~\ref{F: full orbit b} (see the 2021.96 indicators in the figures of this section). The result agrees excellently with our predictions \citep{GRAVITY+21_mass_distribution}. The cases of Figs.~\ref{F: full orbit d} to~\subref{F: full orbit f} would become relevant if the accuracy of GRAVITY could be improved to its performance goal of $10\,\mathrm{\mu as}$. However, based on Table~\ref{Tbl: thresholds} we would only await a further substantial improvement of the dark mass sensitivity due to this improvement once a full orbit were be sampled with that accuracy.

\subsection{Possible reservations}\label{SS: reservations}

We conclude this section with some comments about possible reservations for our above theoretical analysis and the conclusions drawn from it.

Firstly, all model orbits of our preceding investigations have the same initial osculating orbit at pericentre $f=0$ (Eqs.~\eqref{E: ID2}--\eqref{E: ID3}), and the question may be whether a bias results from this particular choice that might endanger the generality of our results. If we had chosen apocentre $f=\pi$ as the common initial location instead, then taking the example of Fig.~\ref{F: om3},  all three curves would indeed coincide in the middle of the apocentre half by construction, and curve I would be farthest apart from curves IIIa, b in the pericentre half. However, this does not change the fact that then also, the slopes of the curves with and without extended mass disagree in the shaded blue region, which cause the deviation in the first place. In this sense, the slopes are more important here in the interpretation than the absolute values, which is also why we overlaid most curves with respective to first derivatives (see for example Figs.~\ref{F: om iia} and~\ref{F: a iia}, in which the influence of the extended mass is isolated). Hence also for $f_0=\pi$, as well as for any other $f_0$, we would have arrived at the same conclusions, that is, that the extended mass drives the orbital change almost exclusively in the apocentre half. We refer to Sect.~\ref{S: mock} below to rout out any remaining doubts. There we double-check all our theoretically obtained results in a mock-data analysis, which is not subject to any bias such as the above.

Secondly, in this section, we investigated the functional form of different quantities with true anomaly $f$ as opposed to time $t$. Because the times at which the star reaches a certain $f$ do not exactly coincide between our models with and without extended mass, this means for the residuals of Figs.~\ref{F: AM plots} and~\ref{F: RV}  that we compared the observables of the stars at slightly different times. The utility of our conclusions drawn from these plots for observation (in particular the detection threshold estimates of Table~\ref{Tbl: thresholds}) might therefore be doubtful because observations are made in the time domain. We first note in this respect that the respective time differences are small (see the time labels in Figs.~\ref{F: omega plots} and~\ref{F: a plots}). Furthermore, the mock-data analysis of Sect.~\ref{S: mock}, which operates in the time domain, serves as an independent verification in this question as well.

Finally, we comment on how we translated astrometric measurement accuracies into obtainable $1\sigma$ detection thresholds in Sect.~\ref{SS: detection thresholds}. This approach is both ad hoc and heuristic, and we gauged $\kappa$ against an observational case that is only approximately represented by Fig.~\ref{F: full orbit a}. In particular, we neglected here the technical but important point of aligning the reference frames of the different instruments \citep{Plewa+15, GRAVITY+20_Schwarzschild_prec}. Furthermore, we only took one apocentre half of NACO mock data in order to make this case directly comparable to the others, and the real data violate our assumption of having the same number of evenly distributed points in both orbital halves. An assessment of the accuracy of our theoretical detection thresholds of Table~\ref{Tbl: thresholds} is thus also part of the mock-data analysis in Sect.~\ref{S: mock}.

\section{Mock data analysis}\label{S: mock}

We now complement the theoretical results of Sect.~\ref{S:  effects} with a mock-data analysis. In Sect.~\ref{SS: mock setup} we lay out the common setup for the following investigations: Firstly, we assess in Sect.~\ref{SS: full orbit} the dark mass sensitivity of increasingly good astrometric datasets for one full orbit, and we assess our theoretically estimated detection thresholds of Table~\ref{Tbl: thresholds} (Sect.~\ref{SS: detection thresholds}). Secondly, we investigate in Sect.~\ref{SS: partial orbit} the dark mass sensitivity of data restricted to one half or three quarters of an orbit. Here we support in particular our theoretical claim that data limited to the pericentre half are not sensitive to a dark mass. In Sect.~\ref{SS: flipped accuracies} we extend the latter analysis by two further cases with again a full orbit of data, but with flipped accuracies compared to the corresponding cases of Sect.~\ref{SS: full orbit}. Finally, we discuss the parameter correlations in Sect.~\ref{SS: correlations}.

\subsection{Setup}\label{SS: mock setup}

Throughout Sect.~\ref{S: mock}, we consider the 1PN + Plummer mass model~\eqref{E: iiia} with a scale parameter $r_0$ of $0.3\mathrm{\,as}$ (Eqs.~\eqref{E: profiles}, \eqref{E: r0}) and we use our Python-based code osculating orbits in General Relativistic environments (OOGRE), which we developed while working on this paper. The code is private for now, but we have plans to make it public. OOGRE is a perturbed Kepler-orbit model code optimised for stars in the galactic centre. It is based on the formalism laid out in Sect.~\ref{S: model} with the following adjustments and additions:

Firstly, we now also took into account the R\o mer delay, the motion of the Solar System, the special relativistic transverse Doppler shift, and the gravitational redshift into the model \citep[Appendix~\ref{App: Roemer}; ][]{KopeikinOzernoy1999, Grould+17}. The component of the relative motion of the Solar System and the MBH parallel to the line of sight thereby introduces a $Z$-velocity parameter $v^Z_0$ in the model. These additional effects would have unnecessarily complicated the discussion for the preceding theoretical investigations, but they are important in order to adequately simulate an observational setting.

Secondly, from the set of six initial osculating orbital elements (Eqs. \eqref{E: ID2}--\eqref{E: ID3}), we swapped the semi-latus rectum $p_0$ and the true anomaly $f_0$ for the orbital period $P_0$ and the time of the (approximately 2018) pericentre passage $T_0$.\footnote{Like all other parameters, $T_0$ also belongs to the initial osculating orbit, that is, it is not the time of pericentre passage of the physical orbit, but of this particular osculating orbit.} We also switched to the opposite inclination angle convention (Appendix~\ref{App: conventions}) such that in total we have $(P_0, T_0, e_0, i_0, \Omega_0, \omega_0)$ as our new set of initial osculating orbital elements. This also represents a complete set of Keplerian orbital elements. No information is lost. We can for instance recover $p_0$ via $P_0$ and $e_0$ using~\eqref{E: semi-major axis} together with Kepler's third law of planetary motion. We made this change of parameters only to adapt OOGRE to our fitting code infrastructure, which interfaces in the same way to our (fully) general relativistic orbit model based on the GYOTO code \citep{Vincent+11, Grould+16, Grould+17, Paumard+19}. The equations integrated at the core of OOGRE remain the same as before (Eqs.~\eqref{E: p dot}--\eqref{E: f dot}).

Finally, also for the sake of consistent interfacing, we considered these elements now as corresponding to the orbit that osculates at the apocentre passage at time $T_0-P_0/2$ (approximately 2010). Based on this, we have a model~\eqref{E: iiia} that, in this section, is specified by the 10 undetermined parameters of the 11-parameter set
\begin{align}\label{E: model}
(P_0, T_0, e_0, i_0, \Omega_0, \omega_0, M_0, R_0, v^Z_0, \rho_0, r_0 = 0.3\,\mathrm{as}) .
\end{align}

To create mock data, we chose the following specific model: For the initial osculating orbital elements, $M_\bullet$, $R_0$, and $v^Z_0$ we again selected the best-fit values of \citet[Table~1]{GRAVITY+20_Schwarzschild_prec}. Except when noted otherwise, we used a density parameter $\rho_0$ of the value given in Eq.~\eqref{E: rho0} such that the profile corresponds to the current $0.1\%$ of $M_\bullet$ upper bound of~\citet{GRAVITY+20_Schwarzschild_prec}. The exceptions are those cases of Sect.~\ref{SS: full orbit} by means of which we examine the detection thresholds of Table~\ref{Tbl: thresholds}, as well as some cases of Sect.~\ref{SS: partial orbit}, in which we increase the extended mass density by a factor $100$. Hence our model for creating the mock data is given by Eq.~\eqref{E: model} with a choice for the extended mass density parameter $\rho_0$, and rounded to two digits,
\begin{align}
M_\bullet&\approx4.26\times10^6\,M_\odot        &       R_0&\approx8246.7\,\mathrm{pc}  &       v^Z_0&\approx-1.6\,\mathrm{km/s}
\label{E: ID1 new} \\
P_0&\approx16.05\,\mathrm{yr}                   &       T_0&\approx2018.38\,\mathrm{yr} &       e_0&\approx0.88
\label{E: ID2 new} \\
i_0&\approx2.35                                         &       \Omega_0&\approx3.98                         &       \omega_0&\approx1.16     .
\label{E: ID3 new}
\end{align}

For each case, we chose $114$ mock observation dates, and for each date, we took one astrometric and one spectroscopic data point $\mathrm{(RA, DEC, RV)}$. The number $114$ was chosen because in the dataset that was used to derive the current upper dark mass bound ($0.1\%$ of $M_\bullet$ enclosed by S2), there are $57$ GRAVITY data points of S2 that are distributed over the pericentre half of the orbit \citep[Sect.~2, Fig.~1]{GRAVITY+20_Schwarzschild_prec}.\footnote{We note that at the time of publication, S2 already entered the apocentre half, such that more data are taken already (see the figures of Sect.~\ref{S: effects} and \citeauthor{GRAVITY+21_mass_distribution}~\citeyear{GRAVITY+21_mass_distribution}).} An equally dense sampling of the apocentre half would yield $114$ data points over one full orbit. For a chosen $\rho_0$, the model orbit was then sampled at the observation dates, and the mock data then follow by offsetting the sample points by errors drawn from normal distributions with standard deviations of $10\,\mathrm{km/s}$ in $\mathrm{RV}$ and case-dependent standard deviations in astrometry. Finally, for each case, we considered an ensemble of $250$ such mock datasets.

We then performed a Markov chain Monte Carlo (MCMC) fit of the ten-parameter model~\eqref{E: model} ($r_0$ being fixed) to each mock dataset.\footnote{Here we used the emcee Python package by \citet{Foreman-Mackey+13}, based on affine-invariant sampling \citep{GoodmanWeare10}.} From Fig.~\ref{F: orientation angles}, the model curves $(\mathrm{RA}(t), \mathrm{DEC}(t), \mathrm{RV}(t))$ are given by the $Y$ and $X$ components of $\mathbf r(t)$ and the $-Z$ component of $\mathbf v(t),$ respectively. They were calculated from Eqs.~\eqref{E: POS} and \eqref{E: VEL} with the osculating orbital elements resulting from the integration of Eqs.~\eqref{E: p dot}--\eqref{E: f dot} for the models~\eqref{E: model} with parameters out of a rectangular domain around Eqs.~\eqref{E: ID1 new}--\eqref{E: ID3 new}. The results are then discussed in particular by means of of the statistics of the best-fit values and error estimates (as given by the median values and $1/2$ times the difference between the $84\%$ and $16\%$ percentiles of the ensemble of $250$ adequately burned-in MCMC chains) of the density parameter $\rho_0$ for each case. We also discuss parameter correlations at the hand of corner plot representations of the posterior distributions in the parameter space. In all cases in which the MCMC fits converge, the chains hit autocorrelation after $500$ to~$1500$ steps with $100$ walkers, after which we burned-in and continued to sample the central mode for another $2500$ steps. This means that our statistics for each case is based on fits to $250$ mock datasets, and parameter chains for each fit with $2500\times100$ mixing samples, that is, with the burn-in samples already discarded.

\subsection{Full orbit of data with increasing sensitivity}\label{SS: full orbit}

\begin{figure}%[H]
\centering
\begin{subfigure}{.495\columnwidth}
        \begin{tikzpicture}[scale=.4]
                \node[anchor=south west,inner sep=0] at (0,0)
                        {\includegraphics[width=\textwidth,clip=true,trim=6.5 8 5 7,scale=.62,left]{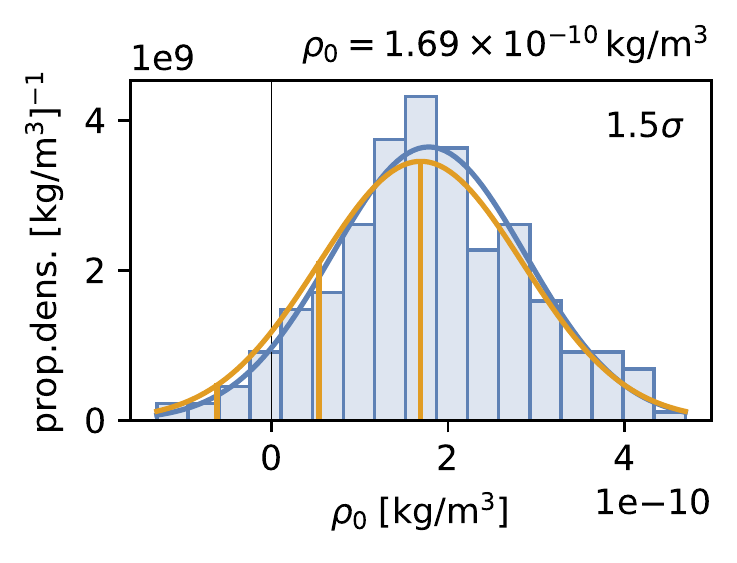}};
                \coordinate (o) at (2.85,6.1);
                \draw (o) circle (1);
                \draw [dashed]($ (o) + (-1,0) $) -- +(2,0);
                \node at ($(o) + (0,.5)$){$\scriptstyle350$};
                \node at ($(o) - (0,.5)$){$\scriptstyle50$};
        \end{tikzpicture}
        \caption{}
        \label{F: case7 hist a}
\end{subfigure}
\begin{subfigure}{.495\columnwidth}
        \begin{tikzpicture}[scale=.4]
                \node[anchor=south west,inner sep=0] at (0,0)
                        {\includegraphics[width=\textwidth,clip=true,trim=6.5 8 5 7,scale=.62,left]{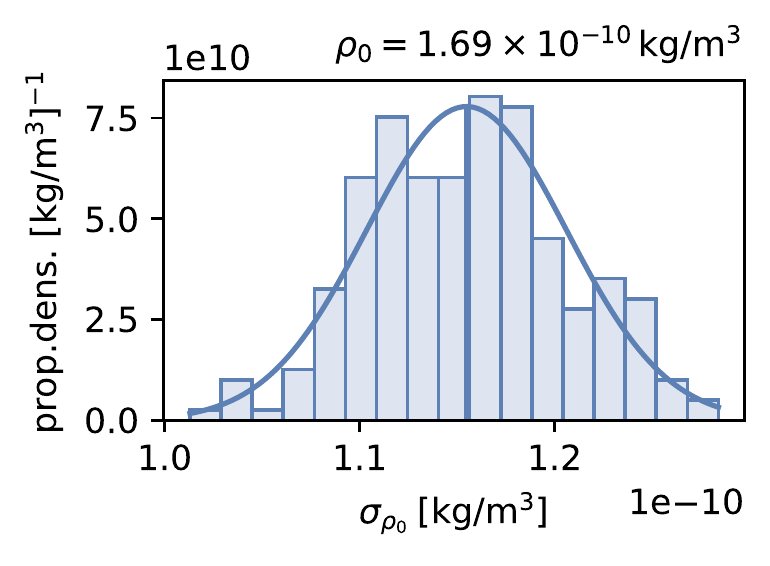}};
        \end{tikzpicture}
        \caption{}
        \label{F: case7 hist b}
\end{subfigure}
\caption{Density parameter statistics of the fits to $250$ mock datasets of the case of Fig.~\ref{F: full orbit a} with histograms of the best-fit parameters (\subref{F: case7 hist a}) and of the corresponding error estimates (\subref{F: case7 hist b}). The blue normal distributions have means and variances according to the data of the histograms. The orange normal distribution in~(\subref{F: case7 hist a}) has a mean of Eq.~\eqref{E: rho0} (i.e. the value the mock data are based on) and a variance equal to the mean of the distribution in~(\subref{F: case7 hist b}). The latter distribution has marks for 1 and 2 one-sided standard deviations. The number of sigmas with which zero is rejected is shown in the top right.}
\label{F: case7 hist}
\end{figure}
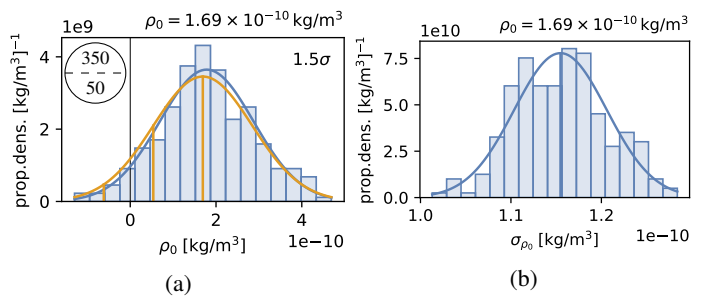

In our first investigation, we fixed $\rho_0$ to the current upper bound value of Eq.~\eqref{E: rho0} and determined how the sensitivity to this particular dark mass increased with the astrometric accuracy along the sequence of cases from Figs.~\ref{F: full orbit a} to~\subref{F: full orbit f}. The $114$ data points are thereby roughly (astrometrically) evenly distributed over the orbit. Fig.~\ref{F: case7 hist} shows the density parameter statistics for the fits to the $250$ mock datasets of the case of Fig.~\ref{F: full orbit a} (as indicated in the top left corner of the first plot).
 %%%%%%%%%%%%%%%%%%%%%%%%%%%%%%%%%%%%%%%%%%%
Panel~(\subref{F: case7 hist a}) shows a histogram of the best-fit values for $\rho_0$. Panel~(\subref{F: case7 hist b}) shows the corresponding histogram of the MCMC error estimates. We note that the error distribution is much narrower (a difference of about two orders of magnitude in standard deviation) than the value distribution. Both histograms are overlaid with a normal distribution plotted in blue, whose means and variances are taken from the data of the respective histograms. Additionally, Fig.~\ref{F: case7 hist a} also shows a normal distribution plotted in orange. Its mean equals Eq.~\eqref{E: rho0}, that is, the value on which the mock data are based, and its variance equals the mean of the error distribution (Fig.~\ref{F: case7 hist b}). For large numbers of mock datasets, the two distributions should converge. In particular, the mean of the blue distribution should converge to the mean of the orange distribution. With our statistics from $250$ mock datasets, these two means are already merely $\sim 9.02\times10^{-12}\mathrm{\,kg/m^3}$ or $\sim 0.078\sigma$ apart. We therefore chose to interpret our results at the hand of the orange distribution that was constructed in this way, although it would not make a difference for the accuracy required for our analysis if we chose to use the blue distribution. In this sense, we conclude that in the case of Fig.~\ref{F: full orbit a} with a $\rho_0$ given by Eq.~\eqref{E: rho0}, the null hypothesis $\rho_0 = 0$ of having no extended mass is rejected with about $1.5\sigma$, as indicated in the top right corner of Fig.~\ref{F: case7 hist a}. This is roughly the expected sensitivity because this case was chosen in coarse analogy to the data from which the current $1\sigma$ upper bound was determined by \citet{GRAVITY+20_Schwarzschild_prec}.

\begin{figure}%[H]
\centering
\begin{subfigure}{.495\columnwidth}
        \begin{tikzpicture}[scale=.4]
                \node[anchor=south west,inner sep=0] at (0,0)
                        {\includegraphics[width=\textwidth,clip=true,trim=6.5 8 5 7,scale=.62,left]{case7_hist_1}};
                \coordinate (o) at (2.85,6.1);
                \draw (o) circle (1);
                \draw [dashed]($ (o) + (-1,0) $) -- +(2,0);
                \node at ($(o) + (0,.5)$){$\scriptstyle350$};
                \node at ($(o) - (0,.5)$){$\scriptstyle50$};
        \end{tikzpicture}
        \caption{}
        \label{F: full orbit p1 a}
\end{subfigure}
\begin{subfigure}{.495\columnwidth}
        \begin{tikzpicture}[scale=.4]
                \node[anchor=south west,inner sep=0] at (0,0)
                        {\includegraphics[width=\textwidth,clip=true,trim=6.5 8 5 7,scale=.62,right]{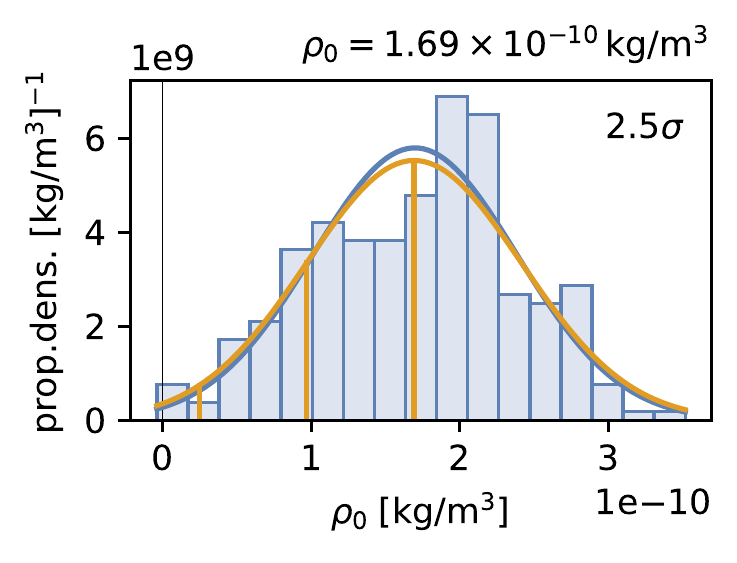}};
                \coordinate (o) at (3.4,6.1);
                \draw (o) circle (1);
                \draw [dashed](o) -- +(0,1);
                \draw [dashed](o) -- +(1,0);
                \node at ($ (o) + (39:.55)$)[align=left]{$\scriptstyle350$};
                \node at ($ (o) + (225:.5)$){$\scriptstyle50$};
        \end{tikzpicture}
        \caption{}
        \label{F: full orbit p1 b}
\end{subfigure}
\\
\begin{subfigure}{.495\columnwidth}
        \begin{tikzpicture}[scale=.4]
                \node[anchor=south west,inner sep=0] at (0,0)
                        {\includegraphics[width=\textwidth,clip=true,trim=6.5 8 5 -5,scale=.60,left]{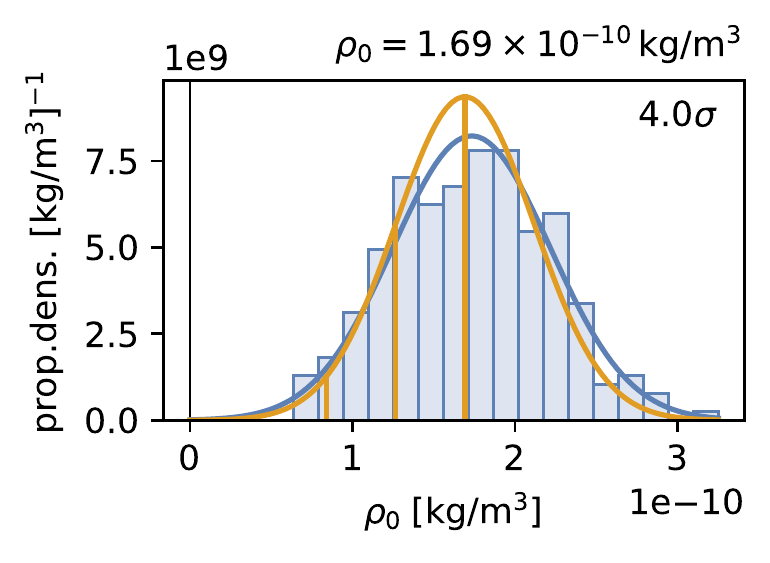}};
                \coordinate (o) at (3.8,5.8);
                \draw (o) circle (1);
                \node at (o){$\scriptstyle50$};
        \end{tikzpicture}
        \caption{}
        \label{F: full orbit p1 c}
\end{subfigure}
\begin{subfigure}{.495\columnwidth}
        \begin{tikzpicture}[scale=.4]
                \node[anchor=south west,inner sep=0] at (0,0)
                        {\includegraphics[width=\textwidth,clip=true,trim=6.5 8 5 -5,scale=.60,right]{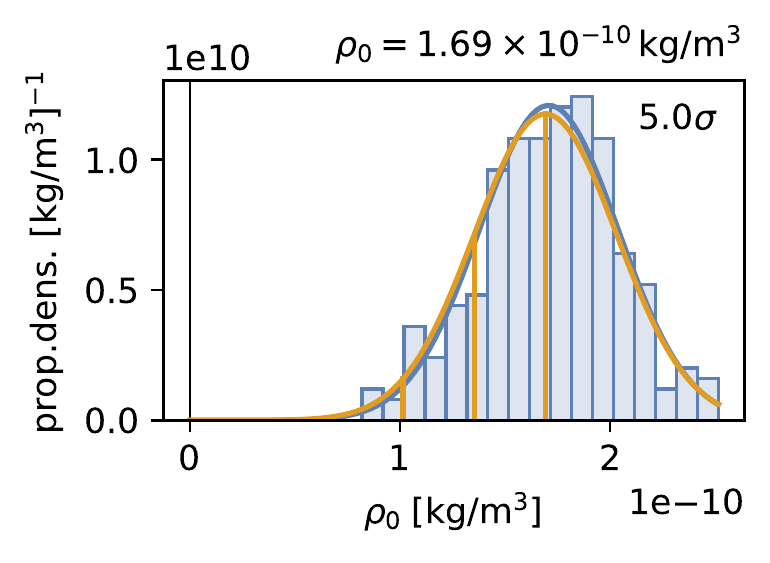}};
                \coordinate (o) at (3.8,5.8);
                \draw (o) circle (1);
                \draw [dashed](o) -- +(0,1);
                \draw [dashed](o) -- +(1,0);
                \node at ($ (o) + (45:.55)$)[align=left]{$\scriptstyle10$};
                \node at ($ (o) + (225:.5)$){$\scriptstyle50$};
        \end{tikzpicture}
        \caption{}
        \label{F: full orbit p1 d}
\end{subfigure}
\\
\begin{subfigure}{.495\columnwidth}
        \begin{tikzpicture}[scale=.4]
                \node[anchor=south west,inner sep=0] at (0,0)
                        {\includegraphics[width=\textwidth,clip=true,trim=6.5 8 5 -5,scale=.61,left]{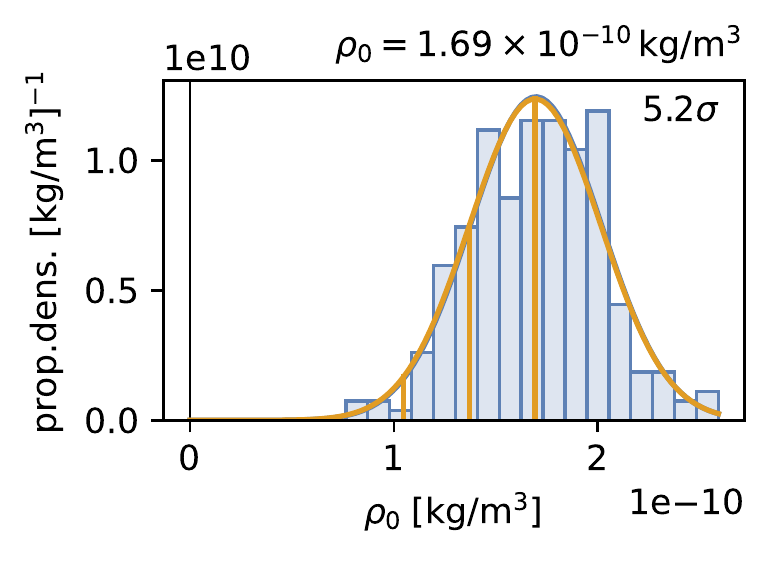}};
                \coordinate (o) at (3.8,5.8);
                \draw (o) circle (1);
                \draw [dashed]($ (o) + (-1,0) $) -- +(2,0);
                \node at ($(o) + (0,.5)$){$\scriptstyle10$};
                \node at ($(o) - (0,.5)$){$\scriptstyle50$};
        \end{tikzpicture}
        \caption{}
        \label{F: full orbit p1 e}
\end{subfigure}
\begin{subfigure}{.495\columnwidth}
        \begin{tikzpicture}[scale=.4]
                \node[anchor=south west,inner sep=0] at (0,0)
                        {\includegraphics[width=\textwidth,clip=true,trim=6.5 8 5 -5,scale=.61,right]{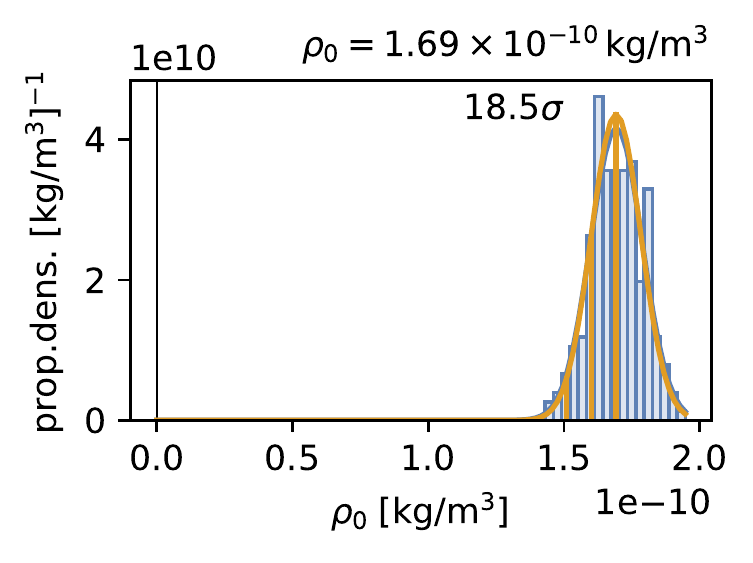}};
                \coordinate (o) at (3.4,5.9);
                \draw (o) circle (1);
                \node at (o){$\scriptstyle10$};
        \end{tikzpicture}
        \caption{}
        \label{F: full orbit p1 f}
\end{subfigure}
\caption{Density parameter statistics for all cases of Fig.~\ref{F: full orbit cases} with the mock data created using $\rho_0=1.69\times10^{-10}\mathrm{\,kg/m^3}$, corresponding to $0.1\%$ of $M_\bullet$ within the apocentre of S2 and to the current upper bound of~\citet{GRAVITY+20_Schwarzschild_prec}.}
\label{F: full orbit p1}
\end{figure}

Repeating the above analysis for the remaining cases of Fig.~\ref{F: full orbit cases} results in Fig.~\ref{F: full orbit p1}, where again the sketches in the top left corner of the plots indicate the respective cases, and the number in the top right corner represents the rejection of the null hypothesis $\rho_0=0$.
 %%%%%%%%%%%%%%%%%%%%%%%%%%%%%%%%%%%%%%%%%%%
The sensitivity of the data to the dark mass of the current upper bound increases with the cases from (\subref{F: full orbit p1 a}) to (\subref{F: full orbit p1 f}), that is, with decreasing astrometric errors. Comparing Figs.~\ref{F: full orbit p1 a} and~\ref{F: full orbit p1 b}, we estimate that a continued monitoring of S2 with GRAVITY at its current performance until its next apocentre passage in 2026 will improve the sensitivity to this particular dark mass by $1\sigma$. The gain achievable with data gathered until the time of publication of the present work (approximately $2021.96$) should consequently lie somewhere between zero and $1\sigma$. From Fig.~\ref{F: full orbit p1 c} we infer that another 1.5 $\sigma$ are gained by monitoring until a full orbit of GRAVITY data is taken in 2033, even without the use of NACO data. At this point, we could think about limiting the dataset to GRAVITY in order to fully take advantage of the instrument's superior systematics (see the discussion in Sect.~\ref{SS: implications}). The remaining Figs.~\ref{F: full orbit p1 d} to~\subref{F: full orbit p1 e} give an impression of how the dark mass sensitivity would improve if GRAVITY would be able to lower its systematic uncertainties to the level of the statistical uncertainties. Notably, from comparing Figs.~\ref{F: full orbit p1 e} and~\subref{F: full orbit p1 f}, we see that the level of improvement would be particularly high if a full orbit were monitored with the increased accuracy.

\begin{figure}%[H]
\centering
\begin{subfigure}{.495\columnwidth}
        \begin{tikzpicture}[scale=.4]
                \node[anchor=south west,inner sep=0] at (0,0)
                        {\includegraphics[width=\textwidth,clip=true,trim=6.5 8 5 7,scale=.62,left]{case7_hist_1}};
                \coordinate (o) at (2.85,6.1);
                \draw (o) circle (1);
                \draw [dashed]($ (o) + (-1,0) $) -- +(2,0);
                \node at ($(o) + (0,.5)$){$\scriptstyle350$};
                \node at ($(o) - (0,.5)$){$\scriptstyle50$};
        \end{tikzpicture}
        \caption{}
        \label{F: full orbit th a}
\end{subfigure}
\begin{subfigure}{.495\columnwidth}
        \begin{tikzpicture}[scale=.4]
                \node[anchor=south west,inner sep=0] at (0,0)
                        {\includegraphics[width=\textwidth,clip=true,trim=6.5 8 5 7,scale=.62,right]{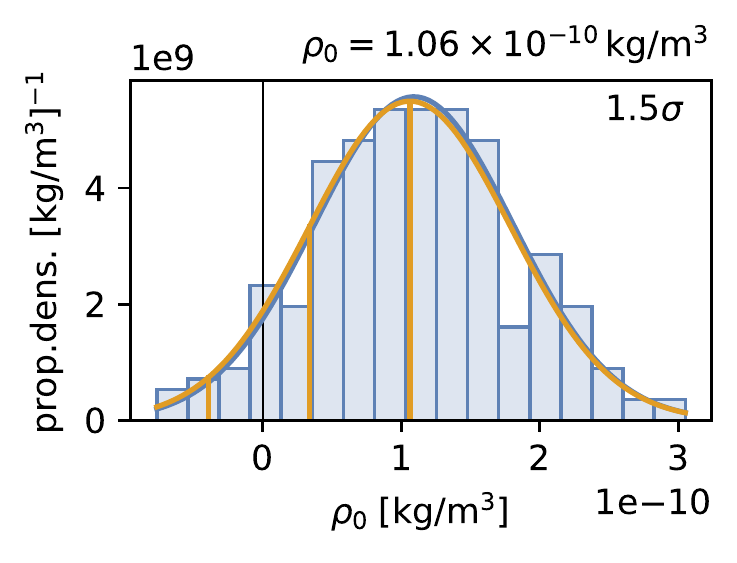}};
                \coordinate (o) at (2.8,6.1);
                \draw (o) circle (1);
                \draw [dashed](o) -- +(0,1);
                \draw [dashed](o) -- +(1,0);
                \node at ($ (o) + (39:.55)$)[align=left]{$\scriptstyle350$};
                \node at ($ (o) + (225:.5)$){$\scriptstyle50$};
        \end{tikzpicture}
        \caption{}
        \label{F: full orbit th b}
\end{subfigure}
\\
\begin{subfigure}{.495\columnwidth}
        \begin{tikzpicture}[scale=.4]
                \node[anchor=south west,inner sep=0] at (0,0)
                        {\includegraphics[width=\textwidth,clip=true,trim=6.5 8 5 -5,scale=.60,left]{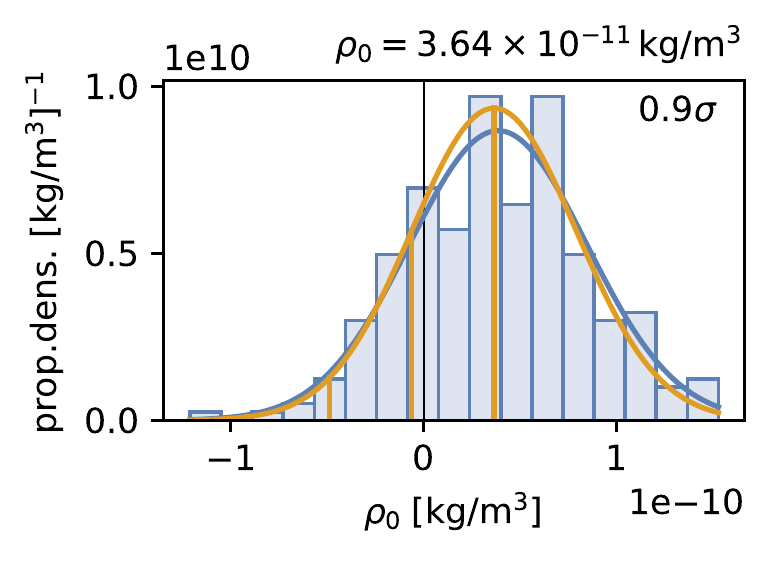}};
                \coordinate (o) at (3.4,5.8);
                \draw (o) circle (1);
                \node at (o){$\scriptstyle50$};
        \end{tikzpicture}
        \caption{}
        \label{F: full orbit th c}
\end{subfigure}
\begin{subfigure}{.495\columnwidth}
        \begin{tikzpicture}[scale=.4]
                \node[anchor=south west,inner sep=0] at (0,0)
                        {\includegraphics[width=\textwidth,clip=true,trim=6.5 8 5 -5,scale=.60,right]{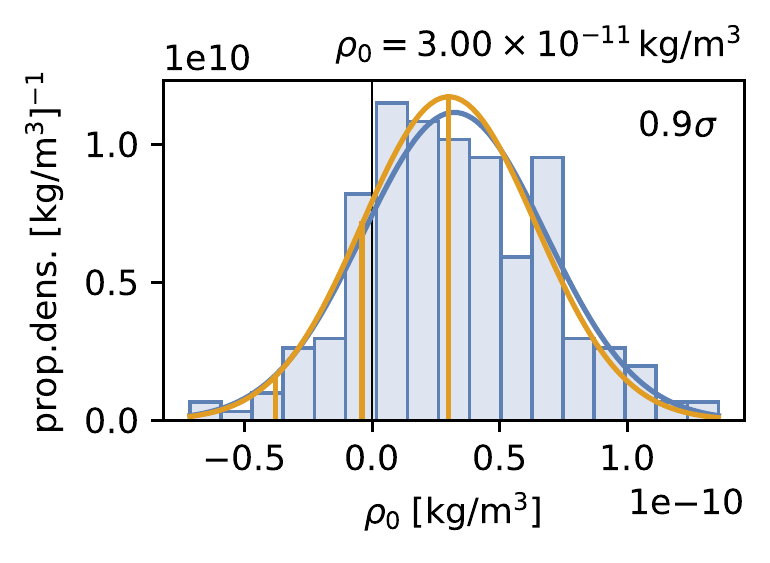}};
                \coordinate (o) at (3.4,5.8);
                \draw (o) circle (1);
                \draw [dashed](o) -- +(0,1);
                \draw [dashed](o) -- +(1,0);
                \node at ($ (o) + (45:.55)$)[align=left]{$\scriptstyle10$};
                \node at ($ (o) + (225:.5)$){$\scriptstyle50$};
        \end{tikzpicture}
        \caption{}
        \label{F: full orbit th d}
\end{subfigure}
\\
\begin{subfigure}{.495\columnwidth}
        \begin{tikzpicture}[scale=.4]
                \node[anchor=south west,inner sep=0] at (0,0)
                        {\includegraphics[width=\textwidth,clip=true,trim=6.5 8 5 -5,scale=.61,left]{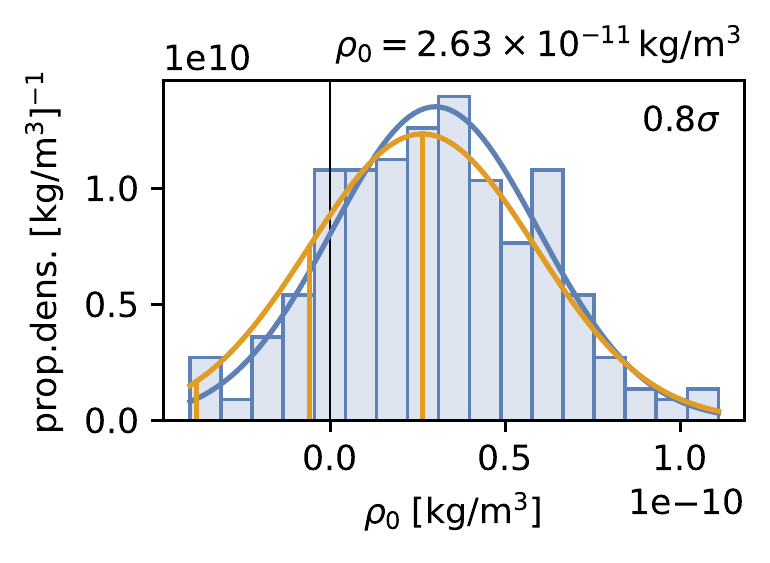}};
                \coordinate (o) at (3.32,5.8);
                \draw (o) circle (1);
                \draw [dashed]($ (o) + (-1,0) $) -- +(2,0);
                \node at ($(o) + (0,.5)$){$\scriptstyle10$};
                \node at ($(o) - (0,.5)$){$\scriptstyle50$};
        \end{tikzpicture}
        \caption{}
        \label{F: full orbit th e}
\end{subfigure}
\begin{subfigure}{.495\columnwidth}
        \begin{tikzpicture}[scale=.4]
                \node[anchor=south west,inner sep=0] at (0,0)
                        {\includegraphics[width=\textwidth,clip=true,trim=6.5 8 5 -5,scale=.61,right]{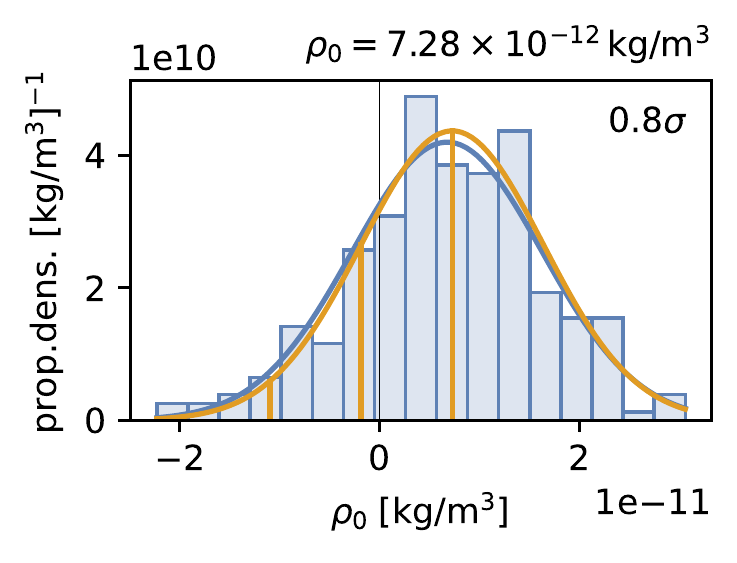}};
                \coordinate (o) at (3,6);
                \draw (o) circle (1);
                \node at (o){$\scriptstyle10$};
        \end{tikzpicture}
        \caption{}
        \label{F: full orbit th f}
\end{subfigure}
\caption{Density parameter statistics for all cases of Fig.~\ref{F: full orbit cases} with the mock data created using the density parameters corresponding to the detection thresholds of Table~\ref{Tbl: thresholds}, which also lists the corresponding masses enclosed by S2.}
\label{F: full orbit th}
\end{figure}

Next, we examined the detection thresholds estimated in Sect.~\ref{SS: detection thresholds}. For this we again used the cases of Fig.~\ref{F: full orbit cases}, but now with the density parameters of Table~\ref{Tbl: thresholds}. Following the same procedure as above results in Fig.~\ref{F: full orbit th}, from which we see that the estimated $1\sigma$ detection thresholds of Table~\ref{Tbl: thresholds} are accurate.
 %%%%%%%%%%%%%%%%%%%%%%%%%%%%%%%%%%%%%%%%%%%

\subsection{Less than one full orbit of data}\label{SS: partial orbit}

\begin{figure}%[H]
\centering
\begin{subfigure}{.495\columnwidth}
        \begin{tikzpicture}[scale=.4]
                \node[anchor=south west,inner sep=0] at (0,0)
                        {\includegraphics[width=\textwidth,clip=true,trim=6.5 8 5 7,scale=.62,left]{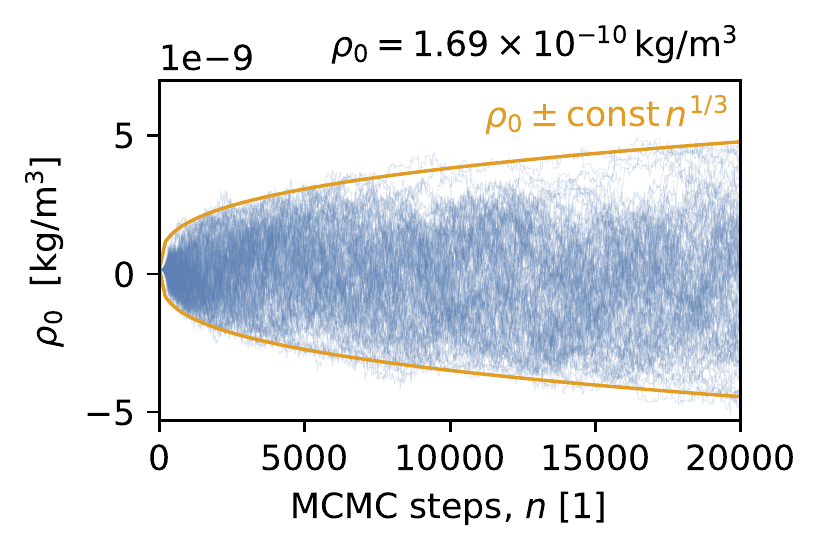}};
                \coordinate (o) at (3.4,6.15);
                \draw ($ (o) + (0.95,0) $) arc (0:-180:0.95);
                \draw ($ (o) + (1.05,0) $) arc (0:-180:1.05);
                \draw [dashed]($ (o) + (-1,0) $) -- ($ (o) + (1,0) $);
                \node at ($ (o) + (0,-.5) $){$\scriptstyle50$};
        \end{tikzpicture}
        \caption{}
        \label{F: remaining results a}
\end{subfigure}
\begin{subfigure}{.495\columnwidth}
        \begin{tikzpicture}[scale=.4]
                \node[anchor=south west,inner sep=0] at (0,0)
                        {\includegraphics[width=\textwidth,clip=true,trim=6.5 8 5 7,scale=.60,right]{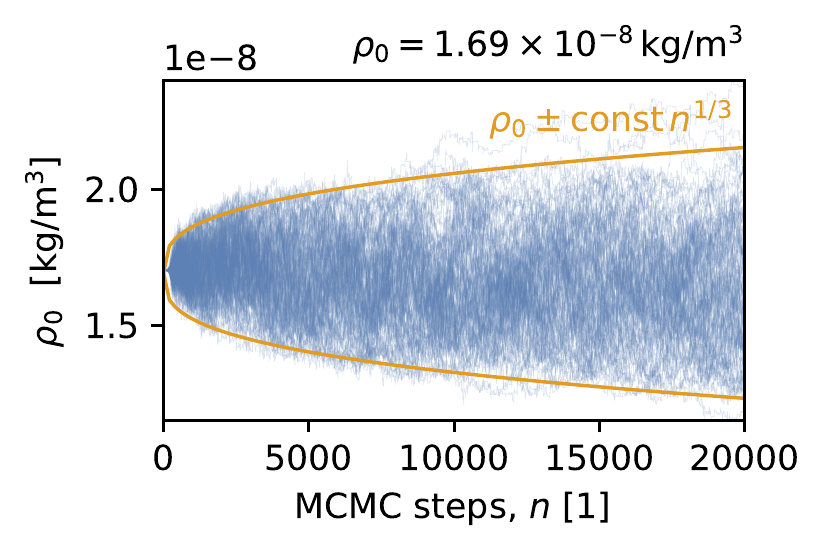}};
                \coordinate (o) at (3.4,6.15);
                \draw ($ (o) + (0.95,0) $) arc (0:-180:0.95);
                \draw ($ (o) + (1.05,0) $) arc (0:-180:1.05);
                \draw [dashed]($ (o) + (-1,0) $) -- ($ (o) + (1,0) $);
                \node at ($ (o) + (0,-.5) $){$\scriptstyle50$};
        \end{tikzpicture}
        \caption{}
        \label{F: remaining results b}
\end{subfigure}
\\
\begin{subfigure}{.495\columnwidth}
        \begin{tikzpicture}[scale=.4]
                \node[anchor=south west,inner sep=0] at (0,0)
                        {\includegraphics[width=\textwidth,clip=true,trim=6.5 8 5 -5,scale=.60,left]{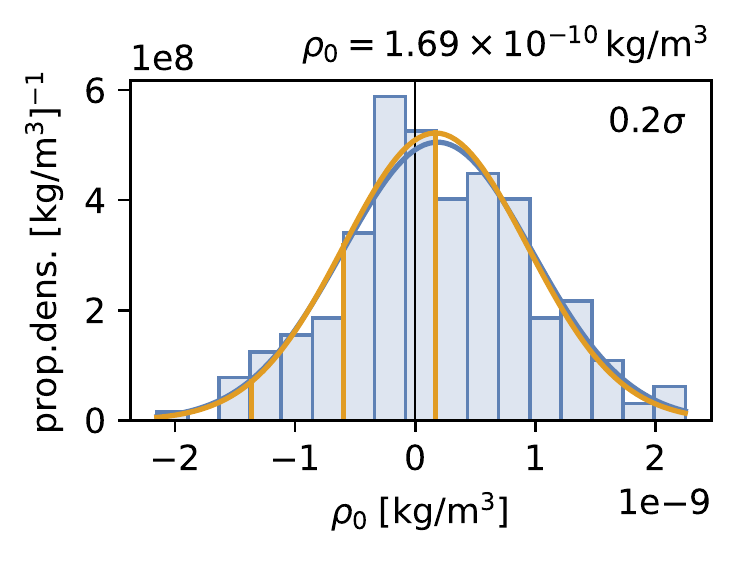}};
                \coordinate (o) at (3.4,6);
                \draw ($ (o) + (0.95,0) $) arc (0:180:0.95);
                \draw ($ (o) + (1.05,0) $) arc (0:180:1.05);
                \draw [dashed]($ (o) + (-1,0) $) -- ($ (o) + (1,0) $);
                \node at ($ (o) + (0,.5) $){$\scriptstyle50$};
        \end{tikzpicture}
        \caption{}
        \label{F: remaining results c}
\end{subfigure}
\begin{subfigure}{.495\columnwidth}
        \begin{tikzpicture}[scale=.4]
                \node[anchor=south west,inner sep=0] at (0,0)
                        {\includegraphics[width=\textwidth,clip=true,trim=6.5 8 5 -5,scale=.60,right]{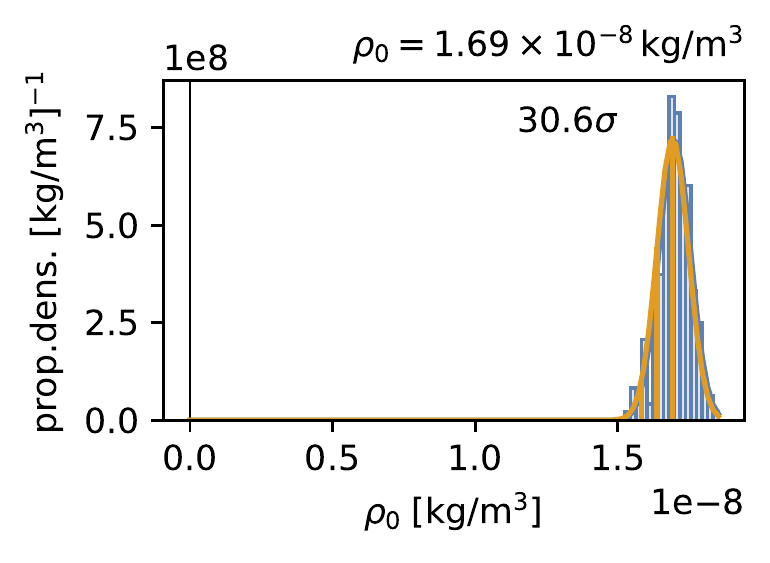}};
                \coordinate (o) at (4,5.7);
                \draw ($ (o) + (0.95,0) $) arc (0:180:0.95);
                \draw ($ (o) + (1.05,0) $) arc (0:180:1.05);
                \draw [dashed]($ (o) + (-1,0) $) -- ($ (o) + (1,0) $);
                \node at ($ (o) + (0,.5) $){$\scriptstyle50$};
        \end{tikzpicture}
        \caption{}
        \label{F: remaining results d}
\end{subfigure}
\\
\begin{subfigure}{.495\columnwidth}
        \begin{tikzpicture}[scale=.4]
                \node[anchor=south west,inner sep=0] at (0,0)
                        {\includegraphics[width=\textwidth,clip=true,trim=6.5 8 5 -5,scale=.60,left]{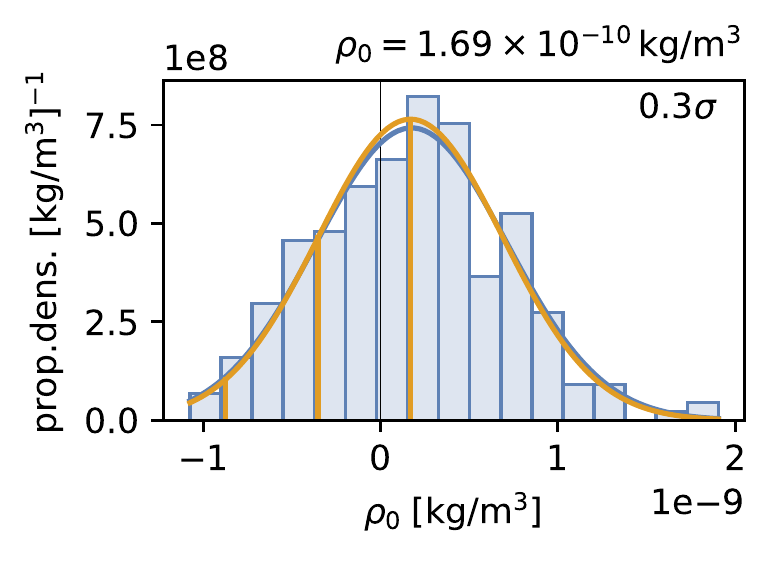}};
                \coordinate (o) at (3.4,5.8);
                \draw ($ (o) + (0,0.95) $) arc (90:270:0.95);
                \draw ($ (o) + (0,1.05) $) arc (90:270:1.05);
                \draw [dashed]($ (o) + (0,-1) $) -- ($ (o) + (0,1) $);
                \node at ($ (o) + (-.5,0) $){$\scriptstyle50$};
        \end{tikzpicture}
        \caption{}
        \label{F: remaining results e}
\end{subfigure}
\begin{subfigure}{.495\columnwidth}
        \begin{tikzpicture}[scale=.4]
                \node[anchor=south west,inner sep=0] at (0,0)
                        {\includegraphics[width=\textwidth,clip=true,trim=6.5 8 5 -5,scale=.62,right]{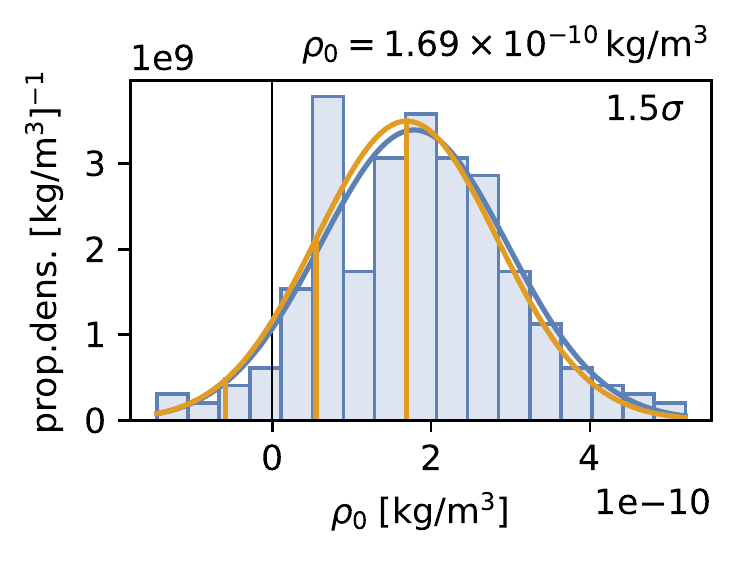}};
                \coordinate (o) at (2.89,6.1);
                \draw ($ (o) + (-1,0) $) arc (180:360:1);
                \draw ($ (o) + (0,0.95) $) arc (90:180:0.95);
                \draw ($ (o) + (0,1.05) $) arc (90:180:1.05);
                \draw [dashed](o) -- +(0,1);
                \draw [dashed](o) -- +(1,0);
                \node at ($ (o) + (225:.5)$){$\scriptstyle50$};
        \end{tikzpicture}
        \caption{}
        \label{F: remaining results f}
\end{subfigure}
\caption{Density parameter statistics for the cases with data limited to less than one full orbit of data, as indicated in the sketches in the top left corner. For data limited to the pericentre half, the MCMC chain for $\rho_0$ does not converge, but spreads out in a random walk. This means that these data are not sensitive to a dark mass, while in contrast, data limited to the apocentre half are; (\subref{F: remaining results c}), (\subref{F: remaining results d}). In panels~(\subref{F: remaining results a}) and~(\subref{F: remaining results d}), the density is increased by a factor $100$ such that $10\%$ of $M_\bullet$ are enclosed by S2. Panels (\subref{F: remaining results e}) and (\subref{F: remaining results f}) can be compared to Fig.~\ref{F: full orbit p1 c}, showing that a doubled sampling of the respective orbital sections cannot make up for the data gap to one full orbit.}
\label{F: remaining results}
\end{figure}
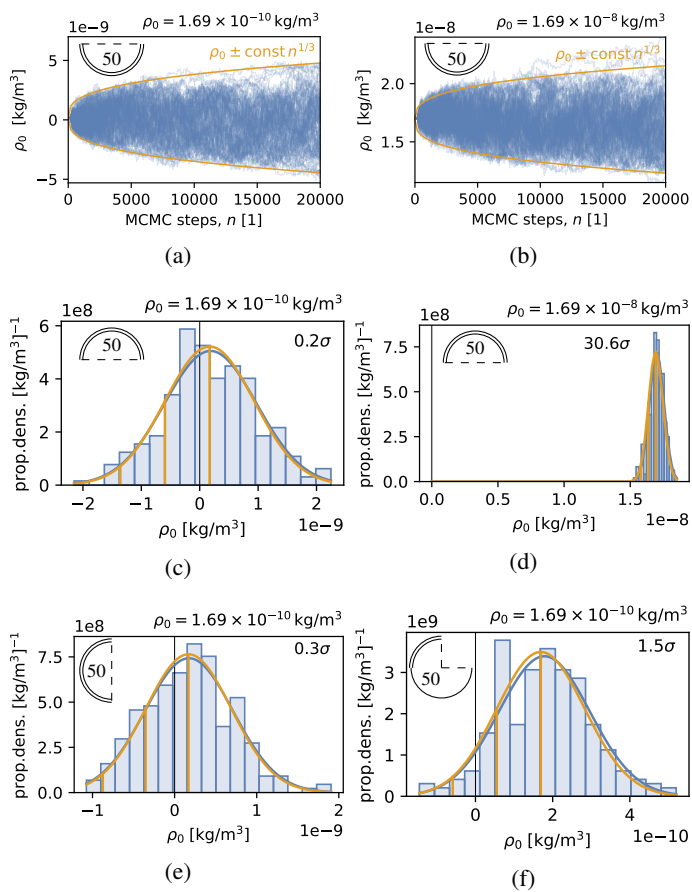

We now explore the dark mass sensitivity of data that are limited to half or three quarters of an orbit. The different cases are indicated in the sketches in the top left corner of the plots of Fig.~\ref{F: remaining results}, which are to be interpreted analogously to the cases of Fig.~\ref{F: full orbit cases}, just that now the data do not cover one full orbit and double lines indicate orbital sections in which the density of data points is doubled.
 %%%%%%%%%%%%%%%%%%%%%%%%%%%%%%%%%%%%%%%%%%%
Due to the latter, we have $114$ data points in total in theses cases as well, which allows a fair comparison of the results to the result obtained with a full orbit of data. Furthermore, it allows us to investigate whether missing data from one orbital section can be made up for by adding more data points in the respective opposite orbital section. The key point we are investigating here, however, is the importance of data in different orbital sections for constraining an extended mass.

We start with the case of $114$ mock observations in the pericentre half of the orbit, with $50\mathrm{\,\mu as}$ and $10\mathrm{\,km/s}$ precision. Fig.~\ref{F: remaining results a} shows that for our $0.1\%$ of $M_\bullet$ extended mass of Eq.~\eqref{E: rho0}, the MCMC chain fails to converge (i.e. to hit autocorrelation) even after $20\,000$ steps with $100$ walkers. As noted above, in all other cases, autocorrelation is reached already after only $500$ to~$1500$ steps. The failure of the MCMC fit to converge matches our theoretical observation of Sect.~\ref{S: effects} that data limited to the pericentre half are not sensitive to a dark mass. We elaborate further on the plausibility of this: Fig.~\ref{F: remaining results a} shows that the MCMC chain for the density parameter spreads out with an envelope of
\begin{align}\label{E: random walk}
\rho_0 \pm \mathrm{const}\,n^{1/3} , \quad \mathrm{const} = 1.7\times10^{-10}\mathrm{\,kg/m^3,}
\end{align}
where  $\rho_0$ denotes the value used to create the mock data and $n$ is the MCMC step. This is reminiscent of random walk behaviour, indicating that the walkers are sampling a flat posterior distribution, or (e.g.) the peak of a very broad Gaussian.\footnote{While an actual random walk spreads out $\propto n^{1/2}$ \citep{Feller1968}, we have to keep in mind that the emcee algorithm does not yield a pure random walk behaviour even when sampling a flat posterior because the walkers are not independent of each other, and so on \citep{GoodmanWeare10, Foreman-Mackey+13}.} We strengthened our case further by increasing the extended mass by a factor $100$, such that the mock data were created with $\rho_0=1.69\times10^{-8}\,\mathrm{kg/m^3}$ (i.e. $10\%$ of $M_\bullet$ or $\sim43\,000\,M_\odot$ within the apocentre), and repeating the analysis. With this density, the retrograde mass precession over one full orbit could easily be observed by naked eye. Despite this, Fig.~\ref{F: remaining results b} reveals that the chain fails to converge here as well and instead spreads out with precisely the same envelope (Eq.~\eqref{E: random walk}) as for the dark mass that is orders of magnitude lower. The fact that this picture did not change even after blowing up the extended mass by a factor $100$ strongly indicates that the posterior is indeed flat in both cases. We conclude that data limited to a pericentre half indeed are not able to constrain a present dark mass, even if it is relatively high. For the analysis of \citet{GRAVITY+20_Schwarzschild_prec}, this means that although its accuracy is orders of magnitude better, the GRAVITY dataset could not have constrained the dark mass on its own, but only in combination with the NACO and/or SINFONI data, simply because by itself at the time, it was limited to the pericentre half (see also the discussion in Sect.~\ref{SS: implications}).

We proceed to analyse the case of data limited to the apocentre half. The fits converge and yield the distributions of Figs.~\ref{F: remaining results c} and~\subref{F: remaining results d} for $0.1\%$ and $10\%$ of $M_\bullet$ extended masses, respectively. The data are strongly sensitive to the higher extended mass and reject $\rho_0=0$ with $30.6\sigma$. The lower extended mass is also constrained, but compared to Fig.~\ref{F: full orbit p1 c}, the constraints are much weaker than in the case of a full orbit of data with the same number of observations, that is, a null rejection of $0.2\sigma$ as opposed to $4.0\sigma$. We conclude that while data limited to the apocentre half are indeed sensitive to a dark mass, they are not strongly sensitive, and the loss of information from the pericentre half cannot be made up for by a denser sampling. We can understand this in the light of our discussion in Sect.~\ref{SS: detection thresholds}: If we do not have samples in the pericentre half, we lack information about the part of the trajectory that plays the role of the hypothetical orbit without an extended mass, and without this, we cannot measure deviations such as Eq.~\eqref{E: absAM} via which we infer the dark mass.

Next, we investigate the dark mass sensitivity of data limited to the orbital half from pericentre to apocentre, which are created with a $\rho_0$ given by Eq.~\eqref{E: rho0}. By intuition, we expect these data to be slightly more sensitive to a dark mass than the previous data because they catche both the peak and a tail of the curve in Fig.~\ref{F: absAM}. Indeed, a comparison of Figs.~\ref{F: remaining results e} and~\ref{F: remaining results c} shows a null rejection of $0.3\sigma$ versus $0.2\sigma$. Although only slightly more sensitive, this difference is significant because the standard deviations of the error statistics are sufficiently small (see the analogous situation in Fig.~\ref{F: case7 hist}). By comparison to Fig.~\ref{F: full orbit p1 c}, it is clear that the missing information from the opposite orbital half cannot be made up for by a denser sampling here either.

Finally, we study the case of three quarters of an orbit of data ending in the next apocentre, where the sampling is doubled in the last quarter. The $\rho_0$ at the base of the mock data is again given by~\eqref{E: rho0}. For this setting, we find a $1.5\sigma$ rejection of $\rho_0=0$ (Fig.~\ref{F: remaining results f}). This is a strong improvement in comparison to the preceding three cases, which only had half an orbit of data. A comparison to Fig.~\ref{F: full orbit p1 b} shows, however, that completing the orbit with data points still constrains the dark mass more strongly than a doubled sampling in the opposite quarter, even when the astrometric accuracy in the completing quarter is much lower. For the current observations of S2 with GRAVITY, this suggests that at least a fraction of NACO data will have to be added to the pool until a full orbit has been sampled with GRAVITY (and MICADO) until about 2033 in order to improve the dark mass constraints. Post $2033,$  GRAVITY data may be used alone to fully take advantage of the instrument's superior systematics (see the discussion in Sect.~\ref{SS: implications}).

\subsection{Flipped accuracies}\label{SS: flipped accuracies}

In the previous subsection, we confirmed amongst other results that data limited to a pericentre half are not sensitive to a dark mass, while data limited to an apocentre half are. However, we also showed that the latter sensitivity is weak, such that a full orbit of data is still much more constraining than an apocentre half with the same number of observations. We understood this phenomenon in the light of our discussion of Sect.~\ref{SS: detection thresholds} by the lack of information to measure discrepancies such as Eq.~\eqref{E: absAM} via which we detect the extended mass.

\begin{figure}%[H]
\centering
\begin{subfigure}{.495\columnwidth}
        \begin{tikzpicture}[scale=.4]
                \node[anchor=south west,inner sep=0] at (0,0)
                        {\includegraphics[width=\textwidth,clip=true,trim=6.5 8 5 7,scale=.60,left]{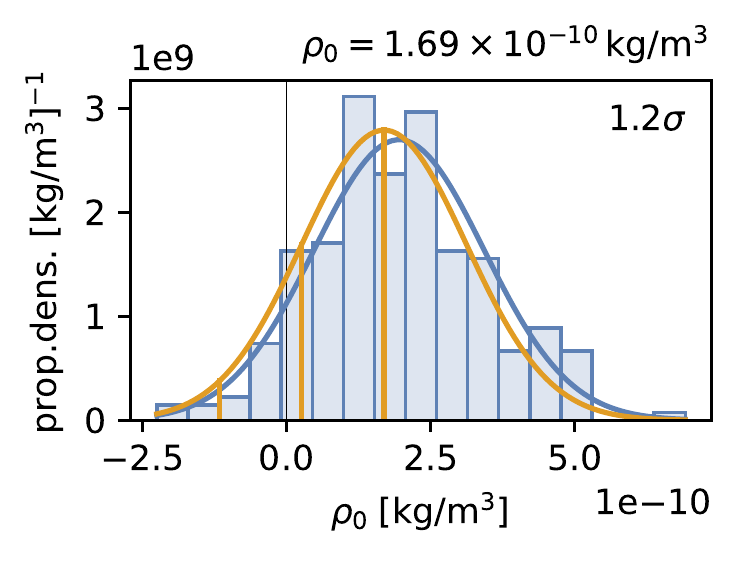}};
                \coordinate (o) at (3,6.1);
                \draw (o) circle (1);
                \draw [dashed]($ (o) + (-1,0) $) -- +(2,0);
                \node at ($(o) + (0,.5)$){$\scriptstyle50$};
                \node at ($(o) - (0,.5)$){$\scriptstyle350$};
        \end{tikzpicture}
        \caption{}
        \label{F: flipped results a}
\end{subfigure}
\begin{subfigure}{.495\columnwidth}
        \begin{tikzpicture}[scale=.4]
                \node[anchor=south west,inner sep=0] at (0,0)
                        {\includegraphics[width=\textwidth,clip=true,trim=6.5 8 5 7,scale=.62,right]{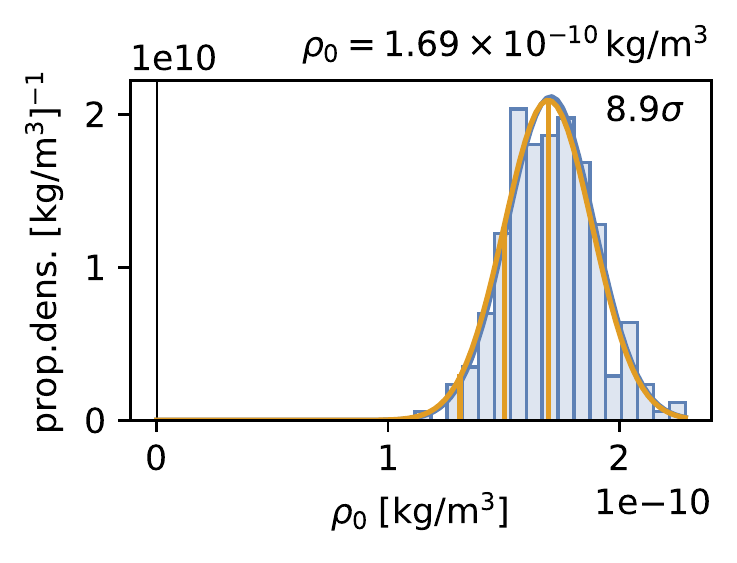}};
                \coordinate (o) at (3.4,6.1);
                \draw (o) circle (1);
                \draw [dashed]($ (o) + (-1,0) $) -- +(2,0);
                \node at ($(o) + (0,.5)$){$\scriptstyle50$};
                \node at ($(o) - (0,.5)$){$\scriptstyle10$};
        \end{tikzpicture}
        \caption{}
        \label{F: flipped results b}
\end{subfigure}
\caption{Density parameter statistics for the two cases indicated by the sketches in the top left corner. A comparison of the resulting null rejections to those of the mirror cases Figs.~\ref{F: full orbit p1 a} and~\ref{F: full orbit p1 e} shows that given one full orbit of data, the data in the pericentre half play the stronger role in constraining the dark mass than the data in the apocentre half.}
\label{F: flipped results}
\end{figure}
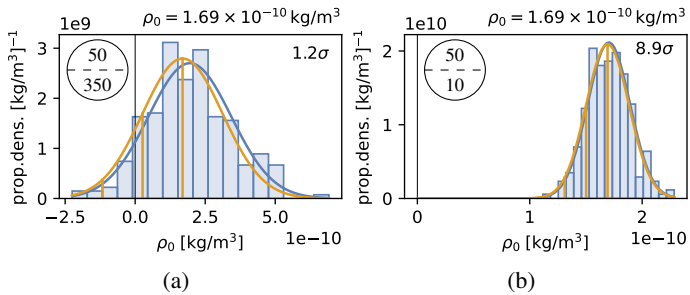

We now extend this investigation and study two further cases with $\rho_0$ given by Eq.~\eqref{E: rho0}, which are related to those of Figs.~\ref{F: full orbit a} and~\ref{F: full orbit e} by flipping the astrometric accuracies between the orbital halves. The results are shown in Fig.~\ref{F: flipped results}.
 %%%%%%%%%%%%%%%%%%%%%%%%%%%%%%%%%%%%%%%%%%%
Comparing Figs.~\ref{F: flipped results a} and~\ref{F: full orbit p1 a} as well as Figs.~\ref{F: flipped results b} and~\ref{F: full orbit p1 e}, we see that if we have a full orbit of data, with different astrometric accuracies in the pericentre and apocentre halves (the accuracies in $\mathrm{RV}$ remain $10\mathrm{\,km/s}$ everywhere), then the cases with better data in the pericentre half constrain the dark mass more strongly than their mirror cases. The difference is larger between the cases with $10\mathrm{\,\mu as}$ and $50\mathrm{\,\mu as}$ errors, but it is present in both comparisons. This result may seem counter-intuitive at first glance because in Sect.~\ref{SS: partial orbit} we clearly confirmed that the pericentre half by itself is not sensitive to a dark mass. Naively, we might therefore expect that the apocentre half is also the more constraining half when a full orbit of data is available. However, from our discussion in Sect.~\ref{SS: detection thresholds}, we understand that data in the pericentre half are necessary to constrain the hypothetical orbit without an extended mass in order to measure discrepancies such as Eq.~\eqref{E: absAM}, by which we determine the dark mass. Although the apocentre half is the half that is directly sensitive to the dark mass density, the pericentre half therefore appears to place the more stringent constraints on the initial osculating orbital elements, which indirectly seems to be the dominant factor in the overall constraining. In Table~\ref{Tbl: OE errors} we list the means of the error distributions (i.e. of the distributions that are analogous to that of Fig.~\ref{F: case7 hist b}) for the initial osculating orbital elements for the cases we just discussed. Indeed, the errors of the elements are greater by a factor~$2$ to~$3$ in the cases with larger errors in the apocentre half than in their mirror cases.
\begin{table*}%[hbt]
\caption{Mean errors for the initial osculating orbital elements for the cases compared in Sect.~\ref{SS: flipped accuracies}.}
\label{Tbl: OE errors}
\centering
\begin{tabular}{l l l l l l l}
\hline\hline
Case    & \multicolumn{6}{l}{Mean errors}               \\
                & $\sigma_{P_0}\,\mathrm{[yr]}$                 & $\sigma_{e_0}\,[1]$
                & $\sigma_{T_0}\,\mathrm{[yr]}$                 & $\sigma_{\Omega_0}\,\mathrm{[rad]}$
                & $\sigma_{\omega_0}\,\mathrm{[rad]}$   & $\sigma_{i_0}\,\mathrm{[rad]}$                                                        \\
\hline
Fig.~\ref{F: full orbit p1 a}   & $3.77\times10^{-3}$   & $4.91\times10^{-5}$   & $2.32\times10^{-3}$     & $3.57\times10^{-4}$
                                                & $4.50\times10^{-4}$   & $3.22\times10^{-4}$     \\
Fig.~\ref{F: flipped results a} & $6.12\times10^{-3}$   & $8.63\times10^{-5}$   & $2.97\times10^{-3}$     & $6.42\times10^{-4}$
                                                & $1.08\times10^{-3}$   & $4.79\times10^{-4}$     \\
\hline
Fig.~\ref{F: flipped results b}& $6.07\times10^{-4}$    & $1.01\times10^{-5}$   & $3.83\times10^{-4}$     & $8.42\times10^{-5}$
                                                & $1.02\times10^{-4}$   & $6.33\times10^{-5}$\\
Fig.~\ref{F: full orbit p1 e}   & $1.24\times10^{-3}$   & $2.84\times10^{-5}$   & $6.70\times10^{-4}$     & $2.39\times10^{-4}$
                                                & $3.41\times10^{-4}$   & $1.13\times10^{-4}$     \\
\hline
\end{tabular}
\tablefoot{Compared to their mirror cases, the errors of the elements are greater by a factor 2 to 3 in the cases with larger errors in the apocentre half.}
\end{table*}

\subsection{Correlations}\label{SS: correlations}

We conclude our analysis with an investigation of parameter correlations at the hand of corner plot representations of the posterior distributions.\footnote{To create the corner plots we use the python based corner module by \citet{Foreman-Mackey16}.} Our posterior distributions show already well-known correlations such as those between $M_\bullet$ and $R_0$, $\omega_0$ and $\Omega_0$ , and $\omega_0$ and $e_0$ \citep[e.g.][]{Gillessen+17, GRAVITY+20_Schwarzschild_prec}. We do not discuss these here, but instead focus on the correlations of the extended mass density $\rho_0$ with other parameters, as well as on the case-dependent overall qualitative features of the corner plots.

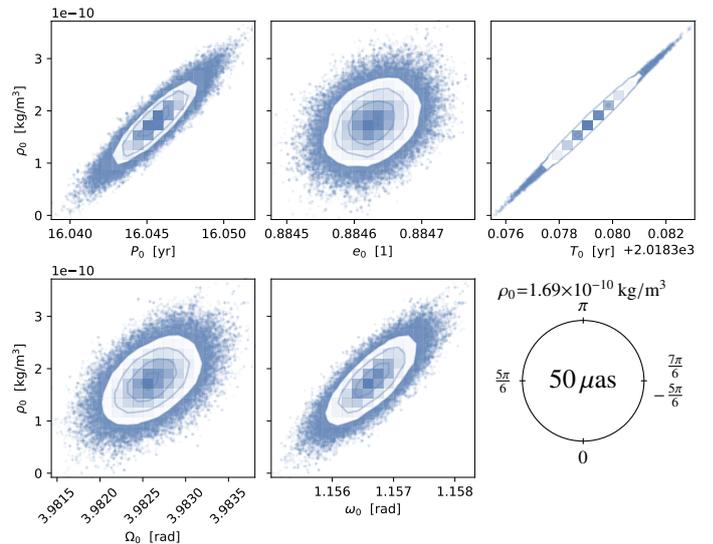
\begin{figure}%[H]
\centering
\begin{tikzpicture}[scale=.8]
        \node[anchor=south west,inner sep=0] at (0,0)
                {\includegraphics[clip=true,trim=7 8 6 7,scale=.49,width=\columnwidth]{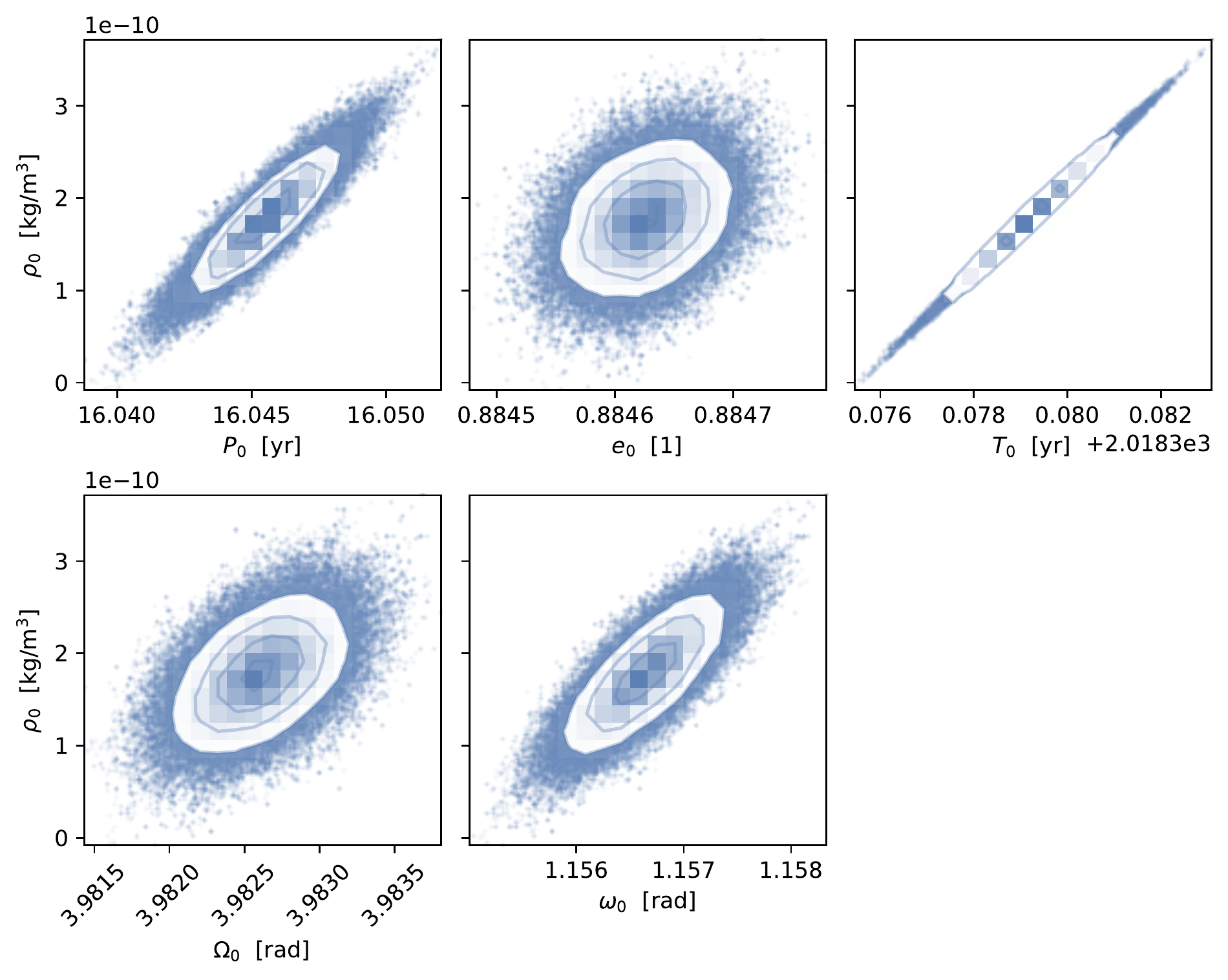}};
        \coordinate (o) at (9.4,2.7);
        \draw (o) circle (1);
        \node at (o){$50\mathrm{\,\mu as}$};
        \draw ($ (o) + (0,-1.05) $) -- ($ (o) + (0,-0.95) $);
        \node at ($ (o) + (0,-1) $)[anchor=north]{$\scriptstyle0$};
        \draw ($ (o) + (-1.05,0) $) -- ($ (o) + (-.95,0) $);
        \node at ($ (o) + (-1,0) $)[anchor=east]{$\scriptstyle\frac{5\pi}{6}$};
        \draw ($ (o) + (0,1.05) $) -- ($ (o) + (0,.95) $);
        \node at ($ (o) + (0,1) $)[anchor=south]{$\scriptstyle\pi$};
        \draw ($ (o) + (1.05,0) $) -- ($ (o) + (.95,0) $);
        \node at ($ (o) + (1,0) $)[anchor=west,align=left]{$\;\,\scriptstyle\frac{7\pi}{6}$\\$\scriptstyle-\frac{5\pi}{6}$};
        \node at ($ (o) + (0,1.5)$){$\scriptstyle\rho_0 = 1.69\times10^{-10}\mathrm{\,kg/m^3}$};
\end{tikzpicture}
\caption{Five correlations between the dark mass density $\rho_0$ and other parameters that are observed in all cases. The particular plots shown here correspond to a fit of the case of Fig.~\ref{F: full orbit p1 c} as indicated by the sketch (see Fig.~\ref{F: corner case1 p1 002} for the full corner plot).}
\label{F: correlations}
\end{figure}

Fig.~\ref{F: correlations} shows five correlations of $\rho_0$ with other parameters at the hand of the respective two-dimensional projections of the posterior distribution.
 %%%%%%%%%%%%%%%%%%%%%%%%%%%%%%%%%%%%%%%%%%%
These five correlations are present in all our cases. The diagrams plotted here correspond to a fit to a mock dataset of the case of Fig.~\ref{F: full orbit p1 c}. The corresponding full corner plot is shown in Fig.~\ref{F: corner case1 p1 002}. Evidently, $\rho_0$ very strongly correlates with the time of pericentre passage of the initial osculating orbit $T_0$. This is clear from the perspective that in order to fit the same data, a later initial time can to a certain extent be counteracted by a stronger retrograde in-plane precession, and thus by a higher dark mass density. In this context, it is also interesting to compare this correlation to that between $T_0$ and the parameter $f_\mathrm{SP}$ in~\citet[Appendix~E]{GRAVITY+20_Schwarzschild_prec}. $f_\mathrm{SP}$ controls the strength of the Schwarzschild precession, which is prograde, and hence these parameters are strongly anti-correlated. The correlation between $\rho_0$ and $P_0$ can also be understood via this link between the precession effect and a time shift. Similarly, we can also interpret the correlation between $\rho_0$ and the argument of pericentre of the initial osculating orbit $\omega_0$: A larger $\omega_0$ shifts the pericentre in the prograde sense, and thus can be countered by a higher retrograde precession, and consequently by a larger $\rho_0$. The weaker correlations of $\rho_0$ with $\Omega_0$ and $\rho_0$ with $e_0$ are carried over from the already well-known correlations of the respective latter parameters with $\omega_0$.

In terms of the overall qualitative features of the corner plots, we do not see variations within each case, but there are variations between the cases. Within the cases with a full orbit of data, these variations are small, but there is a clear contrast to the corner plots for cases with data that are limited to orbital sections, for which almost all parameters correlate mutually strongly. In Appendix~\ref{App: corner plots} we show three corner plots to demonstrate this.

\section{Discussion}\label{S: discussion}

In the following, we conclude with a short summary of the scope of our work (Sect.~\ref{SS: summary}) followed by a collection of our general results (Sect.~\ref{SS: general results}) and the implications and findings of direct relevance to present and future observation (Sect.~\ref{SS: implications}). Finally, we give a short outlook about the utility of our analysis in the wider scope of galactic centre science (Sect.~\ref{SS: outlook}).

\subsection{Summary}\label{SS: summary}

We explored how different precession effects of a star orbiting an MBH can clearly be separated from each other despite their net interference on secular timescales. The key for this was to determine their distinct signatures (in particular with respect to their locations), which they inscribe in the orbit on timescales of a single period. We focused on separating the secular interference of the (prograde) Schwarzschild and (retrograde) mass precessions, the former being the lowest-order relativistic deviation from Keplerian motion, and the latter being caused by a (dark) continuous extended mass distribution around the black hole.

We performed three layers of analysis: Firstly, we examined the impact of both precessions on the osculating orbital elements (in particular, on the argument of pericentre $\omega$) (Sect.~\ref{SS: orbital elements}). Secondly, we investigated the resulting impact on the observables, that is, on astrometry and radial velocity (Sect.~\ref{SS: astrometry and velocity}). Thirdly, we fitted model orbits to mock datasets of a series of cases, which differ by their astrometric accuracies in different orbital sections (Sect.~\ref{S: mock}).

\subsection{General results}\label{SS: general results}

In summary, based on the above analyses, we list our key findings below.
\begin{enumerate}
\item While the mass precession almost exclusively impacts the orbit in the apocentre half, the Schwarzschild precession almost exclusively impacts it in the pericentre half, allowing for a clear separation of the effects (Sects.~\ref{S: effects} and~\ref{S: mock}). \label{I: result 1}
\item Data that are limited to the pericentre half are not sensitive to a dark mass, while data limited to the apocentre half are sensitive to it, but only to a limited extent (Sects.~\ref{S: effects} and~\ref{S: mock}). \label{I: result 2}
\item A full orbit of data is required to substantially constrain a dark mass. We traced this back in part to the circumstance that the pericentre half data are required to properly constrain the base Keplerian osculating orbit elements (Sect.~\ref{S: mock}). \label{I: result 3}
\item Based on a full orbit of astrometric and spectroscopic data, the astrometric component in the pericentre half plays the stronger role in constraining the dark mass than the astrometric component in the apocentre half (Sect.~\ref{S: mock}). \label{I: result 4}
\end{enumerate}
An important general implication of these results is that despite their secular interference, the Schwarzschild and mass precessions can clearly be separated and measured independently of each other, for example, in a fit of a model comprising both components to a full orbit dataset of sufficient precision. Moreover, while these results have been obtained based on the orbit of S2 and with a variety of specific mock datasets, we have no reason to assume that they would not also hold in greater generality.

\subsection{Implications for observation}\label{SS: implications}

Beyond these general statements, we also gathered results that are directly relevant for current and future observations of S2 with VLTI/GRAVITY and with upcoming instruments such as GRAVITY+ \citep{Eisenhauer19} and ELT/MICADO. We repeat and expand on these points below.

Firstly, we interpret the current observational dark mass  $1\sigma$ upper bound of $0.1\%$ of $M_\bullet$ ($\sim4000\,M_\odot$) within the S2 orbit of \citet{GRAVITY+20_Schwarzschild_prec} in the light of our above results. Result~\ref{I: result 2} implies that despite its superior astrometric precision, the corresponding GRAVITY dataset could not have constrained the dark mass unilaterally because it was limited to the pericentre half of the orbit. Based on this and result~\ref{I: result 3}, the combination with NACO and/or SINFONI data was thus crucial. Furthermore, in the context of this combined dataset, we know from result~\ref{I: result 4} that the GRAVITY data have played the dominant role with regard to astrometry in constraining the dark mass.

Secondly, from these conclusions together with results~\ref{I: result 1} and~\ref{I: result 2}, it is clear that the dark mass sensitivity of this dataset will increase substantially as also the apocentre half becomes populated with GRAVITY data points in the coming years.  At the time of publication of this work (approximately $2021.96$), we have entered this domain (see the figures of Sect.~\ref{S: effects} and \citeauthor{GRAVITY+21_mass_distribution}~\citeyear{GRAVITY+21_mass_distribution}). We quantified this predicted gain in sensitivity by estimating the $1\sigma$ detection thresholds that can be achieved with one full orbit of data of different astrometric accuracies (Table~\ref{Tbl: thresholds} and Figs.~\ref{F: full orbit cases} and~\ref{F: full orbit th}). However, our theoretical cases and mock datasets have primarily been designed to extract the general statements of results~\ref{I: result 1}--\ref{I: result 4} and only secondarily to mimic past, present, and future observational datasets and systematics (see also the discussion in Sect.~\ref{SS: reservations}). With this caveat, our estimates are as follows: With GRAVITY data collected until the next apocentre passage (approximately in 2026), at the current performance of the instrument of $50\mathrm{\,\mu as}$, it will be possible to push the $1\sigma$ upper bound down to $\sim 3000\,M_\odot$ enclosed by S2. The bound in reach until the time of publication consequently lies somewhere between this value and the current bound. Once a full orbit of GRAVITY data is gathered (approximately in 2033), the $1\sigma$ upper bound can be pushed down to $\sim 1000\,M_\odot$ within the S2 orbit, without adding any NACO data to the pool.

Despite the importance of a compound of GRAVITY plus NACO and/or SINFONI data until 2033 (see result~\ref{I: result 3}), a post-2033 restriction to GRAVITY as the sole astrometric component would have certain advantages because we would then fully benefit from the edge that interferometry has in systematics compared to classical imaging with 10-metre class single-dish telescopes in the task of the astrometric tracking of S2. Firstly, in particular in the crowded environment of the galactic centre, astrometry with classical imaging is more strongly confusion limited \citep[e.g.][]{Trippe+10}. Secondly, classical imaging suffers from static and variable image distortions \citep{Plewa+15, Service+16}. Thirdly, GRAVITY measures the astrometric position of S2 as a direct offset from Sgr A*, and thus from the MBH, while classical imaging with a 10-metre class telescope does not, and has to bootstrap through the radio reference frame \citep{Plewa+15, Sakai+19}. Overall, while interferometry is certainly not free from systematics, astrometry from classical imaging involves more steps, each of which adds complexity and systematic uncertainties, which need to be modelled with additional free parameters \cite[Sect.~2]{GRAVITY+21_mass_distribution}.

Finally, we also estimated the degree to which the dark mass sensitivity could be further increased if the accuracy of GRAVITY were improved to $\pm10\,\mathrm{\mu as}$ (see Table~\ref{Tbl: thresholds} and Figs.~\ref{F: full orbit p1} and~\ref{F: full orbit th}). We found that this would have a particularly large impact once again one full orbit of data is gathered with this accuracy because by then, also the (in combination) more important pericentre half (see result~\ref{I: result 4}) would also be sampled with maximum precision. Then, the $1\sigma$ upper bound would be pushed down to merely $\sim 200\,M_\odot$. At this low order of magnitude, it is likely that the continuous extended mass model would have to be refined to reflect the possible graininess of a distribution that could be made up of faint stars, stellar remnants, black holes, and/or dark matter. New observational constraints on the latter could come from the GRAVITY+ upgrade as well as from ELT/MICADO starting in approximately 2026, as both instruments will be able to observe the galactic centre to significantly fainter magnitudes \citep[see also][]{GRAVITY+20_faint_stars}.

\subsection{Outlook}\label{SS: outlook}

For the larger picture, we hope that our work will serve as blueprint for analogous studies of the interference and separability of other effects. This should be useful in particular considering the challenging task of detecting the MBH spin via measuring the Lense-Thirring precession, which is the next great milestone in the field. Due to the smallness of the Lense-Thirring precession, it is of vital importance to examine how it interferes with other effects, and to explore how it can be separated from them \citep{WexKopeikin1999, Merritt2013, Zhang+15, Yu+16, Grould+17, Waisberg+18, Qi+21}.

% ########################################## Acknowledgements ###################################

\begin{acknowledgements}
We acknowledge the support of CNRS [PNCG, PNGRAM] and Paris Observatory [CS, PhyFOG].
\end{acknowledgements}

% ########################################## Bibliography ########################################

\bibliography{BibFile}
\bibliographystyle{aa}

% ########################################## Appendix ###########################################

\begin{appendix}

\section{Osculating equations}\label{App: osc equations}

The equations below are given in full generality, such that $m$ denotes the total mass of the two bodies. We recall that in the extreme mass ratio case of an MBH + star, the star can be treated as a test particle, such that formally, $m=M_\bullet$. The osculating equations (Eq.~\eqref{E: abstract osc equations}) in the form of~\citet[Eqs.~(3.64)--(3.66)]{PoissonWill2014} read
\begin{align}
\dot p &= \sqrt{\frac{p^3}{Gm}} \frac{2}{1+e\cos f}\mathcal S \label{E: p dot}\\
\dot e &= \sqrt{\frac{p}{Gm}} \left( \sin f\,\mathcal R + \frac{2\cos f+e(1+\cos^2f)}{1+e\cos f}\mathcal S \right) \\
\dot\iota &= \sqrt{\frac{p}{Gm}} \frac{\cos(\omega+f)}{1+e\cos f}\mathcal W \label{E: i dot}\\
\dot\Omega &= \sqrt{\frac{p}{Gm}} \frac{\sin(\omega+f)}{1+e\cos f} \frac{1}{\sin\iota} \mathcal W \label{E: Om dot} \\
\dot\omega &= \sqrt{\frac{p}{Gm}} \frac{1}{e} \Bigg( -\cos f\,\mathcal R+\frac{2+e\cos f}{1+e\cos f}\sin f\,\mathcal S \notag\\
&\quad- e\cot\iota\frac{\sin(\omega+f)}{1+e\cos f}\mathcal W \Bigg)  \label{E: om dot} \\
\dot f &= \sqrt{\frac{Gm}{p^3}}(1+e\cos f)^2 \notag\\
&\quad+\sqrt{\frac{p}{Gm}}\frac{1}{e}\left( \cos f\,\mathcal R-\frac{2+e\cos f}{1+e\cos f}\sin f\,\mathcal S \right),
 \label{E: f dot}
\end{align}
where a dot denotes a derivative with respect to time $t$, and where $\mathcal{R,S,W}$ denote the Gaussian components of the perturbative acceleration (see Eq.~\eqref{E: Gaussian components} and~Fig.~\ref{F: orientation angles}). The latter are in general functions of the orbital elements themselves.

As mentioned at the end of Sect.~\ref{SS: osculating equations} and shown in detail in \citet[Sect.~3.2.5]{PoissonWill2014}, the transformation law from the osculating orbital elements to $\mathbf r$ and $\mathbf v$ results from an Euler rotation around $\Omega, \iota$ and $\omega + f$,
\begin{align}\label{E: POS}
\mathbf r &= \underbrace{\frac{p}{1 + e\cos f}}_{r} \begin{bmatrix}
\cos\Omega\cos(\omega+f) - \cos\iota\sin\Omega\sin(\omega+f) \\
\sin\Omega\cos(\omega+f) + \cos\iota\cos\Omega\sin(\omega+f) \\
\sin\iota\sin(\omega+f)
\end{bmatrix} ,
\end{align}
\begin{align}
\mathbf v &= -\sqrt{\frac{Gm}{p}} \label{E: VEL}
\left[\begin{matrix}
\cos\Omega(\sin(\omega+f) + e\sin\omega) \\
\sin\Omega(\sin(\omega+f) + e\sin\omega) \\
-\sin\iota(\cos(\omega+f)+e\cos\omega)
\end{matrix}\right. \\ \notag
&\qquad\qquad\qquad\left.\begin{matrix}
+ \cos\iota\sin\Omega(\cos(\omega+f) + e\cos\omega) \\
- \cos\iota\cos\Omega(\cos(\omega+f) + e\cos\omega) \\
\phantom{=}
\end{matrix}\right] .
\end{align}

\section{Remarks about the model}\label{App: model remarks}

Sect.~\ref{S: model} presented our MBH + star + extended mass model in the form of a perturbed Kepler problem with a perturbative acceleration $\mathbf a_p = \mathbf a_\mathrm{1PN} + \mathbf a_\mathrm{XM}$. We thus correct the Newtonian two-body problem relativistically to 1PN order while treating the extended mass in a purely Newtonian manner. We comment on the domain of applicability of such a model below.

Firstly, the truncation at 1PN order for the two-body interaction is legitimate as long as the star orbits the MBH at distances much larger than its Schwarzschild radius. In the case of S2, the pericentre is $\sim1400$ Schwarzschild radii away from the black hole, such that post-Newtonian theory is well applicable.

Secondly, a purely Newtonian treatment of the extended mass distribution is justified as long as the error made by neglecting relativistic corrections for the extended mass is small compared to the modelled effects at play. This is the case as long as the extended mass is much lower than the mass of the MBH because then its 1PN correction will be negligible compared to the 1PN correction of the MBH, which is the only other modelled effect at play. As mentioned in the introduction, current upper bounds lie at the $0.1\%$ level of the MBH mass, and we can conclude that our model is adequate. (We argue here in analogy to \citet[p.~481]{PoissonWill2014}, who laid out the negligibility of PN effects stemming from third-body perturbations in the solar system.)

Finally, we compared our model to that of~\citet{RubilarEckart2001} and \cite{Takamori+20}, who included the extended mass directly into the Newtonian + 1PN two-body terms by replacing the MBH mass by an enclosed mass function of $r$. This is an elegant way to automatically include 1PN corrections for the extended mass in the model. However, it is ad hoc in the sense that it violates a criterion under which the 1PN two-body term has been derived in the first place: namely that the masses are isolated and well separated (\citeauthor{Merritt2013}~\citeyear{Merritt2013}, p.~132; \citeauthor{PoissonWill2014}~\citeyear{PoissonWill2014}, Sect.~9.3).

\section{Orbital angle conventions}\label{App: conventions}
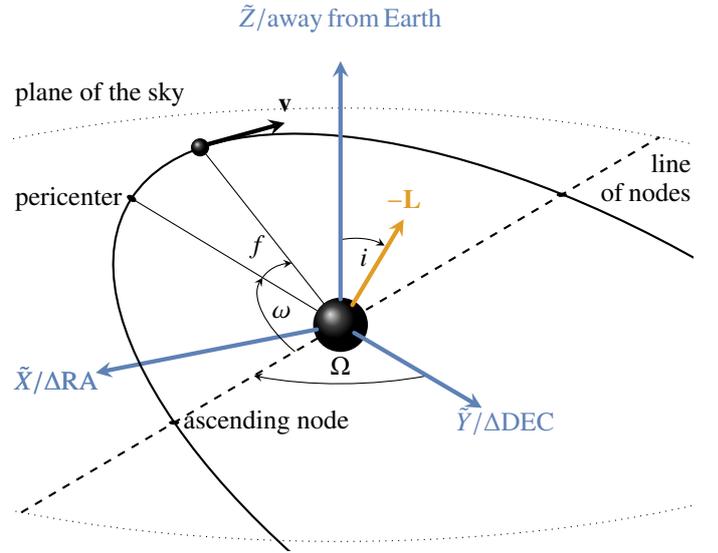
\begin{figure}%[H]

\definecolor{mmblue}{rgb}{0.368417, 0.506779, 0.709798} % Mathematica blue
\definecolor{mmorange}{rgb}{0.880722, 0.611041, 0.142051} % Mathematica orange

\tdplotsetmaincoords{70}{150} % point of view. args: \theta, \phi

% orbital rotation angles
\pgfmathsetmacro{\Om}{-60} % about Z axis
\pgfmathsetmacro{\inc}{-30} % about new X axis
\pgfmathsetmacro{\om}{-75} % about new Z axis
\pgfmathsetmacro{\f}{-35} % about (new) Z axis

\pgfmathsetmacro{\Ompnz}{30} % has to be 90 + \Om
\pgfmathsetmacro{\incmeps}{-34} % has to be \inc - ~3

\pgfmathsetmacro{\drawOm}{150} % has to be 90 - \Om
\pgfmathsetmacro{\drawOmp}{55} % has to be 90 + \Om + 25
\pgfmathsetmacro{\drawom}{165} % has to be 90 - \om
\pgfmathsetmacro{\drawinc}{120} % has to be 90 - \inc
\pgfmathsetmacro{\drawf}{55} % has to be 90 + \f

\pgfmathsetmacro{\aa}{.6} % semi-major axis, a
\pgfmathsetmacro{\bb}{.5} % semi-minor axis, b
\pgfmathsetmacro{\cc}{.33166} % centre to focus, c
\pgfmathsetmacro{\dd}{.26833} % focus to pericenter, c

\pgfmathsetmacro{\rSBH}{.03}    % SBH radius
\pgfmathsetmacro{\rstar}{.01}   % star radius

% def lengths of coordinate axes
\pgfmathsetmacro{\Xlength}{.31} \pgfmathsetmacro{\Xlabel}{.36}
\pgfmathsetmacro{\Ylength}{.31} \pgfmathsetmacro{\Ylabel}{.36}
\pgfmathsetmacro{\Zlength}{.31} \pgfmathsetmacro{\Zlabel}{.36}
\pgfmathsetmacro{\xlength}{.3} \pgfmathsetmacro{\xlabel}{.31}
\pgfmathsetmacro{\ylength}{.3} \pgfmathsetmacro{\ylabel}{.31}
\pgfmathsetmacro{\zlength}{.2} \pgfmathsetmacro{\zlabel}{.21}

\begin{tikzpicture}[scale=12,tdplot_main_coords]

% choose clipping
%\clip[tdplot_screen_coords] (-.72,-.57) rectangle (.83,.32); % clip all
%\clip[tdplot_screen_coords] (-.7,-.32) rectangle (.7,.32); % clip around plane of the sky
%\clip[tdplot_screen_coords] (-.63,-.28) rectangle (.4,.32); % clip around coordinate system
\clip[tdplot_screen_coords] (-.36,-.25) rectangle (.386,.355); % tight clip

% def coordinates
\coordinate (O) at (0,0,0);
\coordinate (X) at (0,\Xlength,0); \coordinate (Xlabel) at (0,\Xlabel,0);
\coordinate (Y) at (\Ylength,0,0); \coordinate (Ylabel) at (\Ylabel,0,0);
\coordinate (Z) at (0,0,\Zlength);  \coordinate (Zlabel) at (0,0,\Zlabel);

% fundamental frame (X,Y,Z)
\begin{scope}[-stealth,ultra thick,mmblue]
\draw (0.05,0,0) -- (Y) node at (Ylabel){$\tilde X/\Delta\mathrm{RA}$};
\end{scope}

% rotation by \Om
\tdplotsetrotatedcoords{\Om}{0}{0}
\begin{scope}[tdplot_rotated_coords]
\tdplotdrawarc[-stealth]{(O)}{.19}{\drawOm}{90}{anchor=south}{$\Omega$} % \Om angle

\draw[dotted] (O) circle (.7);                  % plane of the sky
\node at (319.5:.79) {plane of the sky};
\draw[thick,dashed] (0,.7,0) -- (0,-.7,0);
\node at (0,-.55,0) [anchor=west,align=right] {$\mathrm{line}$\\$\mathrm{of\,nodes}$}; % line of nodes
\draw[fill=black] (0,.365,0) circle (.005) node [anchor=west] {ascending~node};         % ascending node
\draw[fill=black] (0,-.485,0) circle (.005); % descending node
\end{scope}

% rotation by \inc
\tdplotsetrotatedcoords{\Om}{\inc}{0}

% rotate by \om
\tdplotsetrotatedcoords{\Om}{\inc}{\om}
\tdplotdrawarc[tdplot_rotated_coords,-stealth]{(O)}{.1}{\drawom}{90}{anchor=west}{$\omega$} % \om angle

\begin{scope}[tdplot_rotated_coords]
\draw (O) -- (\drawf:.287); % line SBH-star
\tdplotdrawarc[-stealth]{(O)}{.1}{90}{\drawf}{anchor=south east}{$f$} % \f angle
\draw[ultra thick,-stealth] (\drawf:.289) -- (30:.345) node [anchor=south]{$\mathbf v$};

\draw[thick] (0,-\cc,0) ellipse (.5 and \aa); % orbit
\draw[fill=black] (0,\dd,0) circle (.005) node [anchor=east] {$\mathrm{pericenter}$}; % pericenter
\end{scope}

% orbital frame, SBH and star
%\draw[tdplot_rotated_coords,red,thick,-stealth] (-\rSBH,0,0) -- (-\ylength,0,0) node at (-\ylabel,0,0){$y$};
\tdplottransformmainscreen{0}{0}{0}
\begin{scope}[tdplot_screen_coords]
\shade[ball color = black] (\tdplotresx,\tdplotresy) circle (\rSBH);    % SBH
\shade[ball color = black] (-.154,.196) circle (\rstar);                                % star
\end{scope}

\draw [-stealth,ultra thick,mmblue,line cap=round,tdplot_main_coords] (0,\rSBH,0) -- (X) node at (Xlabel){$\tilde Y/\Delta\mathrm{DEC}$};

\begin{scope}[tdplot_rotated_coords]
\draw (0,\rSBH,0) -- (0,\dd,0);
\draw[red,ultra thick,-stealth,mmorange,line cap=round] (0,0,\rSBH) -- (0,0,.16) node [anchor=south] {$-\mathbf L$};
\end{scope}

% draw inclination angle
\tdplotsetthetaplanecoords{0}
\tdplotdrawarc[tdplot_rotated_coords,-stealth]{(0,0,0)}{0.1}{0}{\incmeps}{anchor=north}{$i$} % \inc angle

\begin{scope}[ultra thick,mmblue,line cap=round]
\draw (\rSBH,0,0) -- (0.05,0,0) node at (Ylabel){$\tilde X/\Delta\mathrm{RA}$};
\draw [-stealth] (0,0,\rSBH) -- (Z) node at (Zlabel){$\tilde Z/\mathrm{away\,from\,Earth}$};
\end{scope}

\end{tikzpicture}

\caption{Alternative orbital angle convention. $\mathbf L$ denotes the orbital angular momentum.}
\label{F: MPE convention}
\end{figure}

When we introduced the angles $(\Omega,\iota,\omega)$ that give the orientation of an orbit as shown in Fig.~\ref{F: orientation angles} we followed the convention of modern theory based standard textbooks such as \citet{Merritt2013} or \citet{PoissonWill2014}. In particular, in the observation-based literature, the alternative convention shown in Fig.~\ref{F: MPE convention} is also frequently encountered (see also the literal description in~\citet[Appendix~A]{Paumard+06}).
 %%%%%%%%%%%%%%%%%%%%%%%%%%%%%%%%%%%%%%%%%%%
Both conventions agree with respect to $\Omega$ and $\omega$, but their inclination angles are taken in opposite directions: In Fig.~\ref{F: orientation angles}, $\iota$ is taken in the sense of a right-handed rotation around the line of node axes pointing towards the ascending node, while in Fig.~\ref{F: MPE convention}, $i$ is taken in the left-handed sense. Consequently, both conventions yield the same orbits when $\iota = - i$. Furthermore, in Fig.~\ref{F: orientation angles} (Fig.~\ref{F: MPE convention}) a star transitioning through the ascending node is moving towards (away from) Earth.

\section{Approximated R\o mer delay}\label{App: Roemer}

 Fig.~\ref{F: orientation angles} shows that modulo the time $R_0/c$ that the light travels from the MBH to Earth, the difference between the time of emission $\tilde t_\mathrm e$ and the time of observation on Earth $t_\mathrm o$ is given by
\begin{align}\label{E: Roemer}
t_\mathrm o - \tilde t_\mathrm e = -Z(\tilde t_\mathrm e)/c .
\end{align}
This is the R\o mer equation, which has to be solved implicitly for the $\tilde t_\mathrm e$ with $Z(\tilde t_\mathrm e)$ calculated by the chosen orbit model  if the full effect is taken into account. In practice, this can increase the computational cost and might become a problem in particular in tasks such as MCMC fitting, in which model orbits are created more frequently. However, at least for S2, we obtain an excellent approximation by substituting $Z(\tilde t_\mathrm e)$ by its truncated first-order Taylor series around $t_\mathrm o$, resulting in the approximated R\o mer equation
\begin{align}\label{E: approx Roemer}
t_\mathrm o - t_\mathrm e = -Z(t_\mathrm o)/c - \dot Z(t_\mathrm o) t_\mathrm e/c .
\end{align}
In contrast to Eq.~\eqref{E: Roemer}, Eq.~\eqref{E: approx Roemer} can be solved explicitly for $t_\mathrm e$, while the error $\tilde t_\mathrm e - t_\mathrm e$ is small. For S2, the maximum deviation in astrometry of one orbit calculated with the full versus the approximated R\o mer delay accounts for merely $1.2\times10^{-2}\mathrm{\,\mu as}$, which is orders of magnitude smaller than the effects we consider in this paper.

This approximation, as we used it here, has already been used in previous work \citep[Sect.~A.7]{GRAVITY+18_redshift}. \cite{Takamori+20} dropped the first-order term in Eq.~\eqref{E: approx Roemer}, by which the maximum deviation in astrometry within one orbit of S2 accounts for $\sim 2\mathrm{\,\mu as}$ for S2. However, the strongly increased accuracy by including the first-order term comes at negligible additional computational cost because $\dot Z$ is in any case calculated by the model via Eq.~\eqref{E: VEL}. Moreover, the accuracy might be increased even further by adding the second-order Taylor series term, which comes at negligible additional computational cost as well because the acceleration on the star is given a priori.

\section{Corner plots}\label{App: corner plots}

\begin{figure*}%[H]
\centering
\begin{tikzpicture}%[scale=.49]
        \node[anchor=south west,inner sep=0] at (0,0)
                {\includegraphics[clip=true,trim=7 8 6 6,scale=.49,width=\textwidth]{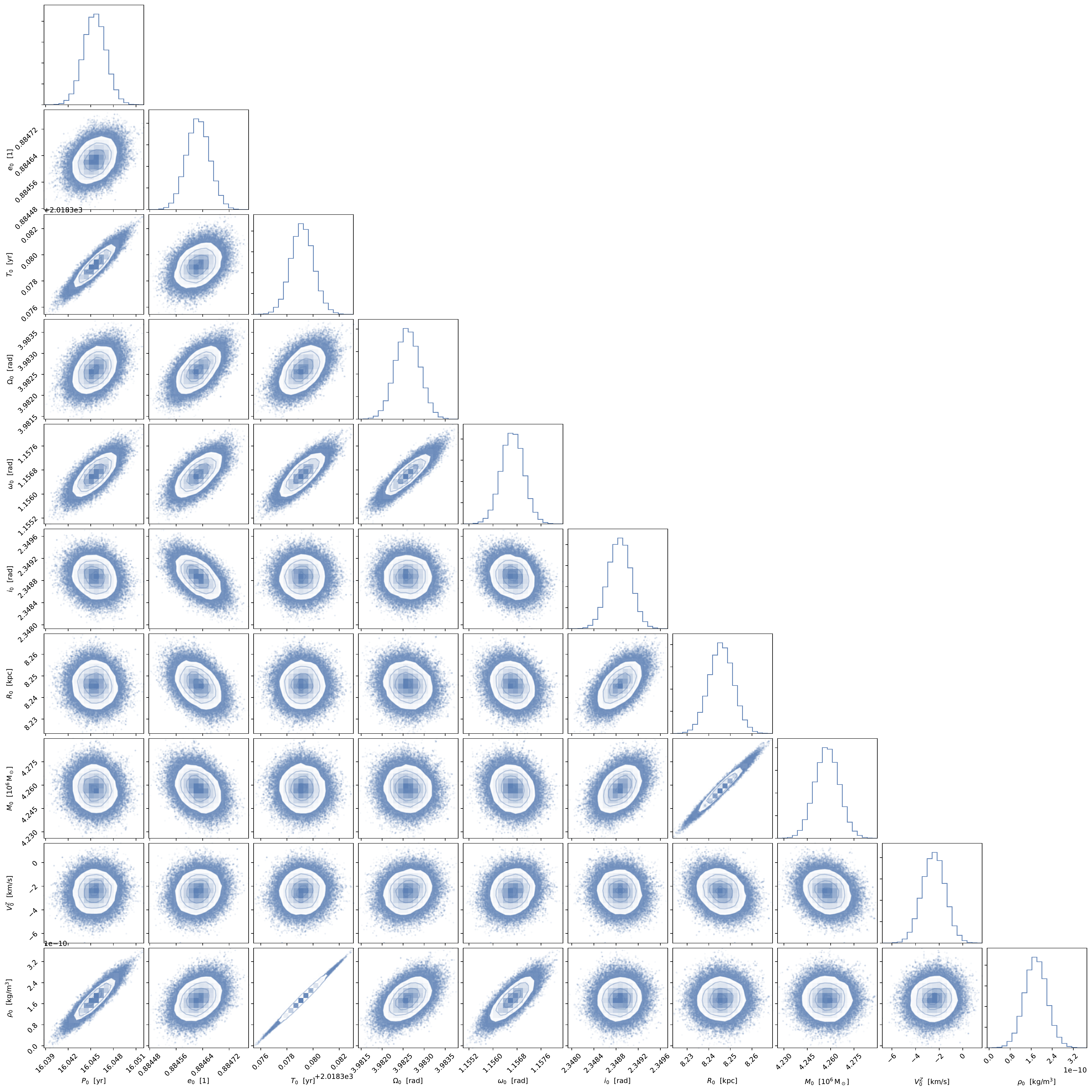}};
        \coordinate (o) at (14,14);
        \draw (o) circle (1);
        \node at (o){\small$50\mathrm{\,\mu as}$};
        \draw ($ (o) + (0,-1.05) $) -- ($ (o) + (0,-0.95) $);
        \node at ($ (o) + (0,-1) $)[anchor=north]{$\scriptstyle0$};
        \draw ($ (o) + (-1.05,0) $) -- ($ (o) + (-.95,0) $);
        \node at ($ (o) + (-1,0) $)[anchor=east]{$\scriptstyle\frac{5\pi}{6}$};
        \draw ($ (o) + (0,1.05) $) -- ($ (o) + (0,.95) $);
        \node at ($ (o) + (0,1) $)[anchor=south]{$\scriptstyle\pi$};
        \draw ($ (o) + (1.05,0) $) -- ($ (o) + (.95,0) $);
        \node at ($ (o) + (1,0) $)[anchor=west,align=left]{$\;\,\scriptstyle\frac{7\pi}{6}$\\$\scriptstyle-\frac{5\pi}{6}$};
        \node at ($ (o) + (0,1.5)$){$\scriptstyle\rho_0 = 1.69\times10^{-10}\mathrm{\,kg/m^3}$};
\end{tikzpicture}
\caption{Corner plot representation of the posterior distribution of a fit to a mock dataset corresponding to the case of Fig.~\ref{F: full orbit p1 c}, as indicated by the sketch in the top right corner.}
\label{F: corner case1 p1 002}
\end{figure*}

In Figs.~\ref{F: corner case1 p1 002}--\ref{F: corner case15 p1 x2 002} we present the corner plot representations of the posterior distributions for the cases of Figs.~\ref{F: full orbit p1 c}, \ref{F: remaining results c}, and~\ref{F: remaining results e}.
 %%%%%%%%%%%%%%%%%%%%%%%%%%%%%%%%%%%%%%%%%%%
\begin{figure*}%[H]
\centering
\begin{tikzpicture}%[scale=.49]
        \node[anchor=south west,inner sep=0] at (0,0)
                {\includegraphics[clip=true,trim=7 8 6 6,scale=.49,width=\textwidth]{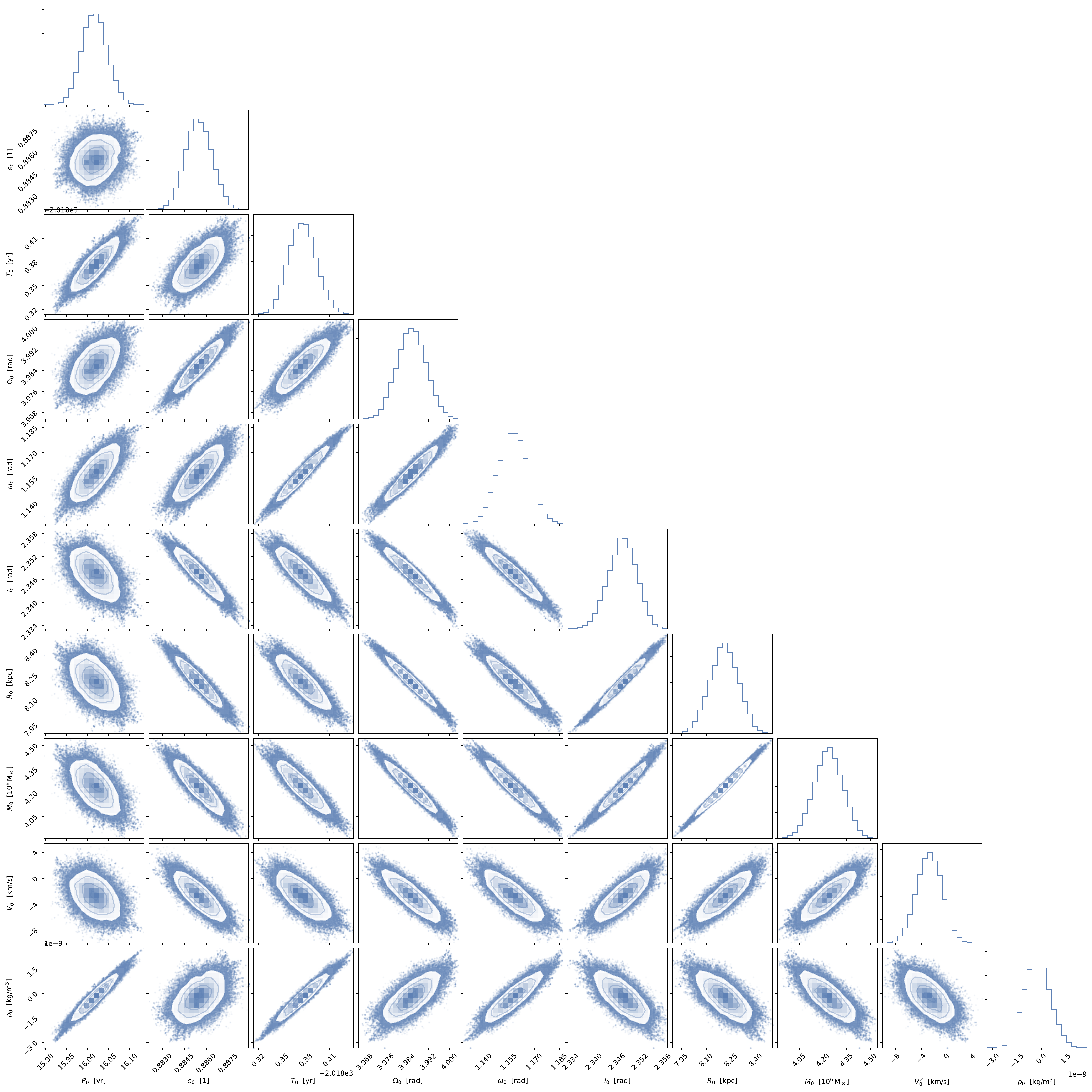}};
        \coordinate (o) at (14,14);
        \draw ($ (o) + (0.95,0) $) arc (0:180:0.95);
        \draw ($ (o) + (1.05,0) $) arc (0:180:1.05);
        \draw [dashed]($ (o) + (-1,0) $) -- ($ (o) + (1,0) $);
        \node at ($ (o) + (0,.4) $){\small$50\mathrm{\,\mu as}$};
        \draw ($ (o) + (-1.05,0) $) -- ($ (o) + (-.95,0) $);
        \node at ($ (o) + (-1,0) $)[anchor=east]{$\scriptstyle\frac{5\pi}{6}$};
        \draw ($ (o) + (0,1.05) $) -- ($ (o) + (0,.95) $);
        \node at ($ (o) + (0,1) $)[anchor=south]{$\scriptstyle\pi$};
        \draw ($ (o) + (1.05,0) $) -- ($ (o) + (.95,0) $);
        \node at ($ (o) + (1,0) $)[anchor=west,align=left]{$\scriptstyle\frac{7\pi}{6}$};
        \node at ($ (o) + (0,1.5)$){$\scriptstyle\rho_0 = 1.69\times10^{-10}\mathrm{\,kg/m^3}$};
\end{tikzpicture}
\caption{Corner plot representation of the posterior distribution of a fit to a mock dataset corresponding to the case of Fig.~\ref{F: remaining results c}, as indicated by the sketch in the top right corner. The plot for the case of Fig.~\ref{F: remaining results d} (i.e. with a density higher by $100$ times) looks qualitatively indistinguishable.}
\label{F: corner case12 p1 x2 002}
\end{figure*}
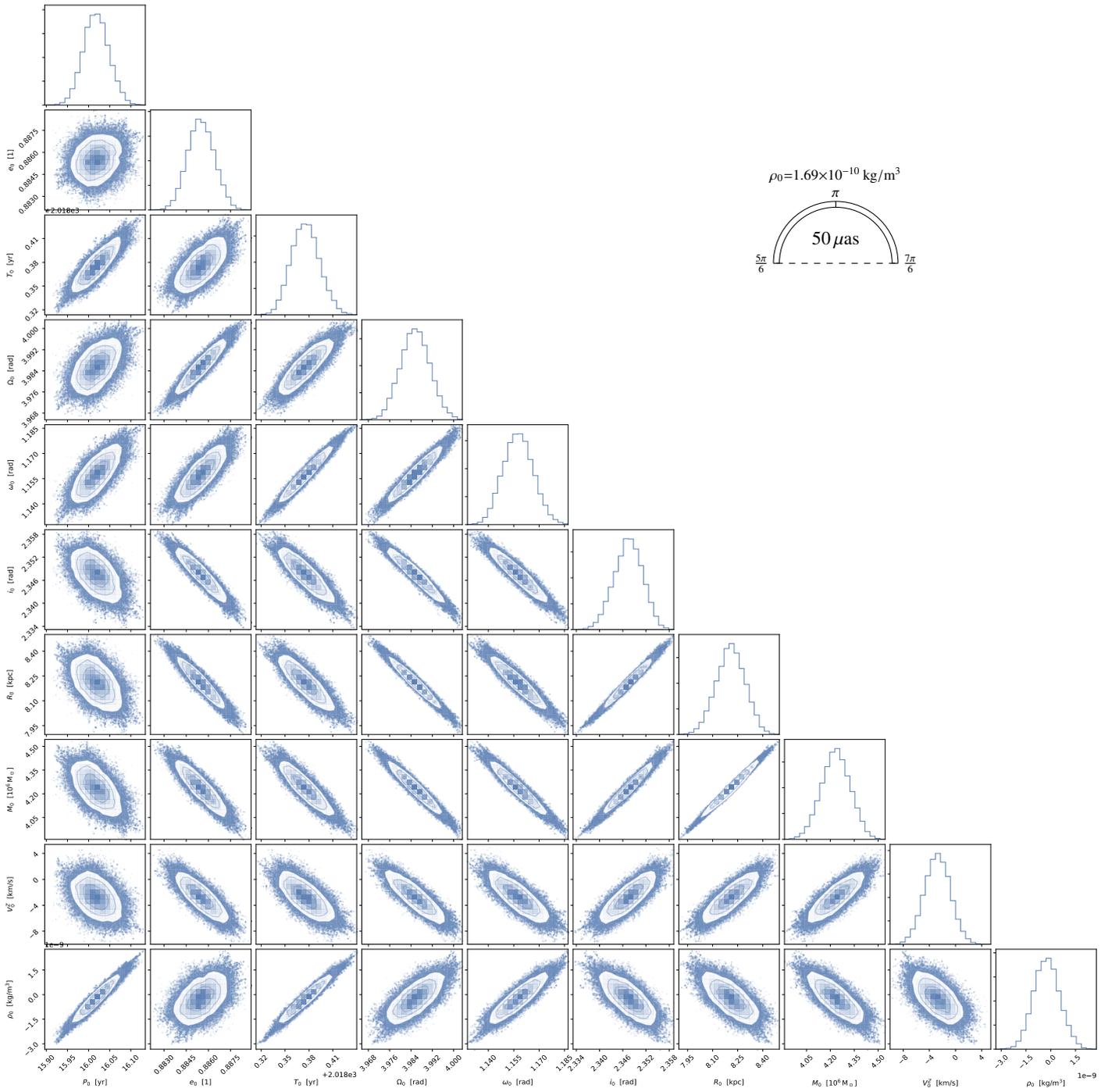 %%%%%%%%%%%%%%%%%%%%%%%%%%%%%%%%%%%%%%%%%%%
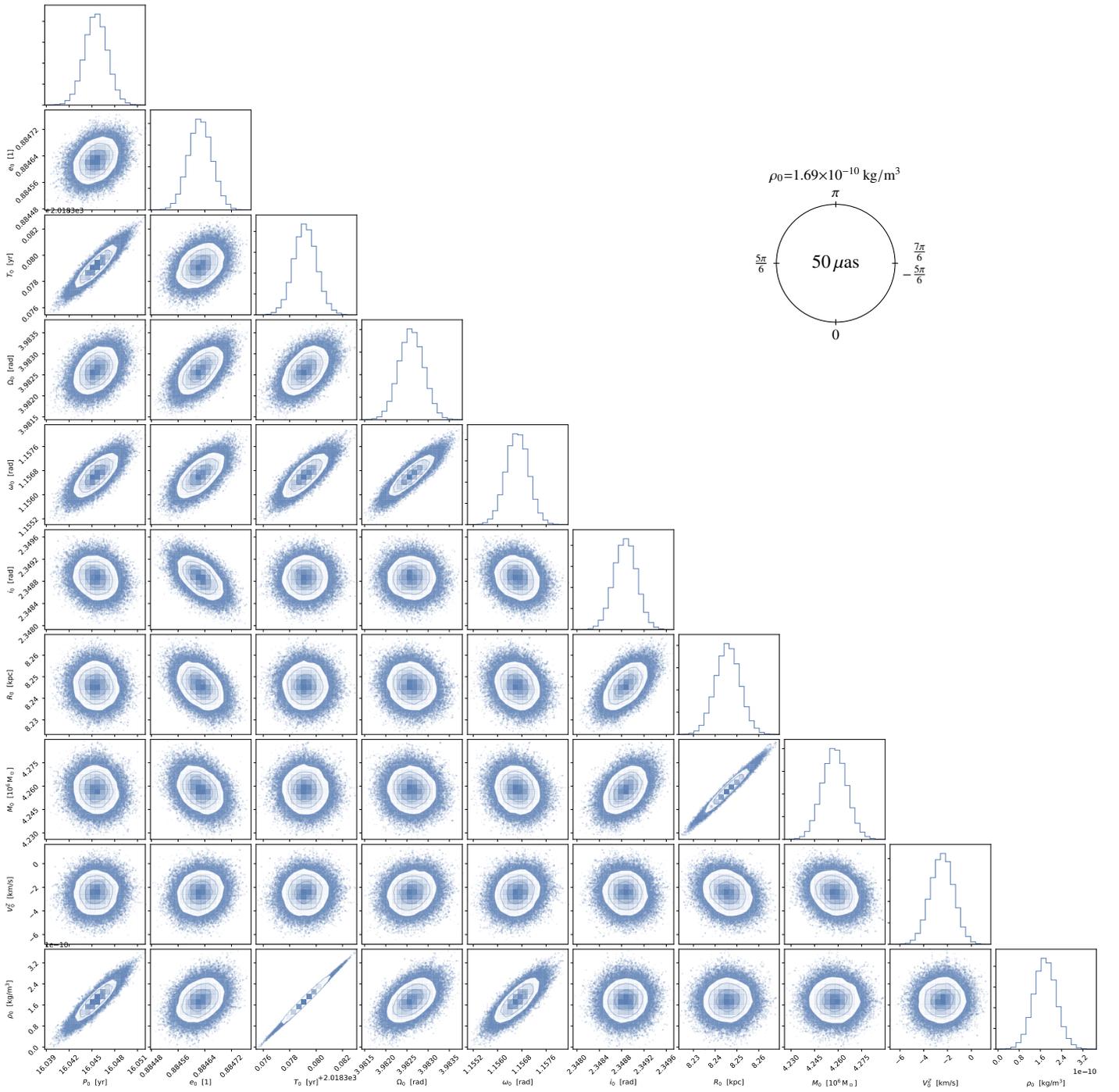
\begin{figure*}%[H]
\centering
\begin{tikzpicture}%[scale=.49]
        \node[anchor=south west,inner sep=0] at (0,0)
                {\includegraphics[clip=true,trim=7 8 6 6,scale=.49,width=\textwidth]{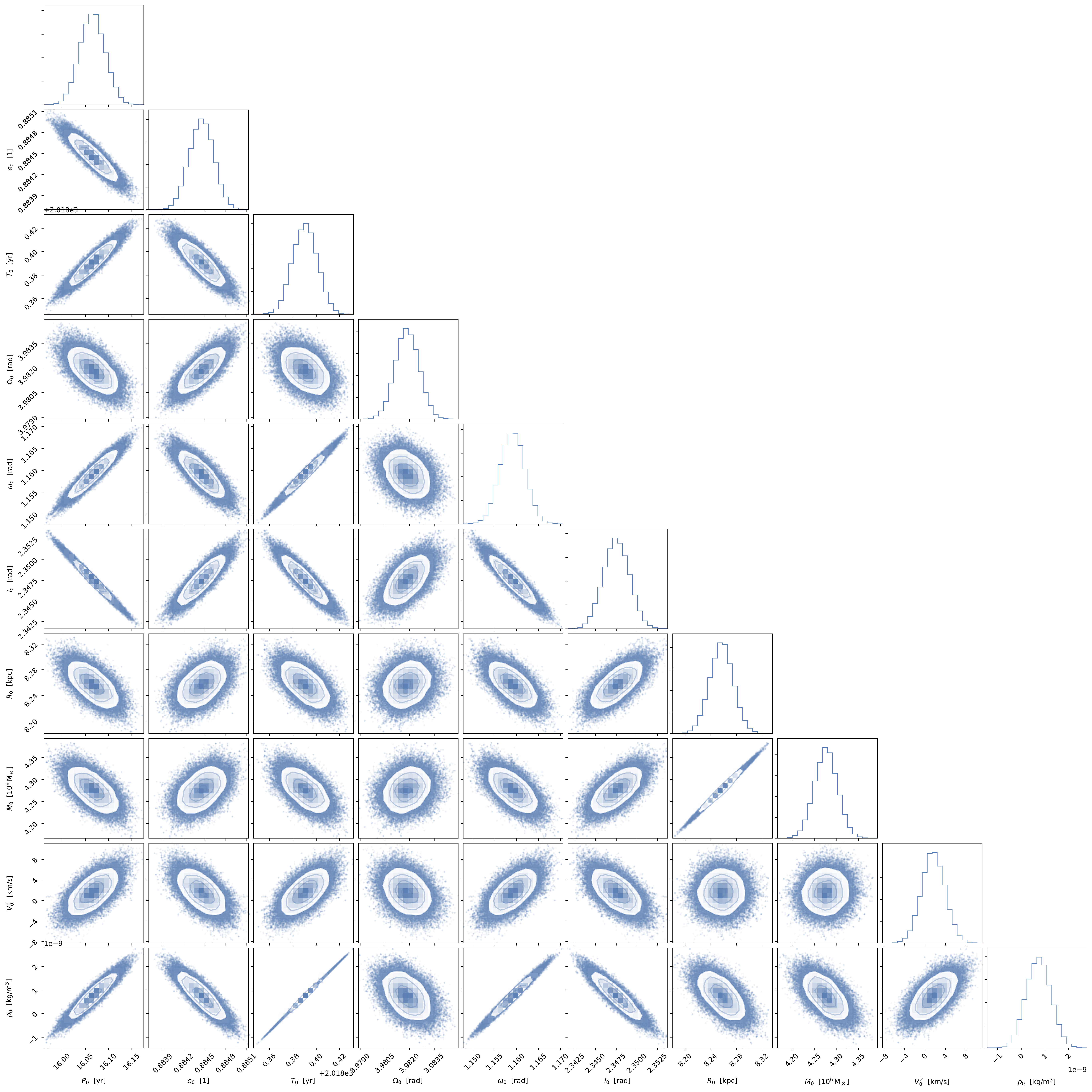}};
        \coordinate (o) at (14,14);
        \draw ($ (o) + (0,0.95) $) arc (90:270:0.95);
        \draw ($ (o) + (0,1.05) $) arc (90:270:1.05);
        \draw [dashed]($ (o) + (0,-1) $) -- ($ (o) + (0,1) $);
        \node at ($ (o) + (-.45,0) $){\small$50\mathrm{\,\mu as}$};
        \draw ($ (o) + (0,-1.05) $) -- ($ (o) + (0,-0.95) $);
        \node at ($ (o) + (0,-1) $)[anchor=north]{$\scriptstyle0$};
        \draw ($ (o) + (-1.05,0) $) -- ($ (o) + (-.95,0) $);
        \node at ($ (o) + (-1,0) $)[anchor=east]{$\scriptstyle\frac{5\pi}{6}$};
        \draw ($ (o) + (0,1.05) $) -- ($ (o) + (0,.95) $);
        \node at ($ (o) + (0,1) $)[anchor=south]{$\scriptstyle\pi$};
        \node at ($ (o) + (0,1.5)$){$\scriptstyle\rho_0 = 1.69\times10^{-10}\mathrm{\,kg/m^3}$};
\end{tikzpicture}
\caption{Corner plot representation of the posterior distribution of a fit to a mock dataset corresponding to the case of Fig.~\ref{F: remaining results e}, as indicated by the sketch in the top right corner.}
\label{F: corner case15 p1 x2 002}
\end{figure*} %%%%%%%%%%%%%%%%%%%%%%%%%%%%%%%%%%%%%%%%%%%
The plot for the cases of Fig.~\ref{F: remaining results d} with a density higher by $100$ times looks qualitatively indistinguishable from its lower-mass counterpart. We note that in these plots, the inclination $i$ follows the convention of Fig.~\ref{F: MPE convention}.

\end{appendix}

\end{document}